# SOME GENERAL EXPRESSIONS FOR THE COEFFICIENT OF THE 14$^{th}$ CHERN FORM

C. C. Briggs

*Center for Academic Computing, Penn State University, University Park, PA 16802*

Friday, April 16, 1999

**Abstract.** Some general expressions are given for the coefficient of the 14$^{th}$ Chern form in terms of the Riemann-Christoffel curvature tensor and some of its concomitants (e.g., Pontrjagin's characteristic tensors) for $n$-dimensional differentiable manifolds having a general linear connection.

PACS numbers: 02.40.-k, 04.20.Cv, 04.20.Fy

This paper presents some general expressions for the coefficient of the 14$^{th}$ Chern form in terms of the Riemann-Christoffel curvature tensor and some of its concomitants (e.g., Pontrjagin's characteristic tensors) for $n$-dimensional differentiable manifolds having a general linear connection.

Figuratively speaking, the $p^{th}$ Chern forms[1] $c_{(p)}$ representing the corresponding $p^{th}$ Chern classes of such a manifold $M$ can be defined by[2-3]

$$c_{(p)} \equiv \begin{cases} 1, & \text{if } p = 0 \\ \frac{i^p}{2^p \pi^p} \Omega_{[i_1}{}^{i_1} \wedge \Omega_{i_2}{}^{i_2} \wedge \ldots \wedge \Omega_{i_p]}{}^{i_p}, & \text{if } p > 0 \end{cases}, \quad (1)$$

where $\Omega_a{}^b$ is the curvature 2-form of $M$ defined by

$$\Omega_a{}^b \equiv \tfrac{1}{2} R_{cda}{}^b \, \omega^c \wedge \omega^d, \quad (2)$$

where $R_{abc}{}^d$ is the Riemann-Christoffel curvature tensor of $M$ defined by[4]

$$R_{abc}{}^d \equiv 2 \, (\partial_{[a} \, \Gamma_{b]}{}^d{}_c + \Gamma_{[a|e|}{}^d \, \Gamma_{b]}{}^e{}_c + \Omega_{a\ b}^{\ e} \, \Gamma_{e}{}^d{}_c), \quad (3)$$

where $\Gamma_a{}^b{}_c$ is the connection coefficient, $\Omega_a{}^b{}_c$ the object of anholonomity, and $\omega^a$ the basis 1-form of $M$. Thus, the $2p$-forms $c_{(p)}$ for $p \geq 1$ are given by

$$c_{(p)} = c_{(p)i_1 i_2 \ldots i_{2p}} \, \omega^{i_1} \wedge \omega^{i_2} \wedge \ldots \wedge \omega^{i_{2p}}, \quad (4)$$

where the coefficients $c_{(p) i_1 i_2 \ldots i_{2p}}$ of $c_{(p)}$ for $p \geq 1$ are given by

$$c_{(p) i_1 i_2 \ldots i_{2p}} = \frac{1}{(2p)!} \langle \mathsf{e}_{i_1} \wedge \mathsf{e}_{i_2} \wedge \ldots \wedge \mathsf{e}_{i_{2p}}, c_{(p)} \rangle \quad (5)$$

$$= \frac{1}{(2p)!} c_{(p) j_1 j_2 \ldots j_{2p}} \langle \mathsf{e}_{i_1} \wedge \mathsf{e}_{i_2} \wedge \ldots \wedge \mathsf{e}_{i_{2p}}, \omega^{j_1} \wedge \omega^{j_2} \wedge \ldots \wedge \omega^{j_{2p}} \rangle$$

$$= \frac{1}{(2p)!} \delta^{j_1 j_2 \ldots j_{2p}}_{i_1 i_2 \ldots i_{2p}} c_{(p) j_1 j_2 \ldots j_{2p}}$$

$$= c_{(p)[i_1 i_2 \ldots i_{2p}]}$$

$$= \frac{i^p}{2^p \pi^p (2p)!} \langle \mathsf{e}_{i_1} \wedge \mathsf{e}_{i_2} \wedge \ldots \wedge \mathsf{e}_{i_{2p}}, \Omega_{[j_1}{}^{j_1} \wedge \Omega_{j_2}{}^{j_2} \wedge \ldots \wedge \Omega_{j_p]}{}^{j_p} \rangle$$

$$= \frac{i^p}{2^{2p} \pi^p (2p)!} R_{k_1 k_2 j_1}{}^{[j_1} R_{k_3 k_4 j_2}{}^{j_2} \ldots R_{k_{2p-1} k_{2p} j_p}{}^{j_p]} \langle \mathsf{e}_{i_1} \wedge \mathsf{e}_{i_2} \wedge \ldots \wedge \mathsf{e}_{i_{2p}}, \omega^{k_1} \wedge \omega^{k_2} \wedge \ldots \wedge \omega^{k_{2p}} \rangle$$

$$= \frac{i^p}{2^{2p} \pi^p (2p)!} \delta^{k_1 k_2 \ldots k_{2p}}_{i_1 i_2 \ldots i_{2p}} R_{k_1 k_2 j_1}{}^{[j_1} R_{k_3 k_4 j_2}{}^{j_2} \ldots R_{k_{2p-1} k_{2p} j_p}{}^{j_p]}$$

$$= \frac{i^p}{2^{2p} \pi^p} R_{[i_1 i_2 |j_1|}{}^{[j_1} R_{i_3 i_4 |j_2|}{}^{j_2} \ldots R_{i_{2p-1} i_{2p}] j_p}{}^{j_p]},$$

where $\mathsf{e}_a$ is the basis tangent vector of $M$ dual to $\omega^a$, i.e. $\langle \mathsf{e}_b, \omega^a \rangle = \delta^a_b$, where $\delta^a_b$ is the Kronecker delta, and

$$\langle \mathsf{e}_{i_1} \wedge \mathsf{e}_{i_2} \wedge \ldots \wedge \mathsf{e}_{i_{2p}}, \omega^{j_1} \wedge \omega^{j_2} \wedge \ldots \wedge \omega^{j_{2p}} \rangle = \delta^{j_1 j_2 \ldots j_{2p}}_{i_1 i_2 \ldots i_{2p}}, \quad (6)$$

where $\delta^{j_1 j_2 \ldots j_{2p}}_{i_1 i_2 \ldots i_{2p}}$ is the generalized Kronecker delta.

Some numerical properties of $c_{(p) i_1 i_2 \ldots i_{2p}}$ for $p = 14$ appear in Table 1. Some general expressions for $c_{(p) i_1 i_2 \ldots i_{2p}}$ for $p = 14$ appear in Eq. (9), the expressions being given (1$^{st}$) in terms of $R_{abc}{}^d$, (2$^{nd}$) in terms of $R_{abc}{}^d$ and Schouten's tensor $V_{ab}$ defined by[5]

$$V_{ab} \equiv R_{abc}{}^c = \nabla_{[a} Q_{b]c}{}^c + S_{ab}{}^d Q_{dc}{}^c, \quad (7)$$

where $S_{ac}{}^b$ is the torsion tensor defined by $S_{ac}{}^b \equiv \Gamma_{[a\ c]}{}^b + \Omega_a{}^b{}_c$ and $Q_a{}^{bc}$ the non-metricity tensor defined by $Q_a{}^{bc} \equiv \nabla_a g^{bc}$, where $g^{ab}$ is the contravariant metric tensor, and (3$^{rd}$) in terms of Pontrjagin's characteristic tensors (from the Russian "характеристические тензоры") $P^{(2p)}{}_{i_1 i_2 \ldots i_{2p}}$ defined by[6-7]

$$P^{(2p)}{}_{i_1 i_2 \ldots i_{2p}} \equiv \frac{1}{2^p} R_{[i_1 i_2 | j_2 |}{}^{j_1} R_{i_3 i_4 | j_3 |}{}^{j_2} \ldots \quad (8)$$

$$\ldots R_{i_{2p-3} i_{2p-2} | j_p |}{}^{j_{p-1}} R_{i_{2p-1} i_{2p}] j_1}{}^{j_p}.$$

TABLE 1. SOME NUMERICAL PROPERTIES OF $c_{(p) i_1 i_2 \ldots i_{2p}}$ FOR $p = 14$

| QUANTITY | ORDER | CURVATURE DEPENDENCE | MINIMUM DIMENSIONALITY | NUMBER OF TERMS | 1$^{st}$ & 2$^{nd}$ OVERALL NUMERICAL FACTORS | 3$^{rd}$ OVERALL NUMERICAL FACTOR |
|---|---|---|---|---|---|---|
| $c_{(p) i_1 i_2 \ldots i_{2p}}$ | $p$ | — | $2p$ | — | $\dfrac{i^p}{2^{2p} \pi^p p!}$ | $\dfrac{i^p}{2^p \pi^p p!}$ |
| $c_{(14) i_1 i_2 i_3 i_4 i_5 i_6 i_7 i_8 i_9 i_{10} i_{11} i_{12} i_{13} i_{14} i_{15} i_{16} i_{17} i_{18} i_{19} i_{20} i_{21} i_{22} i_{23} i_{24} i_{25} i_{26} i_{27} i_{28}}$ | 14 | Quattuordecic | 28 | **135** | $-\dfrac{1}{23{,}401{,}744{,}351{,}572{,}787{,}200 \pi^{14}}$ | $-\dfrac{1}{1{,}428{,}329{,}123{,}020{,}800 \pi^{14}}$ |

## COEFFICIENT OF THE 14$^{th}$ CHERN FORM

The coefficient $c_{(14) i_1 i_2 i_3 i_4 i_5 i_6 i_7 i_8 i_9 i_{10} i_{11} i_{12} i_{13} i_{14} i_{15} i_{16} i_{17} i_{18} i_{19} i_{20} i_{21} i_{22} i_{23} i_{24} i_{25} i_{26} i_{27} i_{28}}$ of the 14$^{th}$ Chern form $c_{(14)}$ is given by

$$c_{(14) i_1 i_2 i_3 i_4 i_5 i_6 i_7 i_8 i_9 i_{10} i_{11} i_{12} i_{13} i_{14} i_{15} i_{16} i_{17} i_{18} i_{19} i_{20} i_{21} i_{22} i_{23} i_{24} i_{25} i_{26} i_{27} i_{28}} = \quad (9)$$

$$= \frac{1}{28!} \langle \mathsf{e}_{i_1} \wedge \mathsf{e}_{i_2} \wedge \mathsf{e}_{i_3} \wedge \mathsf{e}_{i_4} \wedge \mathsf{e}_{i_5} \wedge \mathsf{e}_{i_6} \wedge \mathsf{e}_{i_7} \wedge \mathsf{e}_{i_8} \wedge \mathsf{e}_{i_9} \wedge \mathsf{e}_{i_{10}} \wedge \mathsf{e}_{i_{11}} \wedge \mathsf{e}_{i_{12}} \wedge \mathsf{e}_{i_{13}} \wedge \mathsf{e}_{i_{14}} \wedge \ldots$$

$$\ldots \wedge \mathsf{e}_{i_{15}} \wedge \mathsf{e}_{i_{16}} \wedge \mathsf{e}_{i_{17}} \wedge \mathsf{e}_{i_{18}} \wedge \mathsf{e}_{i_{19}} \wedge \mathsf{e}_{i_{20}} \wedge \mathsf{e}_{i_{21}} \wedge \mathsf{e}_{i_{22}} \wedge \mathsf{e}_{i_{23}} \wedge \mathsf{e}_{i_{24}} \wedge \mathsf{e}_{i_{25}} \wedge \mathsf{e}_{i_{26}} \wedge \mathsf{e}_{i_{27}} \wedge \mathsf{e}_{i_{28}}, c_{(14)} \rangle$$

$$= \frac{i^{14}}{2^{28} \pi^{14} 14!} (- 6{,}227{,}020{,}800 \; R_{[i_1 i_2 | n|}{}^a R_{i_3 i_4 |a|}{}^b R_{i_5 i_6 |b|}{}^c R_{i_7 i_8 |c|}{}^d R_{i_9 i_{10} |d|}{}^e R_{i_{11} i_{12} |e|}{}^f R_{i_{13} i_{14} |f|}{}^g R_{i_{15} i_{16} |g|}{}^h R_{i_{17} i_{18} |h|}{}^i R_{i_{19} i_{20} |i|}{}^j R_{i_{21} i_{22} |j|}{}^k R_{i_{23} i_{24} |k|}{}^l R_{i_{25} i_{26} |l|}{}^m R_{i_{27} i_{28}]m}{}^n +$$

$$+ 6{,}706{,}022{,}400 \; R_{[i_1 i_2 | m|}{}^a R_{i_3 i_4 |a|}{}^b R_{i_5 i_6 |b|}{}^c R_{i_7 i_8 |c|}{}^d R_{i_9 i_{10} |d|}{}^e R_{i_{11} i_{12} |e|}{}^f R_{i_{13} i_{14} |f|}{}^g R_{i_{15} i_{16} |g|}{}^h R_{i_{17} i_{18} |h|}{}^i R_{i_{19} i_{20} |i|}{}^j R_{i_{21} i_{22} |j|}{}^k R_{i_{23} i_{24} |k|}{}^l R_{i_{25} i_{26} |l|}{}^m R_{i_{27} i_{28}]n}{}^n +$$

$$+ 3{,}632{,}428{,}800 \; R_{[i_1 i_2 | l|}{}^a R_{i_3 i_4 |a|}{}^b R_{i_5 i_6 |b|}{}^c R_{i_7 i_8 |c|}{}^d R_{i_9 i_{10} |d|}{}^e R_{i_{11} i_{12} |e|}{}^f R_{i_{13} i_{14} |f|}{}^g R_{i_{15} i_{16} |g|}{}^h R_{i_{17} i_{18} |h|}{}^i R_{i_{19} i_{20} |i|}{}^j R_{i_{21} i_{22} |j|}{}^k R_{i_{23} i_{24} |k|}{}^l R_{i_{25} i_{26} |n|}{}^m R_{i_{27} i_{28}]m}{}^n -$$

$$- 3{,}632{,}428{,}800 \; R_{[i_1 i_2 | l|}{}^a R_{i_3 i_4 |a|}{}^b R_{i_5 i_6 |b|}{}^c R_{i_7 i_8 |c|}{}^d R_{i_9 i_{10} |d|}{}^e R_{i_{11} i_{12} |e|}{}^f R_{i_{13} i_{14} |f|}{}^g R_{i_{15} i_{16} |g|}{}^h R_{i_{17} i_{18} |h|}{}^i R_{i_{19} i_{20} |i|}{}^j R_{i_{21} i_{22} |j|}{}^k R_{i_{23} i_{24} |k|}{}^l R_{i_{25} i_{26} |m|}{}^m R_{i_{27} i_{28}]n}{}^n +$$

$$+ 2{,}641{,}766{,}400 \; R_{[i_1 i_2 | k|}{}^a R_{i_3 i_4 |a|}{}^b R_{i_5 i_6 |b|}{}^c R_{i_7 i_8 |c|}{}^d R_{i_9 i_{10} |d|}{}^e R_{i_{11} i_{12} |e|}{}^f R_{i_{13} i_{14} |f|}{}^g R_{i_{15} i_{16} |g|}{}^h R_{i_{17} i_{18} |h|}{}^i R_{i_{19} i_{20} |i|}{}^j R_{i_{21} i_{22} |j|}{}^k R_{i_{23} i_{24} |n|}{}^l R_{i_{25} i_{26} |l|}{}^m R_{i_{27} i_{28}]m}{}^n -$$

$$- 3{,}962{,}649{,}600 \; R_{[i_1 i_2 | k|}{}^a R_{i_3 i_4 |a|}{}^b R_{i_5 i_6 |b|}{}^c R_{i_7 i_8 |c|}{}^d R_{i_9 i_{10} |d|}{}^e R_{i_{11} i_{12} |e|}{}^f R_{i_{13} i_{14} |f|}{}^g R_{i_{15} i_{16} |g|}{}^h R_{i_{17} i_{18} |h|}{}^i R_{i_{19} i_{20} |i|}{}^j R_{i_{21} i_{22} |j|}{}^k R_{i_{23} i_{24} |m|}{}^l R_{i_{25} i_{26} |l|}{}^m R_{i_{27} i_{28}]n}{}^n +$$

---
[1] Eguchi, T., P. B. Gilkey, and A. J. Hanson, "Gravitation, gauge theories and differential geometry," *Phys. Rep.*, **66** (1980) 213.
[2] Kobayashi, S., and K. Nomizu, *Foundations of Differential Geometry, Vol. II*, John Wiley & Sons, New York (1969), p. 309.
[3] Chern, S. S., "Characteristic classes of Hermitian manifolds," *Annals of Math.*, **47** (1946) 85.
[4] Schouten, J. A., *Ricci Calculus*, Springer-Verlag, Berlin (1954), p. 172.
[5] *Ibid.*, p. 144.
[6] Понтрягин, Л. С., "Некоторые топологические инварианты замкнутых римановых многообразий," *Известия Академии Наук СССР, Серия Математическая*, **13** (1949) 125.
[7] Абрамов, А. А., "О топологических инвариантах римановых пространств, получаемых интегрированием тензорных полей," *Доклады Академии Наук СССР*, **81** (1951) 125; "Формула типа Гаусс-Боннэ для тензорных полей Понтрягина," *ibid.*, **93** (1953) 157.

$$+ 1{,}320{,}883{,}200 \, R_{[i_1 i_2 |k|}{}^a R_{i_3 i_4 |a|}{}^b R_{i_5 i_6 |b|}{}^c R_{i_7 i_8 |c|}{}^d R_{i_9 i_{10} |d|}{}^e R_{i_{11} i_{12} |e|}{}^f R_{i_{13} i_{14} |f|}{}^g R_{i_{15} i_{16} |g|}{}^h R_{i_{17} i_{18} |h|}{}^i R_{i_{19} i_{20} |i|}{}^j R_{i_{21} i_{22} |j|}{}^k R_{i_{23} i_{24} |l|}{}^l R_{i_{25} i_{26} |m|}{}^m R_{i_{27} i_{28}] n}{}^n +$$

$$+ 2{,}179{,}457{,}280 \, R_{[i_1 i_2 |j|}{}^a R_{i_3 i_4 |a|}{}^b R_{i_5 i_6 |b|}{}^c R_{i_7 i_8 |c|}{}^d R_{i_9 i_{10} |d|}{}^e R_{i_{11} i_{12} |e|}{}^f R_{i_{13} i_{14} |f|}{}^g R_{i_{15} i_{16} |g|}{}^h R_{i_{17} i_{18} |h|}{}^i R_{i_{19} i_{20} |i|}{}^j R_{i_{21} i_{22} |n|}{}^k R_{i_{23} i_{24} |k|}{}^l R_{i_{25} i_{26} |l|}{}^m R_{i_{27} i_{28}] m}{}^n -$$

$$- 2{,}905{,}943{,}040 \, R_{[i_1 i_2 |j|}{}^a R_{i_3 i_4 |a|}{}^b R_{i_5 i_6 |b|}{}^c R_{i_7 i_8 |c|}{}^d R_{i_9 i_{10} |d|}{}^e R_{i_{11} i_{12} |e|}{}^f R_{i_{13} i_{14} |f|}{}^g R_{i_{15} i_{16} |g|}{}^h R_{i_{17} i_{18} |h|}{}^i R_{i_{19} i_{20} |i|}{}^j R_{i_{21} i_{22} |m|}{}^k R_{i_{23} i_{24} |k|}{}^l R_{i_{25} i_{26} |l|}{}^m R_{i_{27} i_{28}] n}{}^n -$$

$$- 1{,}089{,}728{,}640 \, R_{[i_1 i_2 |j|}{}^a R_{i_3 i_4 |a|}{}^b R_{i_5 i_6 |b|}{}^c R_{i_7 i_8 |c|}{}^d R_{i_9 i_{10} |d|}{}^e R_{i_{11} i_{12} |e|}{}^f R_{i_{13} i_{14} |f|}{}^g R_{i_{15} i_{16} |g|}{}^h R_{i_{17} i_{18} |h|}{}^i R_{i_{19} i_{20} |i|}{}^j R_{i_{21} i_{22} |k|}{}^k R_{i_{23} i_{24} |l|}{}^l R_{i_{25} i_{26} |n|}{}^m R_{i_{27} i_{28}] m}{}^n +$$

$$+ 2{,}179{,}457{,}280 \, R_{[i_1 i_2 |j|}{}^a R_{i_3 i_4 |a|}{}^b R_{i_5 i_6 |b|}{}^c R_{i_7 i_8 |c|}{}^d R_{i_9 i_{10} |d|}{}^e R_{i_{11} i_{12} |e|}{}^f R_{i_{13} i_{14} |f|}{}^g R_{i_{15} i_{16} |g|}{}^h R_{i_{17} i_{18} |h|}{}^i R_{i_{19} i_{20} |i|}{}^j R_{i_{21} i_{22} |l|}{}^k R_{i_{23} i_{24} |k|}{}^l R_{i_{25} i_{26} |m|}{}^m R_{i_{27} i_{28}] n}{}^n -$$

$$- 363{,}242{,}880 \, R_{[i_1 i_2 |j|}{}^a R_{i_3 i_4 |a|}{}^b R_{i_5 i_6 |b|}{}^c R_{i_7 i_8 |c|}{}^d R_{i_9 i_{10} |d|}{}^e R_{i_{11} i_{12} |e|}{}^f R_{i_{13} i_{14} |f|}{}^g R_{i_{15} i_{16} |g|}{}^h R_{i_{17} i_{18} |h|}{}^i R_{i_{19} i_{20} |i|}{}^j R_{i_{21} i_{22} |k|}{}^k R_{i_{23} i_{24} |l|}{}^l R_{i_{25} i_{26} |m|}{}^m R_{i_{27} i_{28}] n}{}^n +$$

$$+ 1{,}937{,}295{,}360 \, R_{[i_1 i_2 |i|}{}^a R_{i_3 i_4 |a|}{}^b R_{i_5 i_6 |b|}{}^c R_{i_7 i_8 |c|}{}^d R_{i_9 i_{10} |d|}{}^e R_{i_{11} i_{12} |e|}{}^f R_{i_{13} i_{14} |f|}{}^g R_{i_{15} i_{16} |g|}{}^h R_{i_{17} i_{18} |h|}{}^i R_{i_{19} i_{20} |n|}{}^j R_{i_{21} i_{22} |j|}{}^k R_{i_{23} i_{24} |k|}{}^l R_{i_{25} i_{26} |l|}{}^m R_{i_{27} i_{28}] m}{}^n -$$

$$- 2{,}421{,}619{,}200 \, R_{[i_1 i_2 |i|}{}^a R_{i_3 i_4 |a|}{}^b R_{i_5 i_6 |b|}{}^c R_{i_7 i_8 |c|}{}^d R_{i_9 i_{10} |d|}{}^e R_{i_{11} i_{12} |e|}{}^f R_{i_{13} i_{14} |f|}{}^g R_{i_{15} i_{16} |g|}{}^h R_{i_{17} i_{18} |h|}{}^i R_{i_{19} i_{20} |m|}{}^j R_{i_{21} i_{22} |j|}{}^k R_{i_{23} i_{24} |k|}{}^l R_{i_{25} i_{26} |l|}{}^m R_{i_{27} i_{28}] n}{}^n -$$

$$- 1{,}614{,}412{,}800 \, R_{[i_1 i_2 |i|}{}^a R_{i_3 i_4 |a|}{}^b R_{i_5 i_6 |b|}{}^c R_{i_7 i_8 |c|}{}^d R_{i_9 i_{10} |d|}{}^e R_{i_{11} i_{12} |e|}{}^f R_{i_{13} i_{14} |f|}{}^g R_{i_{15} i_{16} |g|}{}^h R_{i_{17} i_{18} |h|}{}^i R_{i_{19} i_{20} |l|}{}^j R_{i_{21} i_{22} |j|}{}^k R_{i_{23} i_{24} |k|}{}^l R_{i_{25} i_{26} |n|}{}^m R_{i_{27} i_{28}] m}{}^n +$$

$$+ 1{,}614{,}412{,}800 \, R_{[i_1 i_2 |i|}{}^a R_{i_3 i_4 |a|}{}^b R_{i_5 i_6 |b|}{}^c R_{i_7 i_8 |c|}{}^d R_{i_9 i_{10} |d|}{}^e R_{i_{11} i_{12} |e|}{}^f R_{i_{13} i_{14} |f|}{}^g R_{i_{15} i_{16} |g|}{}^h R_{i_{17} i_{18} |h|}{}^i R_{i_{19} i_{20} |l|}{}^j R_{i_{21} i_{22} |j|}{}^k R_{i_{23} i_{24} |k|}{}^l R_{i_{25} i_{26} |m|}{}^m R_{i_{27} i_{28}] n}{}^n +$$

$$+ 1{,}210{,}809{,}600 \, R_{[i_1 i_2 |i|}{}^a R_{i_3 i_4 |a|}{}^b R_{i_5 i_6 |b|}{}^c R_{i_7 i_8 |c|}{}^d R_{i_9 i_{10} |d|}{}^e R_{i_{11} i_{12} |e|}{}^f R_{i_{13} i_{14} |f|}{}^g R_{i_{15} i_{16} |g|}{}^h R_{i_{17} i_{18} |h|}{}^i R_{i_{19} i_{20} |k|}{}^j R_{i_{21} i_{22} |j|}{}^k R_{i_{23} i_{24} |m|}{}^l R_{i_{25} i_{26} |l|}{}^m R_{i_{27} i_{28}] n}{}^n -$$

$$- 807{,}206{,}400 \, R_{[i_1 i_2 |i|}{}^a R_{i_3 i_4 |a|}{}^b R_{i_5 i_6 |b|}{}^c R_{i_7 i_8 |c|}{}^d R_{i_9 i_{10} |d|}{}^e R_{i_{11} i_{12} |e|}{}^f R_{i_{13} i_{14} |f|}{}^g R_{i_{15} i_{16} |g|}{}^h R_{i_{17} i_{18} |h|}{}^i R_{i_{19} i_{20} |k|}{}^j R_{i_{21} i_{22} |j|}{}^k R_{i_{23} i_{24} |l|}{}^l R_{i_{25} i_{26} |m|}{}^m R_{i_{27} i_{28}] n}{}^n +$$

$$+ 80{,}720{,}640 \, R_{[i_1 i_2 |i|}{}^a R_{i_3 i_4 |a|}{}^b R_{i_5 i_6 |b|}{}^c R_{i_7 i_8 |c|}{}^d R_{i_9 i_{10} |d|}{}^e R_{i_{11} i_{12} |e|}{}^f R_{i_{13} i_{14} |f|}{}^g R_{i_{15} i_{16} |g|}{}^h R_{i_{17} i_{18} |h|}{}^i R_{i_{19} i_{20} |j|}{}^j R_{i_{21} i_{22} |k|}{}^k R_{i_{23} i_{24} |l|}{}^l R_{i_{25} i_{26} |m|}{}^m R_{i_{27} i_{28}] n}{}^n +$$

$$+ 1{,}816{,}214{,}400 \, R_{[i_1 i_2 |h|}{}^a R_{i_3 i_4 |a|}{}^b R_{i_5 i_6 |b|}{}^c R_{i_7 i_8 |c|}{}^d R_{i_9 i_{10} |d|}{}^e R_{i_{11} i_{12} |e|}{}^f R_{i_{13} i_{14} |f|}{}^g R_{i_{15} i_{16} |g|}{}^h R_{i_{17} i_{18} |n|}{}^i R_{i_{19} i_{20} |i|}{}^j R_{i_{21} i_{22} |j|}{}^k R_{i_{23} i_{24} |k|}{}^l R_{i_{25} i_{26} |l|}{}^m R_{i_{27} i_{28}] m}{}^n -$$

$$- 2{,}179{,}457{,}280 \, R_{[i_1 i_2 |h|}{}^a R_{i_3 i_4 |a|}{}^b R_{i_5 i_6 |b|}{}^c R_{i_7 i_8 |c|}{}^d R_{i_9 i_{10} |d|}{}^e R_{i_{11} i_{12} |e|}{}^f R_{i_{13} i_{14} |f|}{}^g R_{i_{15} i_{16} |g|}{}^h R_{i_{17} i_{18} |m|}{}^i R_{i_{19} i_{20} |i|}{}^j R_{i_{21} i_{22} |j|}{}^k R_{i_{23} i_{24} |k|}{}^l R_{i_{25} i_{26} |l|}{}^m R_{i_{27} i_{28}] n}{}^n -$$

$$- 1{,}362{,}160{,}800 \, R_{[i_1 i_2 |h|}{}^a R_{i_3 i_4 |a|}{}^b R_{i_5 i_6 |b|}{}^c R_{i_7 i_8 |c|}{}^d R_{i_9 i_{10} |d|}{}^e R_{i_{11} i_{12} |e|}{}^f R_{i_{13} i_{14} |f|}{}^g R_{i_{15} i_{16} |g|}{}^h R_{i_{17} i_{18} |l|}{}^i R_{i_{19} i_{20} |i|}{}^j R_{i_{21} i_{22} |j|}{}^k R_{i_{23} i_{24} |k|}{}^l R_{i_{25} i_{26} |n|}{}^m R_{i_{27} i_{28}] m}{}^n +$$

$$+ 1{,}362{,}160{,}800 \, R_{[i_1 i_2 |h|}{}^a R_{i_3 i_4 |a|}{}^b R_{i_5 i_6 |b|}{}^c R_{i_7 i_8 |c|}{}^d R_{i_9 i_{10} |d|}{}^e R_{i_{11} i_{12} |e|}{}^f R_{i_{13} i_{14} |f|}{}^g R_{i_{15} i_{16} |g|}{}^h R_{i_{17} i_{18} |l|}{}^i R_{i_{19} i_{20} |i|}{}^j R_{i_{21} i_{22} |j|}{}^k R_{i_{23} i_{24} |k|}{}^l R_{i_{25} i_{26} |m|}{}^m R_{i_{27} i_{28}] n}{}^n -$$

$$- 605{,}404{,}800 \, R_{[i_1 i_2 |h|}{}^a R_{i_3 i_4 |a|}{}^b R_{i_5 i_6 |b|}{}^c R_{i_7 i_8 |c|}{}^d R_{i_9 i_{10} |d|}{}^e R_{i_{11} i_{12} |e|}{}^f R_{i_{13} i_{14} |f|}{}^g R_{i_{15} i_{16} |g|}{}^h R_{i_{17} i_{18} |k|}{}^i R_{i_{19} i_{20} |i|}{}^j R_{i_{21} i_{22} |j|}{}^k R_{i_{23} i_{24} |l|}{}^l R_{i_{25} i_{26} |n|}{}^m R_{i_{27} i_{28}] m}{}^n +$$

$$+ 1{,}816{,}214{,}400 \, R_{[i_1 i_2 |h|}{}^a R_{i_3 i_4 |a|}{}^b R_{i_5 i_6 |b|}{}^c R_{i_7 i_8 |c|}{}^d R_{i_9 i_{10} |d|}{}^e R_{i_{11} i_{12} |e|}{}^f R_{i_{13} i_{14} |f|}{}^g R_{i_{15} i_{16} |g|}{}^h R_{i_{17} i_{18} |k|}{}^i R_{i_{19} i_{20} |i|}{}^j R_{i_{21} i_{22} |j|}{}^k R_{i_{23} i_{24} |m|}{}^l R_{i_{25} i_{26} |l|}{}^m R_{i_{27} i_{28}] n}{}^n -$$

$$- 605{,}404{,}800 \, R_{[i_1 i_2 |h|}{}^a R_{i_3 i_4 |a|}{}^b R_{i_5 i_6 |b|}{}^c R_{i_7 i_8 |c|}{}^d R_{i_9 i_{10} |d|}{}^e R_{i_{11} i_{12} |e|}{}^f R_{i_{13} i_{14} |f|}{}^g R_{i_{15} i_{16} |g|}{}^h R_{i_{17} i_{18} |k|}{}^i R_{i_{19} i_{20} |i|}{}^j R_{i_{21} i_{22} |j|}{}^k R_{i_{23} i_{24} |l|}{}^l R_{i_{25} i_{26} |m|}{}^m R_{i_{27} i_{28}] n}{}^n +$$

$$+ 227{,}026{,}800 \, R_{[i_1 i_2 |h|}{}^a R_{i_3 i_4 |a|}{}^b R_{i_5 i_6 |b|}{}^c R_{i_7 i_8 |c|}{}^d R_{i_9 i_{10} |d|}{}^e R_{i_{11} i_{12} |e|}{}^f R_{i_{13} i_{14} |f|}{}^g R_{i_{15} i_{16} |g|}{}^h R_{i_{17} i_{18} |j|}{}^i R_{i_{19} i_{20} |i|}{}^j R_{i_{21} i_{22} |k|}{}^k R_{i_{23} i_{24} |l|}{}^l R_{i_{25} i_{26} |n|}{}^m R_{i_{27} i_{28}] m}{}^n -$$

$$- 681{,}080{,}400 \, R_{[i_1 i_2 |h|}{}^a R_{i_3 i_4 |a|}{}^b R_{i_5 i_6 |b|}{}^c R_{i_7 i_8 |c|}{}^d R_{i_9 i_{10} |d|}{}^e R_{i_{11} i_{12} |e|}{}^f R_{i_{13} i_{14} |f|}{}^g R_{i_{15} i_{16} |g|}{}^h R_{i_{17} i_{18} |j|}{}^i R_{i_{19} i_{20} |i|}{}^j R_{i_{21} i_{22} |l|}{}^k R_{i_{23} i_{24} |k|}{}^l R_{i_{25} i_{26} |n|}{}^m R_{i_{27} i_{28}] m}{}^n +$$

$$+ 227{,}026{,}800 \, R_{[i_1 i_2 |h|}{}^a R_{i_3 i_4 |a|}{}^b R_{i_5 i_6 |b|}{}^c R_{i_7 i_8 |c|}{}^d R_{i_9 i_{10} |d|}{}^e R_{i_{11} i_{12} |e|}{}^f R_{i_{13} i_{14} |f|}{}^g R_{i_{15} i_{16} |g|}{}^h R_{i_{17} i_{18} |j|}{}^i R_{i_{19} i_{20} |i|}{}^j R_{i_{21} i_{22} |k|}{}^k R_{i_{23} i_{24} |l|}{}^l R_{i_{25} i_{26} |m|}{}^m R_{i_{27} i_{28}] n}{}^n -$$

$$- 15{,}135{,}120 \, R_{[i_1 i_2 |h|}{}^a R_{i_3 i_4 |a|}{}^b R_{i_5 i_6 |b|}{}^c R_{i_7 i_8 |c|}{}^d R_{i_9 i_{10} |d|}{}^e R_{i_{11} i_{12} |e|}{}^f R_{i_{13} i_{14} |f|}{}^g R_{i_{15} i_{16} |g|}{}^h R_{i_{17} i_{18} |i|}{}^i R_{i_{19} i_{20} |j|}{}^j R_{i_{21} i_{22} |k|}{}^k R_{i_{23} i_{24} |l|}{}^l R_{i_{25} i_{26} |m|}{}^m R_{i_{27} i_{28}] n}{}^n +$$

$$+ 889{,}574{,}400 \, R_{[i_1 i_2 |g|}{}^a R_{i_3 i_4 |a|}{}^b R_{i_5 i_6 |b|}{}^c R_{i_7 i_8 |c|}{}^d R_{i_9 i_{10} |d|}{}^e R_{i_{11} i_{12} |e|}{}^f R_{i_{13} i_{14} |f|}{}^g R_{i_{15} i_{16} |n|}{}^h R_{i_{17} i_{18} |h|}{}^i R_{i_{19} i_{20} |i|}{}^j R_{i_{21} i_{22} |j|}{}^k R_{i_{23} i_{24} |k|}{}^l R_{i_{25} i_{26} |l|}{}^m R_{i_{27} i_{28}] m}{}^n -$$

$$- 2{,}075{,}673{,}600 \, R_{[i_1 i_2 |g|}{}^a R_{i_3 i_4 |a|}{}^b R_{i_5 i_6 |b|}{}^c R_{i_7 i_8 |c|}{}^d R_{i_9 i_{10} |d|}{}^e R_{i_{11} i_{12} |e|}{}^f R_{i_{13} i_{14} |f|}{}^g R_{i_{15} i_{16} |m|}{}^h R_{i_{17} i_{18} |h|}{}^i R_{i_{19} i_{20} |i|}{}^j R_{i_{21} i_{22} |j|}{}^k R_{i_{23} i_{24} |k|}{}^l R_{i_{25} i_{26} |l|}{}^m R_{i_{27} i_{28}] n}{}^n -$$

$$- 1{,}245{,}404{,}160 \, R_{[i_1 i_2 |g|}{}^a R_{i_3 i_4 |a|}{}^b R_{i_5 i_6 |b|}{}^c R_{i_7 i_8 |c|}{}^d R_{i_9 i_{10} |d|}{}^e R_{i_{11} i_{12} |e|}{}^f R_{i_{13} i_{14} |f|}{}^g R_{i_{15} i_{16} |l|}{}^h R_{i_{17} i_{18} |h|}{}^i R_{i_{19} i_{20} |i|}{}^j R_{i_{21} i_{22} |j|}{}^k R_{i_{23} i_{24} |k|}{}^l R_{i_{25} i_{26} |n|}{}^m R_{i_{27} i_{28}] m}{}^n +$$

$$+ 1{,}245{,}404{,}160 \, R_{[i_1 i_2 |g|}{}^a R_{i_3 i_4 |a|}{}^b R_{i_5 i_6 |b|}{}^c R_{i_7 i_8 |c|}{}^d R_{i_9 i_{10} |d|}{}^e R_{i_{11} i_{12} |e|}{}^f R_{i_{13} i_{14} |f|}{}^g R_{i_{15} i_{16} |l|}{}^h R_{i_{17} i_{18} |h|}{}^i R_{i_{19} i_{20} |i|}{}^j R_{i_{21} i_{22} |j|}{}^k R_{i_{23} i_{24} |k|}{}^l R_{i_{25} i_{26} |m|}{}^m R_{i_{27} i_{28}] n}{}^n -$$

$$- 1{,}037{,}836{,}800 \, R_{[i_1 i_2 |g|}{}^a R_{i_3 i_4 |a|}{}^b R_{i_5 i_6 |b|}{}^c R_{i_7 i_8 |c|}{}^d R_{i_9 i_{10} |d|}{}^e R_{i_{11} i_{12} |e|}{}^f R_{i_{13} i_{14} |f|}{}^g R_{i_{15} i_{16} |k|}{}^h R_{i_{17} i_{18} |h|}{}^i R_{i_{19} i_{20} |i|}{}^j R_{i_{21} i_{22} |j|}{}^k R_{i_{23} i_{24} |n|}{}^l R_{i_{25} i_{26} |l|}{}^m R_{i_{27} i_{28}] m}{}^n +$$

$$+ 1{,}556{,}755{,}200 \, R_{[i_1 i_2 |g|}{}^a R_{i_3 i_4 |a|}{}^b R_{i_5 i_6 |b|}{}^c R_{i_7 i_8 |c|}{}^d R_{i_9 i_{10} |d|}{}^e R_{i_{11} i_{12} |e|}{}^f R_{i_{13} i_{14} |f|}{}^g R_{i_{15} i_{16} |k|}{}^h R_{i_{17} i_{18} |h|}{}^i R_{i_{19} i_{20} |i|}{}^j R_{i_{21} i_{22} |j|}{}^k R_{i_{23} i_{24} |m|}{}^l R_{i_{25} i_{26} |l|}{}^m R_{i_{27} i_{28}] n}{}^n -$$

$$- 518{,}918{,}400 \, R_{[i_1 i_2 |g|}{}^a R_{i_3 i_4 |a|}{}^b R_{i_5 i_6 |b|}{}^c R_{i_7 i_8 |c|}{}^d R_{i_9 i_{10} |d|}{}^e R_{i_{11} i_{12} |e|}{}^f R_{i_{13} i_{14} |f|}{}^g R_{i_{15} i_{16} |k|}{}^h R_{i_{17} i_{18} |h|}{}^i R_{i_{19} i_{20} |i|}{}^j R_{i_{21} i_{22} |j|}{}^k R_{i_{23} i_{24} |l|}{}^l R_{i_{25} i_{26} |m|}{}^m R_{i_{27} i_{28}] n}{}^n +$$

$$+ 691{,}891{,}200 \, R_{[i_1 i_2 |g|}{}^a R_{i_3 i_4 |a|}{}^b R_{i_5 i_6 |b|}{}^c R_{i_7 i_8 |c|}{}^d R_{i_9 i_{10} |d|}{}^e R_{i_{11} i_{12} |e|}{}^f R_{i_{13} i_{14} |f|}{}^g R_{i_{15} i_{16} |j|}{}^h R_{i_{17} i_{18} |h|}{}^i R_{i_{19} i_{20} |i|}{}^j R_{i_{21} i_{22} |m|}{}^k R_{i_{23} i_{24} |k|}{}^l R_{i_{25} i_{26} |l|}{}^m R_{i_{27} i_{28}] n}{}^n +$$

$$+ 518{,}918{,}400 \, R_{[i_1 i_2 |g|}{}^a R_{i_3 i_4 |a|}{}^b R_{i_5 i_6 |b|}{}^c R_{i_7 i_8 |c|}{}^d R_{i_9 i_{10} |d|}{}^e R_{i_{11} i_{12} |e|}{}^f R_{i_{13} i_{14} |f|}{}^g R_{i_{15} i_{16} |j|}{}^h R_{i_{17} i_{18} |h|}{}^i R_{i_{19} i_{20} |i|}{}^j R_{i_{21} i_{22} |l|}{}^k R_{i_{23} i_{24} |k|}{}^l R_{i_{25} i_{26} |n|}{}^m R_{i_{27} i_{28}] m}{}^n -$$

$$- 1{,}037{,}836{,}800 \, R_{[i_1 i_2 |g|}{}^a R_{i_3 i_4 |a|}{}^b R_{i_5 i_6 |b|}{}^c R_{i_7 i_8 |c|}{}^d R_{i_9 i_{10} |d|}{}^e R_{i_{11} i_{12} |e|}{}^f R_{i_{13} i_{14} |f|}{}^g R_{i_{15} i_{16} |j|}{}^h R_{i_{17} i_{18} |h|}{}^i R_{i_{19} i_{20} |i|}{}^j R_{i_{21} i_{22} |l|}{}^k R_{i_{23} i_{24} |k|}{}^l R_{i_{25} i_{26} |m|}{}^m R_{i_{27} i_{28}] n}{}^n +$$

$$+ 172{,}972{,}800 \, R_{[i_1 i_2 |g|}{}^a R_{i_3 i_4 |a|}{}^b R_{i_5 i_6 |b|}{}^c R_{i_7 i_8 |c|}{}^d R_{i_9 i_{10} |d|}{}^e R_{i_{11} i_{12} |e|}{}^f R_{i_{13} i_{14} |f|}{}^g R_{i_{15} i_{16} |j|}{}^h R_{i_{17} i_{18} |h|}{}^i R_{i_{19} i_{20} |i|}{}^j R_{i_{21} i_{22} |k|}{}^k R_{i_{23} i_{24} |l|}{}^l R_{i_{25} i_{26} |m|}{}^m R_{i_{27} i_{28}] n}{}^n -$$

$$- 259{,}459{,}200 \, R_{[i_1 i_2 |g|}{}^a R_{i_3 i_4 |a|}{}^b R_{i_5 i_6 |b|}{}^c R_{i_7 i_8 |c|}{}^d R_{i_9 i_{10} |d|}{}^e R_{i_{11} i_{12} |e|}{}^f R_{i_{13} i_{14} |f|}{}^g R_{i_{15} i_{16} |i|}{}^h R_{i_{17} i_{18} |h|}{}^i R_{i_{19} i_{20} |k|}{}^j R_{i_{21} i_{22} |j|}{}^k R_{i_{23} i_{24} |m|}{}^l R_{i_{25} i_{26} |l|}{}^m R_{i_{27} i_{28}] n}{}^n +$$

$$+ 259{,}459{,}200 \, R_{[i_1 i_2 |g|}{}^a R_{i_3 i_4 |a|}{}^b R_{i_5 i_6 |b|}{}^c R_{i_7 i_8 |c|}{}^d R_{i_9 i_{10} |d|}{}^e R_{i_{11} i_{12} |e|}{}^f R_{i_{13} i_{14} |f|}{}^g R_{i_{15} i_{16} |i|}{}^h R_{i_{17} i_{18} |h|}{}^i R_{i_{19} i_{20} |k|}{}^j R_{i_{21} i_{22} |j|}{}^k R_{i_{23} i_{24} |l|}{}^l R_{i_{25} i_{26} |m|}{}^m R_{i_{27} i_{28}] n}{}^n -$$

$$- 51{,}891{,}840 \, R_{[i_1 i_2 |g|}{}^a R_{i_3 i_4 |a|}{}^b R_{i_5 i_6 |b|}{}^c R_{i_7 i_8 |c|}{}^d R_{i_9 i_{10} |d|}{}^e R_{i_{11} i_{12} |e|}{}^f R_{i_{13} i_{14} |f|}{}^g R_{i_{15} i_{16} |i|}{}^h R_{i_{17} i_{18} |h|}{}^i R_{i_{19} i_{20} |j|}{}^j R_{i_{21} i_{22} |k|}{}^k R_{i_{23} i_{24} |l|}{}^l R_{i_{25} i_{26} |m|}{}^m R_{i_{27} i_{28}] n}{}^n +$$

$$+ 2{,}471{,}040 \, R_{[i_1 i_2 |g|}{}^a R_{i_3 i_4 |a|}{}^b R_{i_5 i_6 |b|}{}^c R_{i_7 i_8 |c|}{}^d R_{i_9 i_{10} |d|}{}^e R_{i_{11} i_{12} |e|}{}^f R_{i_{13} i_{14} |f|}{}^g R_{i_{15} i_{16} |h|}{}^h R_{i_{17} i_{18} |i|}{}^i R_{i_{19} i_{20} |j|}{}^j R_{i_{21} i_{22} |k|}{}^k R_{i_{23} i_{24} |l|}{}^l R_{i_{25} i_{26} |m|}{}^m R_{i_{27} i_{28}] n}{}^n -$$

$$- 605{,}404{,}800 \, R_{[i_1 i_2 |f|}{}^a R_{i_3 i_4 |a|}{}^b R_{i_5 i_6 |b|}{}^c R_{i_7 i_8 |c|}{}^d R_{i_9 i_{10} |d|}{}^e R_{i_{11} i_{12} |e|}{}^f R_{i_{13} i_{14} |g|}{}^g R_{i_{15} i_{16} |g|}{}^h R_{i_{17} i_{18} |h|}{}^i R_{i_{19} i_{20} |i|}{}^j R_{i_{21} i_{22} |j|}{}^k R_{i_{23} i_{24} |k|}{}^l R_{i_{25} i_{26} |l|}{}^m R_{i_{27} i_{28}] m}{}^n +$$

$$+ 605{,}404{,}800 \, R_{[i_1 i_2 |f|}{}^a R_{i_3 i_4 |a|}{}^b R_{i_5 i_6 |b|}{}^c R_{i_7 i_8 |c|}{}^d R_{i_9 i_{10} |d|}{}^e R_{i_{11} i_{12} |e|}{}^f R_{i_{13} i_{14} |g|}{}^g R_{i_{15} i_{16} |g|}{}^h R_{i_{17} i_{18} |h|}{}^i R_{i_{19} i_{20} |i|}{}^j R_{i_{21} i_{22} |j|}{}^k R_{i_{23} i_{24} |k|}{}^l R_{i_{25} i_{26} |m|}{}^m R_{i_{27} i_{28}] n}{}^n -$$

$$- 968{,}647{,}680 \, R_{[i_1i_2|f|}{}^a R_{i_3i_4|a|}{}^b R_{i_5i_6|b|}{}^c R_{i_7i_8|c|}{}^d R_{i_9i_{10}|d|}{}^e R_{i_{11}i_{12}|e|}{}^f R_{i_{13}i_{14}|k|}{}^g R_{i_{15}i_{16}|g|}{}^h R_{i_{17}i_{18}|h|}{}^i R_{i_{19}i_{20}|i|}{}^j R_{i_{21}i_{22}|j|}{}^k R_{i_{23}i_{24}|l|}{}^l R_{i_{25}i_{26}|l|}{}^m R_{i_{27}i_{28}]m}{}^n +$$

$$+ 1{,}452{,}971{,}520 \, R_{[i_1i_2|f|}{}^a R_{i_3i_4|a|}{}^b R_{i_5i_6|b|}{}^c R_{i_7i_8|c|}{}^d R_{i_9i_{10}|d|}{}^e R_{i_{11}i_{12}|e|}{}^f R_{i_{13}i_{14}|k|}{}^g R_{i_{15}i_{16}|g|}{}^h R_{i_{17}i_{18}|h|}{}^i R_{i_{19}i_{20}|i|}{}^j R_{i_{21}i_{22}|j|}{}^k R_{i_{23}i_{24}|m|}{}^l R_{i_{25}i_{26}|l|}{}^m R_{i_{27}i_{28}]}{}^n -$$

$$- 484{,}323{,}840 \, R_{[i_1i_2|f|}{}^a R_{i_3i_4|a|}{}^b R_{i_5i_6|b|}{}^c R_{i_7i_8|c|}{}^d R_{i_9i_{10}|d|}{}^e R_{i_{11}i_{12}|e|}{}^f R_{i_{13}i_{14}|k|}{}^g R_{i_{15}i_{16}|g|}{}^h R_{i_{17}i_{18}|h|}{}^i R_{i_{19}i_{20}|i|}{}^j R_{i_{21}i_{22}|j|}{}^k R_{i_{23}i_{24}|l|}{}^l R_{i_{25}i_{26}|m|}{}^m R_{i_{27}i_{28}]}{}^n -$$

$$- 454{,}053{,}600 \, R_{[i_1i_2|f|}{}^a R_{i_3i_4|a|}{}^b R_{i_5i_6|b|}{}^c R_{i_7i_8|c|}{}^d R_{i_9i_{10}|d|}{}^e R_{i_{11}i_{12}|e|}{}^f R_{i_{13}i_{14}|k|}{}^g R_{i_{15}i_{16}|g|}{}^h R_{i_{17}i_{18}|h|}{}^i R_{i_{19}i_{20}|i|}{}^j R_{i_{21}i_{22}|j|}{}^k R_{i_{23}i_{24}|k|}{}^l R_{i_{25}i_{26}|l|}{}^m R_{i_{27}i_{28}]m}{}^n +$$

$$+ 1{,}210{,}809{,}600 \, R_{[i_1i_2|f|}{}^a R_{i_3i_4|a|}{}^b R_{i_5i_6|b|}{}^c R_{i_7i_8|c|}{}^d R_{i_9i_{10}|d|}{}^e R_{i_{11}i_{12}|e|}{}^f R_{i_{13}i_{14}|j|}{}^g R_{i_{15}i_{16}|g|}{}^h R_{i_{17}i_{18}|h|}{}^i R_{i_{19}i_{20}|i|}{}^j R_{i_{21}i_{22}|m|}{}^k R_{i_{23}i_{24}|k|}{}^l R_{i_{25}i_{26}|l|}{}^m R_{i_{27}i_{28}]}{}^n +$$

$$+ 454{,}053{,}600 \, R_{[i_1i_2|f|}{}^a R_{i_3i_4|a|}{}^b R_{i_5i_6|b|}{}^c R_{i_7i_8|c|}{}^d R_{i_9i_{10}|d|}{}^e R_{i_{11}i_{12}|e|}{}^f R_{i_{13}i_{14}|j|}{}^g R_{i_{15}i_{16}|g|}{}^h R_{i_{17}i_{18}|h|}{}^i R_{i_{19}i_{20}|i|}{}^j R_{i_{21}i_{22}|l|}{}^k R_{i_{23}i_{24}|k|}{}^l R_{i_{25}i_{26}|n|}{}^m R_{i_{27}i_{28}]m}{}^n -$$

$$- 908{,}107{,}200 \, R_{[i_1i_2|f|}{}^a R_{i_3i_4|a|}{}^b R_{i_5i_6|b|}{}^c R_{i_7i_8|c|}{}^d R_{i_9i_{10}|d|}{}^e R_{i_{11}i_{12}|e|}{}^f R_{i_{13}i_{14}|j|}{}^g R_{i_{15}i_{16}|g|}{}^h R_{i_{17}i_{18}|h|}{}^i R_{i_{19}i_{20}|i|}{}^j R_{i_{21}i_{22}|l|}{}^k R_{i_{23}i_{24}|k|}{}^l R_{i_{25}i_{26}|m|}{}^m R_{i_{27}i_{28}]}{}^n +$$

$$+ 151{,}351{,}200 \, R_{[i_1i_2|f|}{}^a R_{i_3i_4|a|}{}^b R_{i_5i_6|b|}{}^c R_{i_7i_8|c|}{}^d R_{i_9i_{10}|d|}{}^e R_{i_{11}i_{12}|e|}{}^f R_{i_{13}i_{14}|j|}{}^g R_{i_{15}i_{16}|g|}{}^h R_{i_{17}i_{18}|h|}{}^i R_{i_{19}i_{20}|i|}{}^j R_{i_{21}i_{22}|k|}{}^k R_{i_{23}i_{24}|l|}{}^l R_{i_{25}i_{26}|n|}{}^m R_{i_{27}i_{28}]}{}^n +$$

$$+ 403{,}603{,}200 \, R_{[i_1i_2|f|}{}^a R_{i_3i_4|a|}{}^b R_{i_5i_6|b|}{}^c R_{i_7i_8|c|}{}^d R_{i_9i_{10}|d|}{}^e R_{i_{11}i_{12}|e|}{}^f R_{i_{13}i_{14}|i|}{}^g R_{i_{15}i_{16}|g|}{}^h R_{i_{17}i_{18}|h|}{}^i R_{i_{19}i_{20}|l|}{}^j R_{i_{21}i_{22}|j|}{}^k R_{i_{23}i_{24}|k|}{}^l R_{i_{25}i_{26}|n|}{}^m R_{i_{27}i_{28}]m}{}^n -$$

$$- 403{,}603{,}200 \, R_{[i_1i_2|f|}{}^a R_{i_3i_4|a|}{}^b R_{i_5i_6|b|}{}^c R_{i_7i_8|c|}{}^d R_{i_9i_{10}|d|}{}^e R_{i_{11}i_{12}|e|}{}^f R_{i_{13}i_{14}|i|}{}^g R_{i_{15}i_{16}|g|}{}^h R_{i_{17}i_{18}|h|}{}^i R_{i_{19}i_{20}|l|}{}^j R_{i_{21}i_{22}|j|}{}^k R_{i_{23}i_{24}|k|}{}^l R_{i_{25}i_{26}|m|}{}^m R_{i_{27}i_{28}]}{}^n -$$

$$- 605{,}404{,}800 \, R_{[i_1i_2|f|}{}^a R_{i_3i_4|a|}{}^b R_{i_5i_6|b|}{}^c R_{i_7i_8|c|}{}^d R_{i_9i_{10}|d|}{}^e R_{i_{11}i_{12}|e|}{}^f R_{i_{13}i_{14}|i|}{}^g R_{i_{15}i_{16}|g|}{}^h R_{i_{17}i_{18}|h|}{}^i R_{i_{19}i_{20}|k|}{}^j R_{i_{21}i_{22}|j|}{}^k R_{i_{23}i_{24}|m|}{}^l R_{i_{25}i_{26}|l|}{}^m R_{i_{27}i_{28}]}{}^n +$$

$$+ 403{,}603{,}200 \, R_{[i_1i_2|f|}{}^a R_{i_3i_4|a|}{}^b R_{i_5i_6|b|}{}^c R_{i_7i_8|c|}{}^d R_{i_9i_{10}|d|}{}^e R_{i_{11}i_{12}|e|}{}^f R_{i_{13}i_{14}|i|}{}^g R_{i_{15}i_{16}|g|}{}^h R_{i_{17}i_{18}|h|}{}^i R_{i_{19}i_{20}|k|}{}^j R_{i_{21}i_{22}|j|}{}^k R_{i_{23}i_{24}|l|}{}^l R_{i_{25}i_{26}|n|}{}^m R_{i_{27}i_{28}]m}{}^n -$$

$$- 40{,}360{,}320 \, R_{[i_1i_2|f|}{}^a R_{i_3i_4|a|}{}^b R_{i_5i_6|b|}{}^c R_{i_7i_8|c|}{}^d R_{i_9i_{10}|d|}{}^e R_{i_{11}i_{12}|e|}{}^f R_{i_{13}i_{14}|i|}{}^g R_{i_{15}i_{16}|g|}{}^h R_{i_{17}i_{18}|h|}{}^i R_{i_{19}i_{20}|j|}{}^j R_{i_{21}i_{22}|k|}{}^k R_{i_{23}i_{24}|l|}{}^l R_{i_{25}i_{26}|m|}{}^m R_{i_{27}i_{28}]}{}^n -$$

$$- 37{,}837{,}800 \, R_{[i_1i_2|f|}{}^a R_{i_3i_4|a|}{}^b R_{i_5i_6|b|}{}^c R_{i_7i_8|c|}{}^d R_{i_9i_{10}|d|}{}^e R_{i_{11}i_{12}|e|}{}^f R_{i_{13}i_{14}|h|}{}^g R_{i_{15}i_{16}|g|}{}^h R_{i_{17}i_{18}|j|}{}^i R_{i_{19}i_{20}|i|}{}^j R_{i_{21}i_{22}|k|}{}^k R_{i_{23}i_{24}|k|}{}^l R_{i_{25}i_{26}|n|}{}^m R_{i_{27}i_{28}]m}{}^n +$$

$$+ 151{,}351{,}200 \, R_{[i_1i_2|f|}{}^a R_{i_3i_4|a|}{}^b R_{i_5i_6|b|}{}^c R_{i_7i_8|c|}{}^d R_{i_9i_{10}|d|}{}^e R_{i_{11}i_{12}|e|}{}^f R_{i_{13}i_{14}|h|}{}^g R_{i_{15}i_{16}|g|}{}^h R_{i_{17}i_{18}|j|}{}^i R_{i_{19}i_{20}|i|}{}^j R_{i_{21}i_{22}|l|}{}^k R_{i_{23}i_{24}|k|}{}^l R_{i_{25}i_{26}|n|}{}^m R_{i_{27}i_{28}]m}{}^n -$$

$$- 75{,}675{,}600 \, R_{[i_1i_2|f|}{}^a R_{i_3i_4|a|}{}^b R_{i_5i_6|b|}{}^c R_{i_7i_8|c|}{}^d R_{i_9i_{10}|d|}{}^e R_{i_{11}i_{12}|e|}{}^f R_{i_{13}i_{14}|h|}{}^g R_{i_{15}i_{16}|g|}{}^h R_{i_{17}i_{18}|j|}{}^i R_{i_{19}i_{20}|i|}{}^j R_{i_{21}i_{22}|k|}{}^k R_{i_{23}i_{24}|l|}{}^l R_{i_{25}i_{26}|l|}{}^m R_{i_{27}i_{28}]m}{}^n +$$

$$+ 10{,}090{,}080 \, R_{[i_1i_2|f|}{}^a R_{i_3i_4|a|}{}^b R_{i_5i_6|b|}{}^c R_{i_7i_8|c|}{}^d R_{i_9i_{10}|d|}{}^e R_{i_{11}i_{12}|e|}{}^f R_{i_{13}i_{14}|h|}{}^g R_{i_{15}i_{16}|g|}{}^h R_{i_{17}i_{18}|i|}{}^i R_{i_{19}i_{20}|i|}{}^j R_{i_{21}i_{22}|k|}{}^k R_{i_{23}i_{24}|l|}{}^l R_{i_{25}i_{26}|m|}{}^m R_{i_{27}i_{28}]}{}^n -$$

$$- 360{,}360 \, R_{[i_1i_2|f|}{}^a R_{i_3i_4|a|}{}^b R_{i_5i_6|b|}{}^c R_{i_7i_8|c|}{}^d R_{i_9i_{10}|d|}{}^e R_{i_{11}i_{12}|e|}{}^f R_{i_{13}i_{14}|g|}{}^g R_{i_{15}i_{16}|h|}{}^h R_{i_{17}i_{18}|i|}{}^i R_{i_{19}i_{20}|j|}{}^j R_{i_{21}i_{22}|k|}{}^k R_{i_{23}i_{24}|l|}{}^l R_{i_{25}i_{26}|m|}{}^m R_{i_{27}i_{28}]}{}^n -$$

$$- 435{,}891{,}456 \, R_{[i_1i_2|e|}{}^a R_{i_3i_4|a|}{}^b R_{i_5i_6|b|}{}^c R_{i_7i_8|c|}{}^d R_{i_9i_{10}|d|}{}^e R_{i_{11}i_{12}|j|}{}^f R_{i_{13}i_{14}|f|}{}^g R_{i_{15}i_{16}|g|}{}^h R_{i_{17}i_{18}|h|}{}^i R_{i_{19}i_{20}|i|}{}^j R_{i_{21}i_{22}|n|}{}^k R_{i_{23}i_{24}|k|}{}^l R_{i_{25}i_{26}|l|}{}^m R_{i_{27}i_{28}]m}{}^n +$$

$$+ 581{,}188{,}608 \, R_{[i_1i_2|e|}{}^a R_{i_3i_4|a|}{}^b R_{i_5i_6|b|}{}^c R_{i_7i_8|c|}{}^d R_{i_9i_{10}|d|}{}^e R_{i_{11}i_{12}|j|}{}^f R_{i_{13}i_{14}|f|}{}^g R_{i_{15}i_{16}|g|}{}^h R_{i_{17}i_{18}|h|}{}^i R_{i_{19}i_{20}|i|}{}^j R_{i_{21}i_{22}|m|}{}^k R_{i_{23}i_{24}|k|}{}^l R_{i_{25}i_{26}|l|}{}^m R_{i_{27}i_{28}]}{}^n +$$

$$+ 217{,}945{,}728 \, R_{[i_1i_2|e|}{}^a R_{i_3i_4|a|}{}^b R_{i_5i_6|b|}{}^c R_{i_7i_8|c|}{}^d R_{i_9i_{10}|d|}{}^e R_{i_{11}i_{12}|j|}{}^f R_{i_{13}i_{14}|f|}{}^g R_{i_{15}i_{16}|g|}{}^h R_{i_{17}i_{18}|h|}{}^i R_{i_{19}i_{20}|i|}{}^j R_{i_{21}i_{22}|k|}{}^k R_{i_{23}i_{24}|l|}{}^l R_{i_{25}i_{26}|n|}{}^m R_{i_{27}i_{28}]m}{}^n -$$

$$- 435{,}891{,}456 \, R_{[i_1i_2|e|}{}^a R_{i_3i_4|a|}{}^b R_{i_5i_6|b|}{}^c R_{i_7i_8|c|}{}^d R_{i_9i_{10}|d|}{}^e R_{i_{11}i_{12}|j|}{}^f R_{i_{13}i_{14}|f|}{}^g R_{i_{15}i_{16}|g|}{}^h R_{i_{17}i_{18}|h|}{}^i R_{i_{19}i_{20}|i|}{}^j R_{i_{21}i_{22}|l|}{}^k R_{i_{23}i_{24}|k|}{}^l R_{i_{25}i_{26}|l|}{}^m R_{i_{27}i_{28}]m}{}^n +$$

$$+ 72{,}648{,}576 \, R_{[i_1i_2|e|}{}^a R_{i_3i_4|a|}{}^b R_{i_5i_6|b|}{}^c R_{i_7i_8|c|}{}^d R_{i_9i_{10}|d|}{}^e R_{i_{11}i_{12}|j|}{}^f R_{i_{13}i_{14}|f|}{}^g R_{i_{15}i_{16}|g|}{}^h R_{i_{17}i_{18}|h|}{}^i R_{i_{19}i_{20}|i|}{}^j R_{i_{21}i_{22}|k|}{}^k R_{i_{23}i_{24}|l|}{}^l R_{i_{25}i_{26}|m|}{}^m R_{i_{27}i_{28}]}{}^n +$$

$$+ 544{,}864{,}320 \, R_{[i_1i_2|e|}{}^a R_{i_3i_4|a|}{}^b R_{i_5i_6|b|}{}^c R_{i_7i_8|c|}{}^d R_{i_9i_{10}|d|}{}^e R_{i_{11}i_{12}|i|}{}^f R_{i_{13}i_{14}|f|}{}^g R_{i_{15}i_{16}|g|}{}^h R_{i_{17}i_{18}|h|}{}^i R_{i_{19}i_{20}|m|}{}^j R_{i_{21}i_{22}|j|}{}^k R_{i_{23}i_{24}|k|}{}^l R_{i_{25}i_{26}|l|}{}^m R_{i_{27}i_{28}]}{}^n +$$

$$+ 726{,}485{,}760 \, R_{[i_1i_2|e|}{}^a R_{i_3i_4|a|}{}^b R_{i_5i_6|b|}{}^c R_{i_7i_8|c|}{}^d R_{i_9i_{10}|d|}{}^e R_{i_{11}i_{12}|i|}{}^f R_{i_{13}i_{14}|f|}{}^g R_{i_{15}i_{16}|g|}{}^h R_{i_{17}i_{18}|h|}{}^i R_{i_{19}i_{20}|i|}{}^j R_{i_{21}i_{22}|j|}{}^k R_{i_{23}i_{24}|k|}{}^l R_{i_{25}i_{26}|l|}{}^m R_{i_{27}i_{28}]}{}^n -$$

$$- 726{,}485{,}760 \, R_{[i_1i_2|e|}{}^a R_{i_3i_4|a|}{}^b R_{i_5i_6|b|}{}^c R_{i_7i_8|c|}{}^d R_{i_9i_{10}|d|}{}^e R_{i_{11}i_{12}|i|}{}^f R_{i_{13}i_{14}|f|}{}^g R_{i_{15}i_{16}|g|}{}^h R_{i_{17}i_{18}|h|}{}^i R_{i_{19}i_{20}|l|}{}^j R_{i_{21}i_{22}|j|}{}^k R_{i_{23}i_{24}|k|}{}^l R_{i_{25}i_{26}|n|}{}^m R_{i_{27}i_{28}]m}{}^n -$$

$$- 544{,}864{,}320 \, R_{[i_1i_2|e|}{}^a R_{i_3i_4|a|}{}^b R_{i_5i_6|b|}{}^c R_{i_7i_8|c|}{}^d R_{i_9i_{10}|d|}{}^e R_{i_{11}i_{12}|i|}{}^f R_{i_{13}i_{14}|f|}{}^g R_{i_{15}i_{16}|g|}{}^h R_{i_{17}i_{18}|h|}{}^i R_{i_{19}i_{20}|k|}{}^j R_{i_{21}i_{22}|j|}{}^k R_{i_{23}i_{24}|m|}{}^l R_{i_{25}i_{26}|l|}{}^m R_{i_{27}i_{28}]}{}^n +$$

$$+ 363{,}242{,}880 \, R_{[i_1i_2|e|}{}^a R_{i_3i_4|a|}{}^b R_{i_5i_6|b|}{}^c R_{i_7i_8|c|}{}^d R_{i_9i_{10}|d|}{}^e R_{i_{11}i_{12}|i|}{}^f R_{i_{13}i_{14}|f|}{}^g R_{i_{15}i_{16}|g|}{}^h R_{i_{17}i_{18}|h|}{}^i R_{i_{19}i_{20}|k|}{}^j R_{i_{21}i_{22}|j|}{}^k R_{i_{23}i_{24}|k|}{}^l R_{i_{25}i_{26}|l|}{}^m R_{i_{27}i_{28}]m}{}^n -$$

$$- 36{,}324{,}288 \, R_{[i_1i_2|e|}{}^a R_{i_3i_4|a|}{}^b R_{i_5i_6|b|}{}^c R_{i_7i_8|c|}{}^d R_{i_9i_{10}|d|}{}^e R_{i_{11}i_{12}|i|}{}^f R_{i_{13}i_{14}|f|}{}^g R_{i_{15}i_{16}|g|}{}^h R_{i_{17}i_{18}|h|}{}^i R_{i_{19}i_{20}|j|}{}^j R_{i_{21}i_{22}|k|}{}^k R_{i_{23}i_{24}|l|}{}^l R_{i_{25}i_{26}|m|}{}^m R_{i_{27}i_{28}]}{}^n +$$

$$+ 107{,}627{,}520 \, R_{[i_1i_2|e|}{}^a R_{i_3i_4|a|}{}^b R_{i_5i_6|b|}{}^c R_{i_7i_8|c|}{}^d R_{i_9i_{10}|d|}{}^e R_{i_{11}i_{12}|i|}{}^f R_{i_{13}i_{14}|f|}{}^g R_{i_{15}i_{16}|g|}{}^h R_{i_{17}i_{18}|k|}{}^i R_{i_{19}i_{20}|i|}{}^j R_{i_{21}i_{22}|j|}{}^k R_{i_{23}i_{24}|n|}{}^l R_{i_{25}i_{26}|l|}{}^m R_{i_{27}i_{28}]m}{}^n -$$

$$- 484{,}323{,}840 \, R_{[i_1i_2|e|}{}^a R_{i_3i_4|a|}{}^b R_{i_5i_6|b|}{}^c R_{i_7i_8|c|}{}^d R_{i_9i_{10}|d|}{}^e R_{i_{11}i_{12}|h|}{}^f R_{i_{13}i_{14}|f|}{}^g R_{i_{15}i_{16}|g|}{}^h R_{i_{17}i_{18}|k|}{}^i R_{i_{19}i_{20}|i|}{}^j R_{i_{21}i_{22}|j|}{}^k R_{i_{23}i_{24}|m|}{}^l R_{i_{25}i_{26}|l|}{}^m R_{i_{27}i_{28}]}{}^n +$$

$$+ 161{,}441{,}280 \, R_{[i_1i_2|e|}{}^a R_{i_3i_4|a|}{}^b R_{i_5i_6|b|}{}^c R_{i_7i_8|c|}{}^d R_{i_9i_{10}|d|}{}^e R_{i_{11}i_{12}|h|}{}^f R_{i_{13}i_{14}|f|}{}^g R_{i_{15}i_{16}|g|}{}^h R_{i_{17}i_{18}|k|}{}^i R_{i_{19}i_{20}|i|}{}^j R_{i_{21}i_{22}|j|}{}^k R_{i_{23}i_{24}|l|}{}^l R_{i_{25}i_{26}|l|}{}^m R_{i_{27}i_{28}]m}{}^n -$$

$$- 121{,}080{,}960 \, R_{[i_1i_2|e|}{}^a R_{i_3i_4|a|}{}^b R_{i_5i_6|b|}{}^c R_{i_7i_8|c|}{}^d R_{i_9i_{10}|d|}{}^e R_{i_{11}i_{12}|h|}{}^f R_{i_{13}i_{14}|f|}{}^g R_{i_{15}i_{16}|g|}{}^h R_{i_{17}i_{18}|j|}{}^i R_{i_{19}i_{20}|i|}{}^j R_{i_{21}i_{22}|l|}{}^k R_{i_{23}i_{24}|k|}{}^l R_{i_{25}i_{26}|n|}{}^m R_{i_{27}i_{28}]m}{}^n +$$

$$+ 363{,}242{,}880 \, R_{[i_1i_2|e|}{}^a R_{i_3i_4|a|}{}^b R_{i_5i_6|b|}{}^c R_{i_7i_8|c|}{}^d R_{i_9i_{10}|d|}{}^e R_{i_{11}i_{12}|h|}{}^f R_{i_{13}i_{14}|f|}{}^g R_{i_{15}i_{16}|g|}{}^h R_{i_{17}i_{18}|j|}{}^i R_{i_{19}i_{20}|i|}{}^j R_{i_{21}i_{22}|l|}{}^k R_{i_{23}i_{24}|k|}{}^l R_{i_{25}i_{26}|l|}{}^m R_{i_{27}i_{28}]m}{}^n -$$

$$- 121{,}080{,}960 \, R_{[i_1i_2|e|}{}^a R_{i_3i_4|a|}{}^b R_{i_5i_6|b|}{}^c R_{i_7i_8|c|}{}^d R_{i_9i_{10}|d|}{}^e R_{i_{11}i_{12}|h|}{}^f R_{i_{13}i_{14}|f|}{}^g R_{i_{15}i_{16}|g|}{}^h R_{i_{17}i_{18}|i|}{}^i R_{i_{19}i_{20}|i|}{}^j R_{i_{21}i_{22}|j|}{}^k R_{i_{23}i_{24}|k|}{}^l R_{i_{25}i_{26}|l|}{}^m R_{i_{27}i_{28}]m}{}^n +$$

$$+ 8{,}072{,}064 \, R_{[i_1i_2|e|}{}^a R_{i_3i_4|a|}{}^b R_{i_5i_6|b|}{}^c R_{i_7i_8|c|}{}^d R_{i_9i_{10}|d|}{}^e R_{i_{11}i_{12}|h|}{}^f R_{i_{13}i_{14}|f|}{}^g R_{i_{15}i_{16}|g|}{}^h R_{i_{17}i_{18}|i|}{}^i R_{i_{19}i_{20}|i|}{}^j R_{i_{21}i_{22}|j|}{}^k R_{i_{23}i_{24}|k|}{}^l R_{i_{25}i_{26}|l|}{}^m R_{i_{27}i_{28}]m}{}^n +$$

$$+ 45{,}405{,}360 \, R_{[i_1i_2|e|}{}^a R_{i_3i_4|a|}{}^b R_{i_5i_6|b|}{}^c R_{i_7i_8|c|}{}^d R_{i_9i_{10}|d|}{}^e R_{i_{11}i_{12}|g|}{}^f R_{i_{13}i_{14}|f|}{}^g R_{i_{15}i_{16}|i|}{}^h R_{i_{17}i_{18}|h|}{}^i R_{i_{19}i_{20}|k|}{}^j R_{i_{21}i_{22}|j|}{}^k R_{i_{23}i_{24}|m|}{}^l R_{i_{25}i_{26}|l|}{}^m R_{i_{27}i_{28}]}{}^n -$$

$$- 60{,}540{,}480 \, R_{[i_1i_2|e|}{}^a R_{i_3i_4|a|}{}^b R_{i_5i_6|b|}{}^c R_{i_7i_8|c|}{}^d R_{i_9i_{10}|d|}{}^e R_{i_{11}i_{12}|g|}{}^f R_{i_{13}i_{14}|f|}{}^g R_{i_{15}i_{16}|i|}{}^h R_{i_{17}i_{18}|h|}{}^i R_{i_{19}i_{20}|j|}{}^j R_{i_{21}i_{22}|k|}{}^k R_{i_{23}i_{24}|k|}{}^l R_{i_{25}i_{26}|l|}{}^m R_{i_{27}i_{28}]m}{}^n +$$

$$+ 18{,}162{,}144 \, R_{[i_1i_2|e|}{}^a R_{i_3i_4|a|}{}^b R_{i_5i_6|b|}{}^c R_{i_7i_8|c|}{}^d R_{i_9i_{10}|d|}{}^e R_{i_{11}i_{12}|g|}{}^f R_{i_{13}i_{14}|f|}{}^g R_{i_{15}i_{16}|i|}{}^h R_{i_{17}i_{18}|h|}{}^i R_{i_{19}i_{20}|j|}{}^j R_{i_{21}i_{22}|k|}{}^k R_{i_{23}i_{24}|l|}{}^l R_{i_{25}i_{26}|l|}{}^m R_{i_{27}i_{28}]m}{}^n -$$

$$- 1{,}729{,}728 \, R_{[i_1i_2|e|}{}^a R_{i_3i_4|a|}{}^b R_{i_5i_6|b|}{}^c R_{i_7i_8|c|}{}^d R_{i_9i_{10}|d|}{}^e R_{i_{11}i_{12}|g|}{}^f R_{i_{13}i_{14}|f|}{}^g R_{i_{15}i_{16}|h|}{}^h R_{i_{17}i_{18}|j|}{}^i R_{i_{19}i_{20}|i|}{}^j R_{i_{21}i_{22}|k|}{}^k R_{i_{23}i_{24}|l|}{}^l R_{i_{25}i_{26}|m|}{}^m R_{i_{27}i_{28}]}{}^n +$$

$$+ 48{,}048 \, R_{[i_1i_2|e|}{}^a R_{i_3i_4|a|}{}^b R_{i_5i_6|b|}{}^c R_{i_7i_8|c|}{}^d R_{i_9i_{10}|d|}{}^e R_{i_{11}i_{12}|f|}{}^f R_{i_{13}i_{14}|g|}{}^g R_{i_{15}i_{16}|h|}{}^h R_{i_{17}i_{18}|i|}{}^i R_{i_{19}i_{20}|j|}{}^j R_{i_{21}i_{22}|k|}{}^k R_{i_{23}i_{24}|l|}{}^l R_{i_{25}i_{26}|m|}{}^m R_{i_{27}i_{28}]}{}^n +$$

$$+ 113{,}513{,}400\, R_{[i_1 i_2|d|}{}^a\, R_{i_3 i_4|a|}{}^b\, R_{i_5 i_6|b|}{}^c\, R_{i_7 i_8|c|}{}^d\, R_{i_9 i_{10}|h|}{}^e\, R_{i_{11} i_{12}|e|}{}^f\, R_{i_{13} i_{14}|f|}{}^g\, R_{i_{15} i_{16}|g|}{}^h\, R_{i_{17} i_{18}|l|}{}^i\, R_{i_{19} i_{20}|i|}{}^j\, R_{i_{21} i_{22}|j|}{}^k\, R_{i_{23} i_{24}|k|}{}^l\, R_{i_{25} i_{26}|n|}{}^m\, R_{i_{27} i_{28}]m}{}^n -$$

$$- 113{,}513{,}400\, R_{[i_1 i_2|d|}{}^a\, R_{i_3 i_4|a|}{}^b\, R_{i_5 i_6|b|}{}^c\, R_{i_7 i_8|c|}{}^d\, R_{i_9 i_{10}|h|}{}^e\, R_{i_{11} i_{12}|e|}{}^f\, R_{i_{13} i_{14}|f|}{}^g\, R_{i_{15} i_{16}|g|}{}^h\, R_{i_{17} i_{18}|l|}{}^i\, R_{i_{19} i_{20}|i|}{}^j\, R_{i_{21} i_{22}|j|}{}^k\, R_{i_{23} i_{24}|k|}{}^l\, R_{i_{25} i_{26}|m|}{}^m\, R_{i_{27} i_{28}]n}{}^n +$$

$$+ 151{,}351{,}200\, R_{[i_1 i_2|d|}{}^a\, R_{i_3 i_4|a|}{}^b\, R_{i_5 i_6|b|}{}^c\, R_{i_7 i_8|c|}{}^d\, R_{i_9 i_{10}|h|}{}^e\, R_{i_{11} i_{12}|e|}{}^f\, R_{i_{13} i_{14}|f|}{}^g\, R_{i_{15} i_{16}|g|}{}^h\, R_{i_{17} i_{18}|k|}{}^i\, R_{i_{19} i_{20}|i|}{}^j\, R_{i_{21} i_{22}|j|}{}^k\, R_{i_{23} i_{24}|n|}{}^l\, R_{i_{25} i_{26}|l|}{}^m\, R_{i_{27} i_{28}]m}{}^n -$$

$$- 454{,}053{,}600\, R_{[i_1 i_2|d|}{}^a\, R_{i_3 i_4|a|}{}^b\, R_{i_5 i_6|b|}{}^c\, R_{i_7 i_8|c|}{}^d\, R_{i_9 i_{10}|h|}{}^e\, R_{i_{11} i_{12}|e|}{}^f\, R_{i_{13} i_{14}|f|}{}^g\, R_{i_{15} i_{16}|g|}{}^h\, R_{i_{17} i_{18}|k|}{}^i\, R_{i_{19} i_{20}|i|}{}^j\, R_{i_{21} i_{22}|j|}{}^k\, R_{i_{23} i_{24}|l|}{}^l\, R_{i_{25} i_{26}|n|}{}^m\, R_{i_{27} i_{28}]m}{}^n +$$

$$+ 151{,}351{,}200\, R_{[i_1 i_2|d|}{}^a\, R_{i_3 i_4|a|}{}^b\, R_{i_5 i_6|b|}{}^c\, R_{i_7 i_8|c|}{}^d\, R_{i_9 i_{10}|h|}{}^e\, R_{i_{11} i_{12}|e|}{}^f\, R_{i_{13} i_{14}|f|}{}^g\, R_{i_{15} i_{16}|g|}{}^h\, R_{i_{17} i_{18}|k|}{}^i\, R_{i_{19} i_{20}|i|}{}^j\, R_{i_{21} i_{22}|j|}{}^k\, R_{i_{23} i_{24}|l|}{}^l\, R_{i_{25} i_{26}|m|}{}^m\, R_{i_{27} i_{28}]n}{}^n -$$

$$- 56{,}756{,}700\, R_{[i_1 i_2|d|}{}^a\, R_{i_3 i_4|a|}{}^b\, R_{i_5 i_6|b|}{}^c\, R_{i_7 i_8|c|}{}^d\, R_{i_9 i_{10}|h|}{}^e\, R_{i_{11} i_{12}|e|}{}^f\, R_{i_{13} i_{14}|f|}{}^g\, R_{i_{15} i_{16}|g|}{}^h\, R_{i_{17} i_{18}|j|}{}^i\, R_{i_{19} i_{20}|i|}{}^j\, R_{i_{21} i_{22}|k|}{}^k\, R_{i_{23} i_{24}|l|}{}^l\, R_{i_{25} i_{26}|n|}{}^m\, R_{i_{27} i_{28}]m}{}^n +$$

$$+ 170{,}270{,}100\, R_{[i_1 i_2|d|}{}^a\, R_{i_3 i_4|a|}{}^b\, R_{i_5 i_6|b|}{}^c\, R_{i_7 i_8|c|}{}^d\, R_{i_9 i_{10}|h|}{}^e\, R_{i_{11} i_{12}|e|}{}^f\, R_{i_{13} i_{14}|f|}{}^g\, R_{i_{15} i_{16}|g|}{}^h\, R_{i_{17} i_{18}|j|}{}^i\, R_{i_{19} i_{20}|i|}{}^j\, R_{i_{21} i_{22}|k|}{}^k\, R_{i_{23} i_{24}|l|}{}^l\, R_{i_{25} i_{26}|m|}{}^m\, R_{i_{27} i_{28}]n}{}^n -$$

$$- 56{,}756{,}700\, R_{[i_1 i_2|d|}{}^a\, R_{i_3 i_4|a|}{}^b\, R_{i_5 i_6|b|}{}^c\, R_{i_7 i_8|c|}{}^d\, R_{i_9 i_{10}|h|}{}^e\, R_{i_{11} i_{12}|e|}{}^f\, R_{i_{13} i_{14}|f|}{}^g\, R_{i_{15} i_{16}|g|}{}^h\, R_{i_{17} i_{18}|j|}{}^i\, R_{i_{19} i_{20}|i|}{}^j\, R_{i_{21} i_{22}|k|}{}^k\, R_{i_{23} i_{24}|l|}{}^l\, R_{i_{25} i_{26}|m|}{}^m\, R_{i_{27} i_{28}]n}{}^n +$$

$$+ 3{,}783{,}780\, R_{[i_1 i_2|d|}{}^a\, R_{i_3 i_4|a|}{}^b\, R_{i_5 i_6|b|}{}^c\, R_{i_7 i_8|c|}{}^d\, R_{i_9 i_{10}|h|}{}^e\, R_{i_{11} i_{12}|e|}{}^f\, R_{i_{13} i_{14}|f|}{}^g\, R_{i_{15} i_{16}|g|}{}^h\, R_{i_{17} i_{18}|i|}{}^i\, R_{i_{19} i_{20}|j|}{}^j\, R_{i_{21} i_{22}|k|}{}^k\, R_{i_{23} i_{24}|l|}{}^l\, R_{i_{25} i_{26}|m|}{}^m\, R_{i_{27} i_{28}]n}{}^n -$$

$$- 134{,}534{,}400\, R_{[i_1 i_2|d|}{}^a\, R_{i_3 i_4|a|}{}^b\, R_{i_5 i_6|b|}{}^c\, R_{i_7 i_8|c|}{}^d\, R_{i_9 i_{10}|g|}{}^e\, R_{i_{11} i_{12}|e|}{}^f\, R_{i_{13} i_{14}|f|}{}^g\, R_{i_{15} i_{16}|j|}{}^h\, R_{i_{17} i_{18}|h|}{}^i\, R_{i_{19} i_{20}|i|}{}^j\, R_{i_{21} i_{22}|m|}{}^k\, R_{i_{23} i_{24}|k|}{}^l\, R_{i_{25} i_{26}|l|}{}^m\, R_{i_{27} i_{28}]n}{}^n -$$

$$- 151{,}351{,}200\, R_{[i_1 i_2|d|}{}^a\, R_{i_3 i_4|a|}{}^b\, R_{i_5 i_6|b|}{}^c\, R_{i_7 i_8|c|}{}^d\, R_{i_9 i_{10}|g|}{}^e\, R_{i_{11} i_{12}|e|}{}^f\, R_{i_{13} i_{14}|f|}{}^g\, R_{i_{15} i_{16}|j|}{}^h\, R_{i_{17} i_{18}|h|}{}^i\, R_{i_{19} i_{20}|i|}{}^j\, R_{i_{21} i_{22}|k|}{}^k\, R_{i_{23} i_{24}|l|}{}^l\, R_{i_{25} i_{26}|n|}{}^m\, R_{i_{27} i_{28}]m}{}^n +$$

$$+ 302{,}702{,}400\, R_{[i_1 i_2|d|}{}^a\, R_{i_3 i_4|a|}{}^b\, R_{i_5 i_6|b|}{}^c\, R_{i_7 i_8|c|}{}^d\, R_{i_9 i_{10}|g|}{}^e\, R_{i_{11} i_{12}|e|}{}^f\, R_{i_{13} i_{14}|f|}{}^g\, R_{i_{15} i_{16}|j|}{}^h\, R_{i_{17} i_{18}|h|}{}^i\, R_{i_{19} i_{20}|i|}{}^j\, R_{i_{21} i_{22}|l|}{}^k\, R_{i_{23} i_{24}|k|}{}^l\, R_{i_{25} i_{26}|n|}{}^m\, R_{i_{27} i_{28}]m}{}^n -$$

$$- 50{,}450{,}400\, R_{[i_1 i_2|d|}{}^a\, R_{i_3 i_4|a|}{}^b\, R_{i_5 i_6|b|}{}^c\, R_{i_7 i_8|c|}{}^d\, R_{i_9 i_{10}|g|}{}^e\, R_{i_{11} i_{12}|e|}{}^f\, R_{i_{13} i_{14}|f|}{}^g\, R_{i_{15} i_{16}|j|}{}^h\, R_{i_{17} i_{18}|h|}{}^i\, R_{i_{19} i_{20}|i|}{}^j\, R_{i_{21} i_{22}|l|}{}^k\, R_{i_{23} i_{24}|k|}{}^l\, R_{i_{25} i_{26}|m|}{}^m\, R_{i_{27} i_{28}]n}{}^n +$$

$$+ 151{,}351{,}200\, R_{[i_1 i_2|d|}{}^a\, R_{i_3 i_4|a|}{}^b\, R_{i_5 i_6|b|}{}^c\, R_{i_7 i_8|c|}{}^d\, R_{i_9 i_{10}|g|}{}^e\, R_{i_{11} i_{12}|e|}{}^f\, R_{i_{13} i_{14}|f|}{}^g\, R_{i_{15} i_{16}|i|}{}^h\, R_{i_{17} i_{18}|h|}{}^i\, R_{i_{19} i_{20}|k|}{}^j\, R_{i_{21} i_{22}|j|}{}^k\, R_{i_{23} i_{24}|l|}{}^l\, R_{i_{25} i_{26}|m|}{}^m\, R_{i_{27} i_{28}]n}{}^n -$$

$$- 151{,}351{,}200\, R_{[i_1 i_2|d|}{}^a\, R_{i_3 i_4|a|}{}^b\, R_{i_5 i_6|b|}{}^c\, R_{i_7 i_8|c|}{}^d\, R_{i_9 i_{10}|g|}{}^e\, R_{i_{11} i_{12}|e|}{}^f\, R_{i_{13} i_{14}|f|}{}^g\, R_{i_{15} i_{16}|i|}{}^h\, R_{i_{17} i_{18}|h|}{}^i\, R_{i_{19} i_{20}|k|}{}^j\, R_{i_{21} i_{22}|j|}{}^k\, R_{i_{23} i_{24}|l|}{}^l\, R_{i_{25} i_{26}|m|}{}^m\, R_{i_{27} i_{28}]n}{}^n +$$

$$+ 30{,}270{,}240\, R_{[i_1 i_2|d|}{}^a\, R_{i_3 i_4|a|}{}^b\, R_{i_5 i_6|b|}{}^c\, R_{i_7 i_8|c|}{}^d\, R_{i_9 i_{10}|g|}{}^e\, R_{i_{11} i_{12}|e|}{}^f\, R_{i_{13} i_{14}|f|}{}^g\, R_{i_{15} i_{16}|i|}{}^h\, R_{i_{17} i_{18}|h|}{}^i\, R_{i_{19} i_{20}|j|}{}^j\, R_{i_{21} i_{22}|k|}{}^k\, R_{i_{23} i_{24}|l|}{}^l\, R_{i_{25} i_{26}|m|}{}^m\, R_{i_{27} i_{28}]n}{}^n -$$

$$- 1{,}441{,}440\, R_{[i_1 i_2|d|}{}^a\, R_{i_3 i_4|a|}{}^b\, R_{i_5 i_6|b|}{}^c\, R_{i_7 i_8|c|}{}^d\, R_{i_9 i_{10}|g|}{}^e\, R_{i_{11} i_{12}|e|}{}^f\, R_{i_{13} i_{14}|f|}{}^g\, R_{i_{15} i_{16}|h|}{}^h\, R_{i_{17} i_{18}|i|}{}^i\, R_{i_{19} i_{20}|j|}{}^j\, R_{i_{21} i_{22}|k|}{}^k\, R_{i_{23} i_{24}|l|}{}^l\, R_{i_{25} i_{26}|m|}{}^m\, R_{i_{27} i_{28}]n}{}^n +$$

$$+ 5{,}675{,}670\, R_{[i_1 i_2|d|}{}^a\, R_{i_3 i_4|a|}{}^b\, R_{i_5 i_6|b|}{}^c\, R_{i_7 i_8|c|}{}^d\, R_{i_9 i_{10}|f|}{}^e\, R_{i_{11} i_{12}|e|}{}^f\, R_{i_{13} i_{14}|h|}{}^g\, R_{i_{15} i_{16}|g|}{}^h\, R_{i_{17} i_{18}|i|}{}^i\, R_{i_{19} i_{20}|j|}{}^j\, R_{i_{21} i_{22}|l|}{}^k\, R_{i_{23} i_{24}|k|}{}^l\, R_{i_{25} i_{26}|n|}{}^m\, R_{i_{27} i_{28}]m}{}^n -$$

$$- 28{,}378{,}350\, R_{[i_1 i_2|d|}{}^a\, R_{i_3 i_4|a|}{}^b\, R_{i_5 i_6|b|}{}^c\, R_{i_7 i_8|c|}{}^d\, R_{i_9 i_{10}|f|}{}^e\, R_{i_{11} i_{12}|e|}{}^f\, R_{i_{13} i_{14}|h|}{}^g\, R_{i_{15} i_{16}|g|}{}^h\, R_{i_{17} i_{18}|j|}{}^i\, R_{i_{19} i_{20}|i|}{}^j\, R_{i_{21} i_{22}|l|}{}^k\, R_{i_{23} i_{24}|k|}{}^l\, R_{i_{25} i_{26}|m|}{}^m\, R_{i_{27} i_{28}]n}{}^n +$$

$$+ 18{,}918{,}900\, R_{[i_1 i_2|d|}{}^a\, R_{i_3 i_4|a|}{}^b\, R_{i_5 i_6|b|}{}^c\, R_{i_7 i_8|c|}{}^d\, R_{i_9 i_{10}|f|}{}^e\, R_{i_{11} i_{12}|e|}{}^f\, R_{i_{13} i_{14}|h|}{}^g\, R_{i_{15} i_{16}|g|}{}^h\, R_{i_{17} i_{18}|j|}{}^i\, R_{i_{19} i_{20}|i|}{}^j\, R_{i_{21} i_{22}|l|}{}^k\, R_{i_{23} i_{24}|k|}{}^l\, R_{i_{25} i_{26}|m|}{}^m\, R_{i_{27} i_{28}]n}{}^n -$$

$$- 3{,}783{,}780\, R_{[i_1 i_2|d|}{}^a\, R_{i_3 i_4|a|}{}^b\, R_{i_5 i_6|b|}{}^c\, R_{i_7 i_8|c|}{}^d\, R_{i_9 i_{10}|f|}{}^e\, R_{i_{11} i_{12}|e|}{}^f\, R_{i_{13} i_{14}|h|}{}^g\, R_{i_{15} i_{16}|g|}{}^h\, R_{i_{17} i_{18}|i|}{}^i\, R_{i_{19} i_{20}|j|}{}^j\, R_{i_{21} i_{22}|k|}{}^k\, R_{i_{23} i_{24}|l|}{}^l\, R_{i_{25} i_{26}|m|}{}^m\, R_{i_{27} i_{28}]n}{}^n +$$

$$+ 270{,}270\, R_{[i_1 i_2|d|}{}^a\, R_{i_3 i_4|a|}{}^b\, R_{i_5 i_6|b|}{}^c\, R_{i_7 i_8|c|}{}^d\, R_{i_9 i_{10}|f|}{}^e\, R_{i_{11} i_{12}|e|}{}^f\, R_{i_{13} i_{14}|g|}{}^g\, R_{i_{15} i_{16}|h|}{}^h\, R_{i_{17} i_{18}|i|}{}^i\, R_{i_{19} i_{20}|j|}{}^j\, R_{i_{21} i_{22}|k|}{}^k\, R_{i_{23} i_{24}|l|}{}^l\, R_{i_{25} i_{26}|m|}{}^m\, R_{i_{27} i_{28}]n}{}^n -$$

$$- 6{,}006\, R_{[i_1 i_2|d|}{}^a\, R_{i_3 i_4|a|}{}^b\, R_{i_5 i_6|b|}{}^c\, R_{i_7 i_8|c|}{}^d\, R_{i_9 i_{10}|e|}{}^e\, R_{i_{11} i_{12}|f|}{}^f\, R_{i_{13} i_{14}|g|}{}^g\, R_{i_{15} i_{16}|h|}{}^h\, R_{i_{17} i_{18}|i|}{}^i\, R_{i_{19} i_{20}|j|}{}^j\, R_{i_{21} i_{22}|k|}{}^k\, R_{i_{23} i_{24}|l|}{}^l\, R_{i_{25} i_{26}|m|}{}^m\, R_{i_{27} i_{28}]n}{}^n -$$

$$- 22{,}422{,}400\, R_{[i_1 i_2|c|}{}^a\, R_{i_3 i_4|a|}{}^b\, R_{i_5 i_6|b|}{}^c\, R_{i_7 i_8|f|}{}^d\, R_{i_9 i_{10}|d|}{}^e\, R_{i_{11} i_{12}|e|}{}^f\, R_{i_{13} i_{14}|i|}{}^g\, R_{i_{15} i_{16}|g|}{}^h\, R_{i_{17} i_{18}|h|}{}^i\, R_{i_{19} i_{20}|l|}{}^j\, R_{i_{21} i_{22}|j|}{}^k\, R_{i_{23} i_{24}|k|}{}^l\, R_{i_{25} i_{26}|n|}{}^m\, R_{i_{27} i_{28}]m}{}^n +$$

$$+ 22{,}422{,}400\, R_{[i_1 i_2|c|}{}^a\, R_{i_3 i_4|a|}{}^b\, R_{i_5 i_6|b|}{}^c\, R_{i_7 i_8|f|}{}^d\, R_{i_9 i_{10}|d|}{}^e\, R_{i_{11} i_{12}|e|}{}^f\, R_{i_{13} i_{14}|i|}{}^g\, R_{i_{15} i_{16}|g|}{}^h\, R_{i_{17} i_{18}|h|}{}^i\, R_{i_{19} i_{20}|l|}{}^j\, R_{i_{21} i_{22}|j|}{}^k\, R_{i_{23} i_{24}|k|}{}^l\, R_{i_{25} i_{26}|m|}{}^m\, R_{i_{27} i_{28}]n}{}^n +$$

$$+ 67{,}267{,}200\, R_{[i_1 i_2|c|}{}^a\, R_{i_3 i_4|a|}{}^b\, R_{i_5 i_6|b|}{}^c\, R_{i_7 i_8|f|}{}^d\, R_{i_9 i_{10}|d|}{}^e\, R_{i_{11} i_{12}|e|}{}^f\, R_{i_{13} i_{14}|i|}{}^g\, R_{i_{15} i_{16}|g|}{}^h\, R_{i_{17} i_{18}|h|}{}^i\, R_{i_{19} i_{20}|k|}{}^j\, R_{i_{21} i_{22}|j|}{}^k\, R_{i_{23} i_{24}|m|}{}^l\, R_{i_{25} i_{26}|l|}{}^m\, R_{i_{27} i_{28}]n}{}^n -$$

$$- 44{,}844{,}800\, R_{[i_1 i_2|c|}{}^a\, R_{i_3 i_4|a|}{}^b\, R_{i_5 i_6|b|}{}^c\, R_{i_7 i_8|f|}{}^d\, R_{i_9 i_{10}|d|}{}^e\, R_{i_{11} i_{12}|e|}{}^f\, R_{i_{13} i_{14}|i|}{}^g\, R_{i_{15} i_{16}|g|}{}^h\, R_{i_{17} i_{18}|h|}{}^i\, R_{i_{19} i_{20}|l|}{}^j\, R_{i_{21} i_{22}|j|}{}^k\, R_{i_{23} i_{24}|k|}{}^l\, R_{i_{25} i_{26}|l|}{}^m\, R_{i_{27} i_{28}]n}{}^n +$$

$$+ 4{,}484{,}480\, R_{[i_1 i_2|c|}{}^a\, R_{i_3 i_4|a|}{}^b\, R_{i_5 i_6|b|}{}^c\, R_{i_7 i_8|f|}{}^d\, R_{i_9 i_{10}|d|}{}^e\, R_{i_{11} i_{12}|e|}{}^f\, R_{i_{13} i_{14}|g|}{}^g\, R_{i_{15} i_{16}|g|}{}^h\, R_{i_{17} i_{18}|h|}{}^i\, R_{i_{19} i_{20}|j|}{}^j\, R_{i_{21} i_{22}|k|}{}^k\, R_{i_{23} i_{24}|l|}{}^l\, R_{i_{25} i_{26}|m|}{}^m\, R_{i_{27} i_{28}]n}{}^n +$$

$$+ 12{,}612{,}600\, R_{[i_1 i_2|c|}{}^a\, R_{i_3 i_4|a|}{}^b\, R_{i_5 i_6|b|}{}^c\, R_{i_7 i_8|f|}{}^d\, R_{i_9 i_{10}|d|}{}^e\, R_{i_{11} i_{12}|e|}{}^f\, R_{i_{13} i_{14}|h|}{}^g\, R_{i_{15} i_{16}|g|}{}^h\, R_{i_{17} i_{18}|j|}{}^i\, R_{i_{19} i_{20}|i|}{}^j\, R_{i_{21} i_{22}|l|}{}^k\, R_{i_{23} i_{24}|k|}{}^l\, R_{i_{25} i_{26}|n|}{}^m\, R_{i_{27} i_{28}]m}{}^n -$$

$$- 50{,}450{,}400\, R_{[i_1 i_2|c|}{}^a\, R_{i_3 i_4|a|}{}^b\, R_{i_5 i_6|b|}{}^c\, R_{i_7 i_8|f|}{}^d\, R_{i_9 i_{10}|d|}{}^e\, R_{i_{11} i_{12}|e|}{}^f\, R_{i_{13} i_{14}|h|}{}^g\, R_{i_{15} i_{16}|g|}{}^h\, R_{i_{17} i_{18}|j|}{}^i\, R_{i_{19} i_{20}|i|}{}^j\, R_{i_{21} i_{22}|l|}{}^k\, R_{i_{23} i_{24}|k|}{}^l\, R_{i_{25} i_{26}|m|}{}^m\, R_{i_{27} i_{28}]n}{}^n +$$

$$+ 25{,}225{,}200\, R_{[i_1 i_2|c|}{}^a\, R_{i_3 i_4|a|}{}^b\, R_{i_5 i_6|b|}{}^c\, R_{i_7 i_8|f|}{}^d\, R_{i_9 i_{10}|d|}{}^e\, R_{i_{11} i_{12}|e|}{}^f\, R_{i_{13} i_{14}|h|}{}^g\, R_{i_{15} i_{16}|g|}{}^h\, R_{i_{17} i_{18}|i|}{}^i\, R_{i_{19} i_{20}|j|}{}^j\, R_{i_{21} i_{22}|k|}{}^k\, R_{i_{23} i_{24}|l|}{}^l\, R_{i_{25} i_{26}|m|}{}^m\, R_{i_{27} i_{28}]n}{}^n -$$

$$- 3{,}363{,}360\, R_{[i_1 i_2|c|}{}^a\, R_{i_3 i_4|a|}{}^b\, R_{i_5 i_6|b|}{}^c\, R_{i_7 i_8|f|}{}^d\, R_{i_9 i_{10}|d|}{}^e\, R_{i_{11} i_{12}|e|}{}^f\, R_{i_{13} i_{14}|h|}{}^g\, R_{i_{15} i_{16}|g|}{}^h\, R_{i_{17} i_{18}|i|}{}^i\, R_{i_{19} i_{20}|j|}{}^j\, R_{i_{21} i_{22}|k|}{}^k\, R_{i_{23} i_{24}|l|}{}^l\, R_{i_{25} i_{26}|m|}{}^m\, R_{i_{27} i_{28}]n}{}^n +$$

$$+ 120{,}120\, R_{[i_1 i_2|c|}{}^a\, R_{i_3 i_4|a|}{}^b\, R_{i_5 i_6|b|}{}^c\, R_{i_7 i_8|f|}{}^d\, R_{i_9 i_{10}|d|}{}^e\, R_{i_{11} i_{12}|e|}{}^f\, R_{i_{13} i_{14}|g|}{}^g\, R_{i_{15} i_{16}|h|}{}^h\, R_{i_{17} i_{18}|i|}{}^i\, R_{i_{19} i_{20}|j|}{}^j\, R_{i_{21} i_{22}|k|}{}^k\, R_{i_{23} i_{24}|l|}{}^l\, R_{i_{25} i_{26}|m|}{}^m\, R_{i_{27} i_{28}]n}{}^n -$$

$$- 7{,}567{,}560\, R_{[i_1 i_2|c|}{}^a\, R_{i_3 i_4|a|}{}^b\, R_{i_5 i_6|b|}{}^c\, R_{i_7 i_8|e|}{}^d\, R_{i_9 i_{10}|d|}{}^e\, R_{i_{11} i_{12}|g|}{}^f\, R_{i_{13} i_{14}|f|}{}^g\, R_{i_{15} i_{16}|i|}{}^h\, R_{i_{17} i_{18}|h|}{}^i\, R_{i_{19} i_{20}|k|}{}^j\, R_{i_{21} i_{22}|j|}{}^k\, R_{i_{23} i_{24}|m|}{}^l\, R_{i_{25} i_{26}|l|}{}^m\, R_{i_{27} i_{28}]n}{}^n +$$

$$+ 12{,}612{,}600\, R_{[i_1 i_2|c|}{}^a\, R_{i_3 i_4|a|}{}^b\, R_{i_5 i_6|b|}{}^c\, R_{i_7 i_8|e|}{}^d\, R_{i_9 i_{10}|d|}{}^e\, R_{i_{11} i_{12}|g|}{}^f\, R_{i_{13} i_{14}|f|}{}^g\, R_{i_{15} i_{16}|i|}{}^h\, R_{i_{17} i_{18}|h|}{}^i\, R_{i_{19} i_{20}|j|}{}^j\, R_{i_{21} i_{22}|k|}{}^k\, R_{i_{23} i_{24}|l|}{}^l\, R_{i_{25} i_{26}|m|}{}^m\, R_{i_{27} i_{28}]n}{}^n -$$

$$- 5{,}045{,}040\, R_{[i_1 i_2|c|}{}^a\, R_{i_3 i_4|a|}{}^b\, R_{i_5 i_6|b|}{}^c\, R_{i_7 i_8|e|}{}^d\, R_{i_9 i_{10}|d|}{}^e\, R_{i_{11} i_{12}|g|}{}^f\, R_{i_{13} i_{14}|f|}{}^g\, R_{i_{15} i_{16}|i|}{}^h\, R_{i_{17} i_{18}|h|}{}^i\, R_{i_{19} i_{20}|j|}{}^j\, R_{i_{21} i_{22}|k|}{}^k\, R_{i_{23} i_{24}|l|}{}^l\, R_{i_{25} i_{26}|m|}{}^m\, R_{i_{27} i_{28}]n}{}^n +$$

$$+ 720{,}720\, R_{[i_1 i_2|c|}{}^a\, R_{i_3 i_4|a|}{}^b\, R_{i_5 i_6|b|}{}^c\, R_{i_7 i_8|e|}{}^d\, R_{i_9 i_{10}|d|}{}^e\, R_{i_{11} i_{12}|g|}{}^f\, R_{i_{13} i_{14}|f|}{}^g\, R_{i_{15} i_{16}|h|}{}^h\, R_{i_{17} i_{18}|i|}{}^i\, R_{i_{19} i_{20}|j|}{}^j\, R_{i_{21} i_{22}|k|}{}^k\, R_{i_{23} i_{24}|l|}{}^l\, R_{i_{25} i_{26}|m|}{}^m\, R_{i_{27} i_{28}]n}{}^n -$$

$$- 40{,}040\, R_{[i_1 i_2|c|}{}^a\, R_{i_3 i_4|a|}{}^b\, R_{i_5 i_6|b|}{}^c\, R_{i_7 i_8|e|}{}^d\, R_{i_9 i_{10}|d|}{}^e\, R_{i_{11} i_{12}|f|}{}^f\, R_{i_{13} i_{14}|g|}{}^g\, R_{i_{15} i_{16}|h|}{}^h\, R_{i_{17} i_{18}|i|}{}^i\, R_{i_{19} i_{20}|j|}{}^j\, R_{i_{21} i_{22}|k|}{}^k\, R_{i_{23} i_{24}|l|}{}^l\, R_{i_{25} i_{26}|m|}{}^m\, R_{i_{27} i_{28}]n}{}^n +$$

$$+ 728\, R_{[i_1 i_2|c|}{}^a\, R_{i_3 i_4|a|}{}^b\, R_{i_5 i_6|b|}{}^c\, R_{i_7 i_8|d|}{}^d\, R_{i_9 i_{10}|e|}{}^e\, R_{i_{11} i_{12}|f|}{}^f\, R_{i_{13} i_{14}|g|}{}^g\, R_{i_{15} i_{16}|h|}{}^h\, R_{i_{17} i_{18}|i|}{}^i\, R_{i_{19} i_{20}|j|}{}^j\, R_{i_{21} i_{22}|k|}{}^k\, R_{i_{23} i_{24}|l|}{}^l\, R_{i_{25} i_{26}|m|}{}^m\, R_{i_{27} i_{28}]n}{}^n -$$

$$- 135{,}135\, R_{[i_1 i_2|b|}{}^a\, R_{i_3 i_4|a|}{}^b\, R_{i_5 i_6|d|}{}^c\, R_{i_7 i_8|c|}{}^d\, R_{i_9 i_{10}|f|}{}^e\, R_{i_{11} i_{12}|e|}{}^f\, R_{i_{13} i_{14}|h|}{}^g\, R_{i_{15} i_{16}|g|}{}^h\, R_{i_{17} i_{18}|j|}{}^i\, R_{i_{19} i_{20}|i|}{}^j\, R_{i_{21} i_{22}|l|}{}^k\, R_{i_{23} i_{24}|k|}{}^l\, R_{i_{25} i_{26}|n|}{}^m\, R_{i_{27} i_{28}]m}{}^n +$$

$$+ 945{,}945\, R_{[i_1 i_2|b|}{}^a\, R_{i_3 i_4|a|}{}^b\, R_{i_5 i_6|d|}{}^c\, R_{i_7 i_8|c|}{}^d\, R_{i_9 i_{10}|f|}{}^e\, R_{i_{11} i_{12}|e|}{}^f\, R_{i_{13} i_{14}|h|}{}^g\, R_{i_{15} i_{16}|g|}{}^h\, R_{i_{17} i_{18}|j|}{}^i\, R_{i_{19} i_{20}|i|}{}^j\, R_{i_{21} i_{22}|l|}{}^k\, R_{i_{23} i_{24}|k|}{}^l\, R_{i_{25} i_{26}|m|}{}^m\, R_{i_{27} i_{28}]n}{}^n -$$

$$\begin{aligned}
&- 945{,}945\, R_{[i_1i_2|b|}{}^a R_{i_3i_4|a|}{}^b R_{i_5i_6|d|}{}^c R_{i_7i_8|c|}{}^d R_{i_9i_{10}|f|}{}^e R_{i_{11}i_{12}|e|}{}^f R_{i_{13}i_{14}|h|}{}^g R_{i_{15}i_{16}|g|}{}^h R_{i_{17}i_{18}|j|}{}^i R_{i_{19}i_{20}|i|}{}^j R_{i_{21}i_{22}|k|}{}^k R_{i_{23}i_{24}|l|}{}^l R_{i_{25}i_{26}|m|}{}^m R_{i_{27}i_{28}]}{}^n + \\
&+ 315{,}315\, R_{[i_1i_2|b|}{}^a R_{i_3i_4|a|}{}^b R_{i_5i_6|d|}{}^c R_{i_7i_8|c|}{}^d R_{i_9i_{10}|f|}{}^e R_{i_{11}i_{12}|e|}{}^f R_{i_{13}i_{14}|g|}{}^g R_{i_{15}i_{16}|h|}{}^h R_{i_{17}i_{18}|i|}{}^i R_{i_{19}i_{20}|j|}{}^j R_{i_{21}i_{22}|k|}{}^k R_{i_{23}i_{24}|l|}{}^l R_{i_{25}i_{26}|m|}{}^m R_{i_{27}i_{28}]}{}^n - \\
&- 45{,}045\, R_{[i_1i_2|b|}{}^a R_{i_3i_4|a|}{}^b R_{i_5i_6|d|}{}^c R_{i_7i_8|c|}{}^d R_{i_9i_{10}|f|}{}^e R_{i_{11}i_{12}|e|}{}^f R_{i_{13}i_{14}|g|}{}^g R_{i_{15}i_{16}|h|}{}^h R_{i_{17}i_{18}|i|}{}^i R_{i_{19}i_{20}|j|}{}^j R_{i_{21}i_{22}|k|}{}^k R_{i_{23}i_{24}|l|}{}^l R_{i_{25}i_{26}|m|}{}^m R_{i_{27}i_{28}]}{}^n + \\
&+ 3{,}003\, R_{[i_1i_2|b|}{}^a R_{i_3i_4|a|}{}^b R_{i_5i_6|d|}{}^c R_{i_7i_8|c|}{}^d R_{i_9i_{10}|e|}{}^e R_{i_{11}i_{12}|f|}{}^f R_{i_{13}i_{14}|g|}{}^g R_{i_{15}i_{16}|h|}{}^h R_{i_{17}i_{18}|i|}{}^i R_{i_{19}i_{20}|j|}{}^j R_{i_{21}i_{22}|k|}{}^k R_{i_{23}i_{24}|l|}{}^l R_{i_{25}i_{26}|m|}{}^m R_{i_{27}i_{28}]}{}^n - \\
&- 91\, R_{[i_1i_2|b|}{}^a R_{i_3i_4|a|}{}^b R_{i_5i_6|c|}{}^c R_{i_7i_8|d|}{}^d R_{i_9i_{10}|e|}{}^e R_{i_{11}i_{12}|f|}{}^f R_{i_{13}i_{14}|g|}{}^g R_{i_{15}i_{16}|h|}{}^h R_{i_{17}i_{18}|i|}{}^i R_{i_{19}i_{20}|j|}{}^j R_{i_{21}i_{22}|k|}{}^k R_{i_{23}i_{24}|l|}{}^l R_{i_{25}i_{26}|m|}{}^m R_{i_{27}i_{28}]}{}^n + \\
&+ R_{[i_1i_2|a|}{}^a R_{i_3i_4|b|}{}^b R_{i_5i_6|c|}{}^c R_{i_7i_8|d|}{}^d R_{i_9i_{10}|e|}{}^e R_{i_{11}i_{12}|f|}{}^f R_{i_{13}i_{14}|g|}{}^g R_{i_{15}i_{16}|h|}{}^h R_{i_{17}i_{18}|i|}{}^i R_{i_{19}i_{20}|j|}{}^j R_{i_{21}i_{22}|k|}{}^k R_{i_{23}i_{24}|l|}{}^l R_{i_{25}i_{26}|m|}{}^m R_{i_{27}i_{28}]}{}^n )
\end{aligned}$$

$$= \frac{i^{14}}{2^{28}\pi^{14}14!}\Big( - 6{,}227{,}020{,}800\, R_{[i_1i_2|n|}{}^a R_{i_3i_4|a|}{}^b R_{i_5i_6|b|}{}^c R_{i_7i_8|c|}{}^d R_{i_9i_{10}|d|}{}^e R_{i_{11}i_{12}|e|}{}^f R_{i_{13}i_{14}|f|}{}^g R_{i_{15}i_{16}|g|}{}^h R_{i_{17}i_{18}|h|}{}^i R_{i_{19}i_{20}|i|}{}^j R_{i_{21}i_{22}|j|}{}^k R_{i_{23}i_{24}|k|}{}^l R_{i_{25}i_{26}|l|}{}^m R_{i_{27}i_{28}]}{}^m + $$

$$+ 6{,}706{,}022{,}400\, R_{[i_1i_2|m|}{}^a R_{i_3i_4|a|}{}^b R_{i_5i_6|b|}{}^c R_{i_7i_8|c|}{}^d R_{i_9i_{10}|d|}{}^e R_{i_{11}i_{12}|e|}{}^f R_{i_{13}i_{14}|f|}{}^g R_{i_{15}i_{16}|g|}{}^h R_{i_{17}i_{18}|h|}{}^i R_{i_{19}i_{20}|i|}{}^j R_{i_{21}i_{22}|j|}{}^k R_{i_{23}i_{24}|k|}{}^l R_{i_{25}i_{26}|l|}{}^m V_{i_{27}i_{28}]} + $$

$$+ 3{,}632{,}428{,}800\, R_{[i_1i_2|l|}{}^a R_{i_3i_4|a|}{}^b R_{i_5i_6|b|}{}^c R_{i_7i_8|c|}{}^d R_{i_9i_{10}|d|}{}^e R_{i_{11}i_{12}|e|}{}^f R_{i_{13}i_{14}|f|}{}^g R_{i_{15}i_{16}|g|}{}^h R_{i_{17}i_{18}|h|}{}^i R_{i_{19}i_{20}|i|}{}^j R_{i_{21}i_{22}|j|}{}^k R_{i_{23}i_{24}|k|}{}^l R_{i_{25}i_{26}|n|}{}^m R_{i_{27}i_{28}]}{}^m - $$

$$- 3{,}632{,}428{,}800\, R_{[i_1i_2|l|}{}^a R_{i_3i_4|a|}{}^b R_{i_5i_6|b|}{}^c R_{i_7i_8|c|}{}^d R_{i_9i_{10}|d|}{}^e R_{i_{11}i_{12}|e|}{}^f R_{i_{13}i_{14}|f|}{}^g R_{i_{15}i_{16}|g|}{}^h R_{i_{17}i_{18}|h|}{}^i R_{i_{19}i_{20}|i|}{}^j R_{i_{21}i_{22}|j|}{}^k R_{i_{23}i_{24}|k|}{}^l V_{i_{25}i_{26}} V_{i_{27}i_{28}]} + $$

$$+ 2{,}641{,}766{,}400\, R_{[i_1i_2|k|}{}^a R_{i_3i_4|a|}{}^b R_{i_5i_6|b|}{}^c R_{i_7i_8|c|}{}^d R_{i_9i_{10}|d|}{}^e R_{i_{11}i_{12}|e|}{}^f R_{i_{13}i_{14}|f|}{}^g R_{i_{15}i_{16}|g|}{}^h R_{i_{17}i_{18}|h|}{}^i R_{i_{19}i_{20}|i|}{}^j R_{i_{21}i_{22}|j|}{}^k R_{i_{23}i_{24}|n|}{}^l R_{i_{25}i_{26}|l|}{}^m R_{i_{27}i_{28}]}{}^m - $$

$$- 3{,}962{,}649{,}600\, R_{[i_1i_2|k|}{}^a R_{i_3i_4|a|}{}^b R_{i_5i_6|b|}{}^c R_{i_7i_8|c|}{}^d R_{i_9i_{10}|d|}{}^e R_{i_{11}i_{12}|e|}{}^f R_{i_{13}i_{14}|f|}{}^g R_{i_{15}i_{16}|g|}{}^h R_{i_{17}i_{18}|h|}{}^i R_{i_{19}i_{20}|i|}{}^j R_{i_{21}i_{22}|j|}{}^k R_{i_{23}i_{24}|m|}{}^l R_{i_{25}i_{26}|l|}{}^m V_{i_{27}i_{28}]} + $$

$$+ 1{,}320{,}883{,}200\, R_{[i_1i_2|k|}{}^a R_{i_3i_4|a|}{}^b R_{i_5i_6|b|}{}^c R_{i_7i_8|c|}{}^d R_{i_9i_{10}|d|}{}^e R_{i_{11}i_{12}|e|}{}^f R_{i_{13}i_{14}|f|}{}^g R_{i_{15}i_{16}|g|}{}^h R_{i_{17}i_{18}|h|}{}^i R_{i_{19}i_{20}|i|}{}^j R_{i_{21}i_{22}|j|}{}^k V_{i_{23}i_{24}} V_{i_{25}i_{26}} V_{i_{27}i_{28}]} + $$

$$+ 2{,}179{,}457{,}280\, R_{[i_1i_2|j|}{}^a R_{i_3i_4|a|}{}^b R_{i_5i_6|b|}{}^c R_{i_7i_8|c|}{}^d R_{i_9i_{10}|d|}{}^e R_{i_{11}i_{12}|e|}{}^f R_{i_{13}i_{14}|f|}{}^g R_{i_{15}i_{16}|g|}{}^h R_{i_{17}i_{18}|h|}{}^i R_{i_{19}i_{20}|i|}{}^j R_{i_{21}i_{22}|n|}{}^k R_{i_{23}i_{24}|k|}{}^l R_{i_{25}i_{26}|l|}{}^m R_{i_{27}i_{28}]}{}^m - $$

$$- 2{,}905{,}943{,}040\, R_{[i_1i_2|j|}{}^a R_{i_3i_4|a|}{}^b R_{i_5i_6|b|}{}^c R_{i_7i_8|c|}{}^d R_{i_9i_{10}|d|}{}^e R_{i_{11}i_{12}|e|}{}^f R_{i_{13}i_{14}|f|}{}^g R_{i_{15}i_{16}|g|}{}^h R_{i_{17}i_{18}|h|}{}^i R_{i_{19}i_{20}|i|}{}^j R_{i_{21}i_{22}|m|}{}^k R_{i_{23}i_{24}|k|}{}^l R_{i_{25}i_{26}|l|}{}^m V_{i_{27}i_{28}]} - $$

$$- 1{,}089{,}728{,}640\, R_{[i_1i_2|j|}{}^a R_{i_3i_4|a|}{}^b R_{i_5i_6|b|}{}^c R_{i_7i_8|c|}{}^d R_{i_9i_{10}|d|}{}^e R_{i_{11}i_{12}|e|}{}^f R_{i_{13}i_{14}|f|}{}^g R_{i_{15}i_{16}|g|}{}^h R_{i_{17}i_{18}|h|}{}^i R_{i_{19}i_{20}|i|}{}^j R_{i_{21}i_{22}|l|}{}^k R_{i_{23}i_{24}|k|}{}^l R_{i_{25}i_{26}|n|}{}^m R_{i_{27}i_{28}]}{}^m + $$

$$+ 2{,}179{,}457{,}280\, R_{[i_1i_2|j|}{}^a R_{i_3i_4|a|}{}^b R_{i_5i_6|b|}{}^c R_{i_7i_8|c|}{}^d R_{i_9i_{10}|d|}{}^e R_{i_{11}i_{12}|e|}{}^f R_{i_{13}i_{14}|f|}{}^g R_{i_{15}i_{16}|g|}{}^h R_{i_{17}i_{18}|h|}{}^i R_{i_{19}i_{20}|i|}{}^j R_{i_{21}i_{22}|l|}{}^k R_{i_{23}i_{24}|k|}{}^l V_{i_{25}i_{26}} V_{i_{27}i_{28}]} - $$

$$- 363{,}242{,}880\, R_{[i_1i_2|j|}{}^a R_{i_3i_4|a|}{}^b R_{i_5i_6|b|}{}^c R_{i_7i_8|c|}{}^d R_{i_9i_{10}|d|}{}^e R_{i_{11}i_{12}|e|}{}^f R_{i_{13}i_{14}|f|}{}^g R_{i_{15}i_{16}|g|}{}^h R_{i_{17}i_{18}|h|}{}^i R_{i_{19}i_{20}|i|}{}^j V_{i_{21}i_{22}} V_{i_{23}i_{24}} V_{i_{25}i_{26}} V_{i_{27}i_{28}]} + $$

$$+ 1{,}937{,}295{,}360\, R_{[i_1i_2|i|}{}^a R_{i_3i_4|a|}{}^b R_{i_5i_6|b|}{}^c R_{i_7i_8|c|}{}^d R_{i_9i_{10}|d|}{}^e R_{i_{11}i_{12}|e|}{}^f R_{i_{13}i_{14}|f|}{}^g R_{i_{15}i_{16}|g|}{}^h R_{i_{17}i_{18}|h|}{}^i R_{i_{19}i_{20}|n|}{}^j R_{i_{21}i_{22}|j|}{}^k R_{i_{23}i_{24}|k|}{}^l R_{i_{25}i_{26}|l|}{}^m R_{i_{27}i_{28}]}{}^m - $$

$$- 2{,}421{,}619{,}200\, R_{[i_1i_2|i|}{}^a R_{i_3i_4|a|}{}^b R_{i_5i_6|b|}{}^c R_{i_7i_8|c|}{}^d R_{i_9i_{10}|d|}{}^e R_{i_{11}i_{12}|e|}{}^f R_{i_{13}i_{14}|f|}{}^g R_{i_{15}i_{16}|g|}{}^h R_{i_{17}i_{18}|h|}{}^i R_{i_{19}i_{20}|m|}{}^j R_{i_{21}i_{22}|j|}{}^k R_{i_{23}i_{24}|k|}{}^l R_{i_{25}i_{26}|l|}{}^m V_{i_{27}i_{28}]} - $$

$$- 1{,}614{,}412{,}800\, R_{[i_1i_2|i|}{}^a R_{i_3i_4|a|}{}^b R_{i_5i_6|b|}{}^c R_{i_7i_8|c|}{}^d R_{i_9i_{10}|d|}{}^e R_{i_{11}i_{12}|e|}{}^f R_{i_{13}i_{14}|f|}{}^g R_{i_{15}i_{16}|g|}{}^h R_{i_{17}i_{18}|h|}{}^i R_{i_{19}i_{20}|l|}{}^j R_{i_{21}i_{22}|j|}{}^k R_{i_{23}i_{24}|k|}{}^l R_{i_{25}i_{26}|n|}{}^m R_{i_{27}i_{28}]}{}^m + $$

$$+ 1{,}614{,}412{,}800\, R_{[i_1i_2|i|}{}^a R_{i_3i_4|a|}{}^b R_{i_5i_6|b|}{}^c R_{i_7i_8|c|}{}^d R_{i_9i_{10}|d|}{}^e R_{i_{11}i_{12}|e|}{}^f R_{i_{13}i_{14}|f|}{}^g R_{i_{15}i_{16}|g|}{}^h R_{i_{17}i_{18}|h|}{}^i R_{i_{19}i_{20}|l|}{}^j R_{i_{21}i_{22}|j|}{}^k R_{i_{23}i_{24}|k|}{}^l V_{i_{25}i_{26}} V_{i_{27}i_{28}]} + $$

$$+ 1{,}210{,}809{,}600\, R_{[i_1i_2|i|}{}^a R_{i_3i_4|a|}{}^b R_{i_5i_6|b|}{}^c R_{i_7i_8|c|}{}^d R_{i_9i_{10}|d|}{}^e R_{i_{11}i_{12}|e|}{}^f R_{i_{13}i_{14}|f|}{}^g R_{i_{15}i_{16}|g|}{}^h R_{i_{17}i_{18}|h|}{}^i R_{i_{19}i_{20}|k|}{}^j R_{i_{21}i_{22}|j|}{}^k R_{i_{23}i_{24}|m|}{}^l R_{i_{25}i_{26}|l|}{}^m V_{i_{27}i_{28}]} - $$

$$- 807{,}206{,}400\, R_{[i_1i_2|i|}{}^a R_{i_3i_4|a|}{}^b R_{i_5i_6|b|}{}^c R_{i_7i_8|c|}{}^d R_{i_9i_{10}|d|}{}^e R_{i_{11}i_{12}|e|}{}^f R_{i_{13}i_{14}|f|}{}^g R_{i_{15}i_{16}|g|}{}^h R_{i_{17}i_{18}|h|}{}^i R_{i_{19}i_{20}|k|}{}^j R_{i_{21}i_{22}|j|}{}^k V_{i_{23}i_{24}} V_{i_{25}i_{26}} V_{i_{27}i_{28}]} + $$

$$+ 80{,}720{,}640\, R_{[i_1i_2|i|}{}^a R_{i_3i_4|a|}{}^b R_{i_5i_6|b|}{}^c R_{i_7i_8|c|}{}^d R_{i_9i_{10}|d|}{}^e R_{i_{11}i_{12}|e|}{}^f R_{i_{13}i_{14}|f|}{}^g R_{i_{15}i_{16}|g|}{}^h R_{i_{17}i_{18}|h|}{}^i V_{i_{19}i_{20}} V_{i_{21}i_{22}} V_{i_{23}i_{24}} V_{i_{25}i_{26}} V_{i_{27}i_{28}]} + $$

$$+ 1{,}816{,}214{,}400\, R_{[i_1i_2|h|}{}^a R_{i_3i_4|a|}{}^b R_{i_5i_6|b|}{}^c R_{i_7i_8|c|}{}^d R_{i_9i_{10}|d|}{}^e R_{i_{11}i_{12}|e|}{}^f R_{i_{13}i_{14}|f|}{}^g R_{i_{15}i_{16}|g|}{}^h R_{i_{17}i_{18}|n|}{}^i R_{i_{19}i_{20}|i|}{}^j R_{i_{21}i_{22}|j|}{}^k R_{i_{23}i_{24}|k|}{}^l R_{i_{25}i_{26}|l|}{}^m R_{i_{27}i_{28}]}{}^m - $$

$$- 2{,}179{,}457{,}280\, R_{[i_1i_2|h|}{}^a R_{i_3i_4|a|}{}^b R_{i_5i_6|b|}{}^c R_{i_7i_8|c|}{}^d R_{i_9i_{10}|d|}{}^e R_{i_{11}i_{12}|e|}{}^f R_{i_{13}i_{14}|f|}{}^g R_{i_{15}i_{16}|g|}{}^h R_{i_{17}i_{18}|m|}{}^i R_{i_{19}i_{20}|i|}{}^j R_{i_{21}i_{22}|j|}{}^k R_{i_{23}i_{24}|k|}{}^l R_{i_{25}i_{26}|l|}{}^m V_{i_{27}i_{28}]} - $$

$$- 1{,}362{,}160{,}800\, R_{[i_1i_2|h|}{}^a R_{i_3i_4|a|}{}^b R_{i_5i_6|b|}{}^c R_{i_7i_8|c|}{}^d R_{i_9i_{10}|d|}{}^e R_{i_{11}i_{12}|e|}{}^f R_{i_{13}i_{14}|f|}{}^g R_{i_{15}i_{16}|g|}{}^h R_{i_{17}i_{18}|l|}{}^i R_{i_{19}i_{20}|i|}{}^j R_{i_{21}i_{22}|j|}{}^k R_{i_{23}i_{24}|k|}{}^l R_{i_{25}i_{26}|n|}{}^m R_{i_{27}i_{28}]}{}^m + $$

$$+ 1{,}362{,}160{,}800\, R_{[i_1i_2|h|}{}^a R_{i_3i_4|a|}{}^b R_{i_5i_6|b|}{}^c R_{i_7i_8|c|}{}^d R_{i_9i_{10}|d|}{}^e R_{i_{11}i_{12}|e|}{}^f R_{i_{13}i_{14}|f|}{}^g R_{i_{15}i_{16}|g|}{}^h R_{i_{17}i_{18}|l|}{}^i R_{i_{19}i_{20}|i|}{}^j R_{i_{21}i_{22}|j|}{}^k R_{i_{23}i_{24}|k|}{}^l V_{i_{25}i_{26}} V_{i_{27}i_{28}]} - $$

$$- 605{,}404{,}800\, R_{[i_1i_2|h|}{}^a R_{i_3i_4|a|}{}^b R_{i_5i_6|b|}{}^c R_{i_7i_8|c|}{}^d R_{i_9i_{10}|d|}{}^e R_{i_{11}i_{12}|e|}{}^f R_{i_{13}i_{14}|f|}{}^g R_{i_{15}i_{16}|g|}{}^h R_{i_{17}i_{18}|k|}{}^i R_{i_{19}i_{20}|i|}{}^j R_{i_{21}i_{22}|j|}{}^k R_{i_{23}i_{24}|n|}{}^l R_{i_{25}i_{26}|l|}{}^m R_{i_{27}i_{28}]}{}^m + $$

$$+ 1{,}816{,}214{,}400\, R_{[i_1i_2|h|}{}^a R_{i_3i_4|a|}{}^b R_{i_5i_6|b|}{}^c R_{i_7i_8|c|}{}^d R_{i_9i_{10}|d|}{}^e R_{i_{11}i_{12}|e|}{}^f R_{i_{13}i_{14}|f|}{}^g R_{i_{15}i_{16}|g|}{}^h R_{i_{17}i_{18}|k|}{}^i R_{i_{19}i_{20}|i|}{}^j R_{i_{21}i_{22}|j|}{}^k R_{i_{23}i_{24}|m|}{}^l R_{i_{25}i_{26}|l|}{}^m V_{i_{27}i_{28}]} - $$

$$- 605{,}404{,}800\, R_{[i_1i_2|h|}{}^a R_{i_3i_4|a|}{}^b R_{i_5i_6|b|}{}^c R_{i_7i_8|c|}{}^d R_{i_9i_{10}|d|}{}^e R_{i_{11}i_{12}|e|}{}^f R_{i_{13}i_{14}|f|}{}^g R_{i_{15}i_{16}|g|}{}^h R_{i_{17}i_{18}|k|}{}^i R_{i_{19}i_{20}|i|}{}^j R_{i_{21}i_{22}|j|}{}^k V_{i_{23}i_{24}} V_{i_{25}i_{26}} V_{i_{27}i_{28}]} + $$

$$+ 227{,}026{,}800\, R_{[i_1i_2|h|}{}^a R_{i_3i_4|a|}{}^b R_{i_5i_6|b|}{}^c R_{i_7i_8|c|}{}^d R_{i_9i_{10}|d|}{}^e R_{i_{11}i_{12}|e|}{}^f R_{i_{13}i_{14}|f|}{}^g R_{i_{15}i_{16}|g|}{}^h R_{i_{17}i_{18}|j|}{}^i R_{i_{19}i_{20}|i|}{}^j R_{i_{21}i_{22}|l|}{}^k R_{i_{23}i_{24}|k|}{}^l R_{i_{25}i_{26}|n|}{}^m R_{i_{27}i_{28}]}{}^m - $$

$$- 681{,}080{,}400\, R_{[i_1i_2|h|}{}^a R_{i_3i_4|a|}{}^b R_{i_5i_6|b|}{}^c R_{i_7i_8|c|}{}^d R_{i_9i_{10}|d|}{}^e R_{i_{11}i_{12}|e|}{}^f R_{i_{13}i_{14}|f|}{}^g R_{i_{15}i_{16}|g|}{}^h R_{i_{17}i_{18}|j|}{}^i R_{i_{19}i_{20}|i|}{}^j R_{i_{21}i_{22}|l|}{}^k R_{i_{23}i_{24}|k|}{}^l V_{i_{25}i_{26}} V_{i_{27}i_{28}]} + $$

$$+ 227{,}026{,}800\, R_{[i_1i_2|h|}{}^a R_{i_3i_4|a|}{}^b R_{i_5i_6|b|}{}^c R_{i_7i_8|c|}{}^d R_{i_9i_{10}|d|}{}^e R_{i_{11}i_{12}|e|}{}^f R_{i_{13}i_{14}|f|}{}^g R_{i_{15}i_{16}|g|}{}^h R_{i_{17}i_{18}|j|}{}^i R_{i_{19}i_{20}|i|}{}^j V_{i_{21}i_{22}} V_{i_{23}i_{24}} V_{i_{25}i_{26}} V_{i_{27}i_{28}]} - $$

$$- 15{,}135{,}120\, R_{[i_1i_2|h|}{}^a R_{i_3i_4|a|}{}^b R_{i_5i_6|b|}{}^c R_{i_7i_8|c|}{}^d R_{i_9i_{10}|d|}{}^e R_{i_{11}i_{12}|e|}{}^f R_{i_{13}i_{14}|f|}{}^g R_{i_{15}i_{16}|g|}{}^h V_{i_{17}i_{18}} V_{i_{19}i_{20}} V_{i_{21}i_{22}} V_{i_{23}i_{24}} V_{i_{25}i_{26}} V_{i_{27}i_{28}]} + $$

$$+ 889{,}574{,}400\, R_{[i_1i_2|g|}{}^a R_{i_3i_4|a|}{}^b R_{i_5i_6|b|}{}^c R_{i_7i_8|c|}{}^d R_{i_9i_{10}|d|}{}^e R_{i_{11}i_{12}|e|}{}^f R_{i_{13}i_{14}|f|}{}^g R_{i_{15}i_{16}|n|}{}^h R_{i_{17}i_{18}|h|}{}^i R_{i_{19}i_{20}|i|}{}^j R_{i_{21}i_{22}|j|}{}^k R_{i_{23}i_{24}|k|}{}^l R_{i_{25}i_{26}|l|}{}^m R_{i_{27}i_{28}]}{}^m - $$

$$- 2{,}075{,}673{,}600\, R_{[i_1i_2|g|}{}^a R_{i_3i_4|a|}{}^b R_{i_5i_6|b|}{}^c R_{i_7i_8|c|}{}^d R_{i_9i_{10}|d|}{}^e R_{i_{11}i_{12}|e|}{}^f R_{i_{13}i_{14}|f|}{}^g R_{i_{15}i_{16}|m|}{}^h R_{i_{17}i_{18}|h|}{}^i R_{i_{19}i_{20}|i|}{}^j R_{i_{21}i_{22}|j|}{}^k R_{i_{23}i_{24}|k|}{}^l R_{i_{25}i_{26}|l|}{}^m V_{i_{27}i_{28}]} - $$

$$- 1{,}245{,}404{,}160\, R_{[i_1i_2|g|}{}^a R_{i_3i_4|a|}{}^b R_{i_5i_6|b|}{}^c R_{i_7i_8|c|}{}^d R_{i_9i_{10}|d|}{}^e R_{i_{11}i_{12}|e|}{}^f R_{i_{13}i_{14}|f|}{}^g R_{i_{15}i_{16}|l|}{}^h R_{i_{17}i_{18}|h|}{}^i R_{i_{19}i_{20}|i|}{}^j R_{i_{21}i_{22}|j|}{}^k R_{i_{23}i_{24}|k|}{}^l R_{i_{25}i_{26}|n|}{}^m R_{i_{27}i_{28}]}{}^m + $$

$$+ 1{,}245{,}404{,}160\, R_{[i_1i_2|g|}{}^a R_{i_3i_4|a|}{}^b R_{i_5i_6|b|}{}^c R_{i_7i_8|c|}{}^d R_{i_9i_{10}|d|}{}^e R_{i_{11}i_{12}|e|}{}^f R_{i_{13}i_{14}|f|}{}^g R_{i_{15}i_{16}|l|}{}^h R_{i_{17}i_{18}|h|}{}^i R_{i_{19}i_{20}|i|}{}^j R_{i_{21}i_{22}|j|}{}^k R_{i_{23}i_{24}|k|}{}^l V_{i_{25}i_{26}} V_{i_{27}i_{28}]} - $$

$$- 1{,}037{,}836{,}800\, R_{[i_1i_2|g|}{}^a R_{i_3i_4|a|}{}^b R_{i_5i_6|b|}{}^c R_{i_7i_8|c|}{}^d R_{i_9i_{10}|d|}{}^e R_{i_{11}i_{12}|e|}{}^f R_{i_{13}i_{14}|f|}{}^g R_{i_{15}i_{16}|k|}{}^h R_{i_{17}i_{18}|h|}{}^i R_{i_{19}i_{20}|i|}{}^j R_{i_{21}i_{22}|j|}{}^k R_{i_{23}i_{24}|n|}{}^l R_{i_{25}i_{26}|l|}{}^m R_{i_{27}i_{28}]}{}^m + $$

$$+ 1{,}556{,}755{,}200\, R_{[i_1 i_2|g|}{}^a R_{i_3 i_4|a|}{}^b R_{i_5 i_6|b|}{}^c R_{i_7 i_8|c|}{}^d R_{i_9 i_{10}|d|}{}^e R_{i_{11} i_{12}|e|}{}^f R_{i_{13} i_{14}|f|}{}^g R_{i_{15} i_{16}|k|}{}^h R_{i_{17} i_{18}|h|}{}^i R_{i_{19} i_{20}|i|}{}^j R_{i_{21} i_{22}|j|}{}^k R_{i_{23} i_{24}|m|}{}^l R_{i_{25} i_{26}|l|}{}^m V_{i_{27} i_{28}]} -$$

$$- 518{,}918{,}400\, R_{[i_1 i_2|g|}{}^a R_{i_3 i_4|a|}{}^b R_{i_5 i_6|b|}{}^c R_{i_7 i_8|c|}{}^d R_{i_9 i_{10}|d|}{}^e R_{i_{11} i_{12}|e|}{}^f R_{i_{13} i_{14}|f|}{}^g R_{i_{15} i_{16}|k|}{}^h R_{i_{17} i_{18}|h|}{}^i R_{i_{19} i_{20}|i|}{}^j R_{i_{21} i_{22}|j|}{}^k V_{i_{23} i_{24}} V_{i_{25} i_{26}} V_{i_{27} i_{28}]} +$$

$$+ 691{,}891{,}200\, R_{[i_1 i_2|g|}{}^a R_{i_3 i_4|a|}{}^b R_{i_5 i_6|b|}{}^c R_{i_7 i_8|c|}{}^d R_{i_9 i_{10}|d|}{}^e R_{i_{11} i_{12}|e|}{}^f R_{i_{13} i_{14}|f|}{}^g R_{i_{15} i_{16}|j|}{}^h R_{i_{17} i_{18}|h|}{}^i R_{i_{19} i_{20}|i|}{}^j R_{i_{21} i_{22}|m|}{}^k R_{i_{23} i_{24}|k|}{}^l R_{i_{25} i_{26}|l|}{}^m V_{i_{27} i_{28}]} +$$

$$+ 518{,}918{,}400\, R_{[i_1 i_2|g|}{}^a R_{i_3 i_4|a|}{}^b R_{i_5 i_6|b|}{}^c R_{i_7 i_8|c|}{}^d R_{i_9 i_{10}|d|}{}^e R_{i_{11} i_{12}|e|}{}^f R_{i_{13} i_{14}|f|}{}^g R_{i_{15} i_{16}|j|}{}^h R_{i_{17} i_{18}|h|}{}^i R_{i_{19} i_{20}|i|}{}^j R_{i_{21} i_{22}|j|}{}^k R_{i_{23} i_{24}|k|}{}^l R_{i_{25} i_{26}|n|}{}^m R_{i_{27} i_{28}]}{}^n -$$

$$- 1{,}037{,}836{,}800\, R_{[i_1 i_2|g|}{}^a R_{i_3 i_4|a|}{}^b R_{i_5 i_6|b|}{}^c R_{i_7 i_8|c|}{}^d R_{i_9 i_{10}|d|}{}^e R_{i_{11} i_{12}|e|}{}^f R_{i_{13} i_{14}|f|}{}^g R_{i_{15} i_{16}|j|}{}^h R_{i_{17} i_{18}|h|}{}^i R_{i_{19} i_{20}|i|}{}^j R_{i_{21} i_{22}|l|}{}^k R_{i_{23} i_{24}|k|}{}^l V_{i_{25} i_{26}} V_{i_{27} i_{28}]} +$$

$$+ 172{,}972{,}800\, R_{[i_1 i_2|g|}{}^a R_{i_3 i_4|a|}{}^b R_{i_5 i_6|b|}{}^c R_{i_7 i_8|c|}{}^d R_{i_9 i_{10}|d|}{}^e R_{i_{11} i_{12}|e|}{}^f R_{i_{13} i_{14}|f|}{}^g R_{i_{15} i_{16}|j|}{}^h R_{i_{17} i_{18}|h|}{}^i R_{i_{19} i_{20}|i|}{}^j V_{i_{21} i_{22}} V_{i_{23} i_{24}} V_{i_{25} i_{26}} V_{i_{27} i_{28}]} -$$

$$- 259{,}459{,}200\, R_{[i_1 i_2|g|}{}^a R_{i_3 i_4|a|}{}^b R_{i_5 i_6|b|}{}^c R_{i_7 i_8|c|}{}^d R_{i_9 i_{10}|d|}{}^e R_{i_{11} i_{12}|e|}{}^f R_{i_{13} i_{14}|f|}{}^g R_{i_{15} i_{16}|i|}{}^h R_{i_{17} i_{18}|h|}{}^i R_{i_{19} i_{20}|k|}{}^j R_{i_{21} i_{22}|j|}{}^k R_{i_{23} i_{24}|m|}{}^l R_{i_{25} i_{26}|l|}{}^m V_{i_{27} i_{28}]} +$$

$$+ 259{,}459{,}200\, R_{[i_1 i_2|g|}{}^a R_{i_3 i_4|a|}{}^b R_{i_5 i_6|b|}{}^c R_{i_7 i_8|c|}{}^d R_{i_9 i_{10}|d|}{}^e R_{i_{11} i_{12}|e|}{}^f R_{i_{13} i_{14}|f|}{}^g R_{i_{15} i_{16}|i|}{}^h R_{i_{17} i_{18}|h|}{}^i R_{i_{19} i_{20}|k|}{}^j R_{i_{21} i_{22}|j|}{}^k V_{i_{23} i_{24}} V_{i_{25} i_{26}} V_{i_{27} i_{28}]} -$$

$$- 51{,}891{,}840\, R_{[i_1 i_2|g|}{}^a R_{i_3 i_4|a|}{}^b R_{i_5 i_6|b|}{}^c R_{i_7 i_8|c|}{}^d R_{i_9 i_{10}|d|}{}^e R_{i_{11} i_{12}|e|}{}^f R_{i_{13} i_{14}|f|}{}^g R_{i_{15} i_{16}|i|}{}^h R_{i_{17} i_{18}|h|}{}^i V_{i_{19} i_{20}} V_{i_{21} i_{22}} V_{i_{23} i_{24}} V_{i_{25} i_{26}} V_{i_{27} i_{28}]} +$$

$$+ 2{,}471{,}040\, R_{[i_1 i_2|g|}{}^a R_{i_3 i_4|a|}{}^b R_{i_5 i_6|b|}{}^c R_{i_7 i_8|c|}{}^d R_{i_9 i_{10}|d|}{}^e R_{i_{11} i_{12}|e|}{}^f R_{i_{13} i_{14}|f|}{}^g V_{i_{15} i_{16}} V_{i_{17} i_{18}} V_{i_{19} i_{20}} V_{i_{21} i_{22}} V_{i_{23} i_{24}} V_{i_{25} i_{26}} V_{i_{27} i_{28}]} -$$

$$- 605{,}404{,}800\, R_{[i_1 i_2|f|}{}^a R_{i_3 i_4|a|}{}^b R_{i_5 i_6|b|}{}^c R_{i_7 i_8|c|}{}^d R_{i_9 i_{10}|d|}{}^e R_{i_{11} i_{12}|e|}{}^f R_{i_{13} i_{14}|l|}{}^g R_{i_{15} i_{16}|g|}{}^h R_{i_{17} i_{18}|h|}{}^i R_{i_{19} i_{20}|i|}{}^j R_{i_{21} i_{22}|j|}{}^k R_{i_{23} i_{24}|k|}{}^l R_{i_{25} i_{26}|n|}{}^m R_{i_{27} i_{28}]}{}^n +$$

$$+ 605{,}404{,}800\, R_{[i_1 i_2|f|}{}^a R_{i_3 i_4|a|}{}^b R_{i_5 i_6|b|}{}^c R_{i_7 i_8|c|}{}^d R_{i_9 i_{10}|d|}{}^e R_{i_{11} i_{12}|e|}{}^f R_{i_{13} i_{14}|l|}{}^g R_{i_{15} i_{16}|g|}{}^h R_{i_{17} i_{18}|h|}{}^i R_{i_{19} i_{20}|i|}{}^j R_{i_{21} i_{22}|j|}{}^k R_{i_{23} i_{24}|k|}{}^l V_{i_{25} i_{26}} V_{i_{27} i_{28}]} -$$

$$- 968{,}647{,}680\, R_{[i_1 i_2|f|}{}^a R_{i_3 i_4|a|}{}^b R_{i_5 i_6|b|}{}^c R_{i_7 i_8|c|}{}^d R_{i_9 i_{10}|d|}{}^e R_{i_{11} i_{12}|e|}{}^f R_{i_{13} i_{14}|k|}{}^g R_{i_{15} i_{16}|g|}{}^h R_{i_{17} i_{18}|h|}{}^i R_{i_{19} i_{20}|i|}{}^j R_{i_{21} i_{22}|j|}{}^k R_{i_{23} i_{24}|n|}{}^l R_{i_{25} i_{26}|l|}{}^m R_{i_{27} i_{28}]}{}^n +$$

$$+ 1{,}452{,}971{,}520\, R_{[i_1 i_2|f|}{}^a R_{i_3 i_4|a|}{}^b R_{i_5 i_6|b|}{}^c R_{i_7 i_8|c|}{}^d R_{i_9 i_{10}|d|}{}^e R_{i_{11} i_{12}|e|}{}^f R_{i_{13} i_{14}|k|}{}^g R_{i_{15} i_{16}|g|}{}^h R_{i_{17} i_{18}|h|}{}^i R_{i_{19} i_{20}|i|}{}^j R_{i_{21} i_{22}|j|}{}^k R_{i_{23} i_{24}|m|}{}^l R_{i_{25} i_{26}|l|}{}^m V_{i_{27} i_{28}]} -$$

$$- 484{,}323{,}840\, R_{[i_1 i_2|f|}{}^a R_{i_3 i_4|a|}{}^b R_{i_5 i_6|b|}{}^c R_{i_7 i_8|c|}{}^d R_{i_9 i_{10}|d|}{}^e R_{i_{11} i_{12}|e|}{}^f R_{i_{13} i_{14}|k|}{}^g R_{i_{15} i_{16}|g|}{}^h R_{i_{17} i_{18}|h|}{}^i R_{i_{19} i_{20}|i|}{}^j R_{i_{21} i_{22}|j|}{}^k V_{i_{23} i_{24}} V_{i_{25} i_{26}} V_{i_{27} i_{28}]} -$$

$$- 454{,}053{,}600\, R_{[i_1 i_2|f|}{}^a R_{i_3 i_4|a|}{}^b R_{i_5 i_6|b|}{}^c R_{i_7 i_8|c|}{}^d R_{i_9 i_{10}|d|}{}^e R_{i_{11} i_{12}|e|}{}^f R_{i_{13} i_{14}|j|}{}^g R_{i_{15} i_{16}|g|}{}^h R_{i_{17} i_{18}|h|}{}^i R_{i_{19} i_{20}|i|}{}^j R_{i_{21} i_{22}|n|}{}^k R_{i_{23} i_{24}|k|}{}^l R_{i_{25} i_{26}|l|}{}^m R_{i_{27} i_{28}]}{}^m +$$

$$+ 1{,}210{,}809{,}600\, R_{[i_1 i_2|f|}{}^a R_{i_3 i_4|a|}{}^b R_{i_5 i_6|b|}{}^c R_{i_7 i_8|c|}{}^d R_{i_9 i_{10}|d|}{}^e R_{i_{11} i_{12}|e|}{}^f R_{i_{13} i_{14}|j|}{}^g R_{i_{15} i_{16}|g|}{}^h R_{i_{17} i_{18}|h|}{}^i R_{i_{19} i_{20}|i|}{}^j R_{i_{21} i_{22}|m|}{}^k R_{i_{23} i_{24}|k|}{}^l R_{i_{25} i_{26}|l|}{}^m V_{i_{27} i_{28}]} +$$

$$+ 454{,}053{,}600\, R_{[i_1 i_2|f|}{}^a R_{i_3 i_4|a|}{}^b R_{i_5 i_6|b|}{}^c R_{i_7 i_8|c|}{}^d R_{i_9 i_{10}|d|}{}^e R_{i_{11} i_{12}|e|}{}^f R_{i_{13} i_{14}|j|}{}^g R_{i_{15} i_{16}|g|}{}^h R_{i_{17} i_{18}|h|}{}^i R_{i_{19} i_{20}|i|}{}^j R_{i_{21} i_{22}|j|}{}^k R_{i_{23} i_{24}|k|}{}^l R_{i_{25} i_{26}|n|}{}^m R_{i_{27} i_{28}]}{}^m -$$

$$- 908{,}107{,}200\, R_{[i_1 i_2|f|}{}^a R_{i_3 i_4|a|}{}^b R_{i_5 i_6|b|}{}^c R_{i_7 i_8|c|}{}^d R_{i_9 i_{10}|d|}{}^e R_{i_{11} i_{12}|e|}{}^f R_{i_{13} i_{14}|j|}{}^g R_{i_{15} i_{16}|g|}{}^h R_{i_{17} i_{18}|h|}{}^i R_{i_{19} i_{20}|i|}{}^j R_{i_{21} i_{22}|l|}{}^k R_{i_{23} i_{24}|k|}{}^l V_{i_{25} i_{26}} V_{i_{27} i_{28}]} +$$

$$+ 151{,}351{,}200\, R_{[i_1 i_2|f|}{}^a R_{i_3 i_4|a|}{}^b R_{i_5 i_6|b|}{}^c R_{i_7 i_8|c|}{}^d R_{i_9 i_{10}|d|}{}^e R_{i_{11} i_{12}|e|}{}^f R_{i_{13} i_{14}|j|}{}^g R_{i_{15} i_{16}|g|}{}^h R_{i_{17} i_{18}|h|}{}^i R_{i_{19} i_{20}|i|}{}^j V_{i_{21} i_{22}} V_{i_{23} i_{24}} V_{i_{25} i_{26}} V_{i_{27} i_{28}]} +$$

$$+ 403{,}603{,}200\, R_{[i_1 i_2|f|}{}^a R_{i_3 i_4|a|}{}^b R_{i_5 i_6|b|}{}^c R_{i_7 i_8|c|}{}^d R_{i_9 i_{10}|d|}{}^e R_{i_{11} i_{12}|e|}{}^f R_{i_{13} i_{14}|i|}{}^g R_{i_{15} i_{16}|g|}{}^h R_{i_{17} i_{18}|h|}{}^i R_{i_{19} i_{20}|l|}{}^j R_{i_{21} i_{22}|j|}{}^k R_{i_{23} i_{24}|k|}{}^l R_{i_{25} i_{26}|n|}{}^m R_{i_{27} i_{28}]}{}^m -$$

$$- 403{,}603{,}200\, R_{[i_1 i_2|f|}{}^a R_{i_3 i_4|a|}{}^b R_{i_5 i_6|b|}{}^c R_{i_7 i_8|c|}{}^d R_{i_9 i_{10}|d|}{}^e R_{i_{11} i_{12}|e|}{}^f R_{i_{13} i_{14}|i|}{}^g R_{i_{15} i_{16}|g|}{}^h R_{i_{17} i_{18}|h|}{}^i R_{i_{19} i_{20}|l|}{}^j R_{i_{21} i_{22}|j|}{}^k R_{i_{23} i_{24}|k|}{}^l V_{i_{25} i_{26}} V_{i_{27} i_{28}]} -$$

$$- 605{,}404{,}800\, R_{[i_1 i_2|f|}{}^a R_{i_3 i_4|a|}{}^b R_{i_5 i_6|b|}{}^c R_{i_7 i_8|c|}{}^d R_{i_9 i_{10}|d|}{}^e R_{i_{11} i_{12}|e|}{}^f R_{i_{13} i_{14}|i|}{}^g R_{i_{15} i_{16}|g|}{}^h R_{i_{17} i_{18}|h|}{}^i R_{i_{19} i_{20}|k|}{}^j R_{i_{21} i_{22}|j|}{}^k R_{i_{23} i_{24}|m|}{}^l R_{i_{25} i_{26}|l|}{}^m V_{i_{27} i_{28}]} +$$

$$+ 403{,}603{,}200\, R_{[i_1 i_2|f|}{}^a R_{i_3 i_4|a|}{}^b R_{i_5 i_6|b|}{}^c R_{i_7 i_8|c|}{}^d R_{i_9 i_{10}|d|}{}^e R_{i_{11} i_{12}|e|}{}^f R_{i_{13} i_{14}|i|}{}^g R_{i_{15} i_{16}|g|}{}^h R_{i_{17} i_{18}|h|}{}^i R_{i_{19} i_{20}|k|}{}^j R_{i_{21} i_{22}|j|}{}^k V_{i_{23} i_{24}} V_{i_{25} i_{26}} V_{i_{27} i_{28}]} -$$

$$- 40{,}360{,}320\, R_{[i_1 i_2|f|}{}^a R_{i_3 i_4|a|}{}^b R_{i_5 i_6|b|}{}^c R_{i_7 i_8|c|}{}^d R_{i_9 i_{10}|d|}{}^e R_{i_{11} i_{12}|e|}{}^f R_{i_{13} i_{14}|i|}{}^g R_{i_{15} i_{16}|g|}{}^h R_{i_{17} i_{18}|h|}{}^i V_{i_{19} i_{20}} V_{i_{21} i_{22}} V_{i_{23} i_{24}} V_{i_{25} i_{26}} V_{i_{27} i_{28}]} -$$

$$- 37{,}837{,}800\, R_{[i_1 i_2|f|}{}^a R_{i_3 i_4|a|}{}^b R_{i_5 i_6|b|}{}^c R_{i_7 i_8|c|}{}^d R_{i_9 i_{10}|d|}{}^e R_{i_{11} i_{12}|e|}{}^f R_{i_{13} i_{14}|h|}{}^g R_{i_{15} i_{16}|g|}{}^h R_{i_{17} i_{18}|j|}{}^i R_{i_{19} i_{20}|i|}{}^j R_{i_{21} i_{22}|l|}{}^k R_{i_{23} i_{24}|k|}{}^l R_{i_{25} i_{26}|n|}{}^m R_{i_{27} i_{28}]}{}^n +$$

$$+ 151{,}351{,}200\, R_{[i_1 i_2|f|}{}^a R_{i_3 i_4|a|}{}^b R_{i_5 i_6|b|}{}^c R_{i_7 i_8|c|}{}^d R_{i_9 i_{10}|d|}{}^e R_{i_{11} i_{12}|e|}{}^f R_{i_{13} i_{14}|h|}{}^g R_{i_{15} i_{16}|g|}{}^h R_{i_{17} i_{18}|j|}{}^i R_{i_{19} i_{20}|i|}{}^j R_{i_{21} i_{22}|l|}{}^k R_{i_{23} i_{24}|k|}{}^l V_{i_{25} i_{26}} V_{i_{27} i_{28}]} -$$

$$- 75{,}675{,}600\, R_{[i_1 i_2|f|}{}^a R_{i_3 i_4|a|}{}^b R_{i_5 i_6|b|}{}^c R_{i_7 i_8|c|}{}^d R_{i_9 i_{10}|d|}{}^e R_{i_{11} i_{12}|e|}{}^f R_{i_{13} i_{14}|h|}{}^g R_{i_{15} i_{16}|g|}{}^h R_{i_{17} i_{18}|j|}{}^i R_{i_{19} i_{20}|i|}{}^j V_{i_{21} i_{22}} V_{i_{23} i_{24}} V_{i_{25} i_{26}} V_{i_{27} i_{28}]} +$$

$$+ 10{,}090{,}080\, R_{[i_1 i_2|f|}{}^a R_{i_3 i_4|a|}{}^b R_{i_5 i_6|b|}{}^c R_{i_7 i_8|c|}{}^d R_{i_9 i_{10}|d|}{}^e R_{i_{11} i_{12}|e|}{}^f R_{i_{13} i_{14}|h|}{}^g R_{i_{15} i_{16}|g|}{}^h V_{i_{17} i_{18}} V_{i_{19} i_{20}} V_{i_{21} i_{22}} V_{i_{23} i_{24}} V_{i_{25} i_{26}} V_{i_{27} i_{28}]} -$$

$$- 360{,}360\, R_{[i_1 i_2|f|}{}^a R_{i_3 i_4|a|}{}^b R_{i_5 i_6|b|}{}^c R_{i_7 i_8|c|}{}^d R_{i_9 i_{10}|d|}{}^e R_{i_{11} i_{12}|e|}{}^f V_{i_{13} i_{14}} V_{i_{15} i_{16}} V_{i_{17} i_{18}} V_{i_{19} i_{20}} V_{i_{21} i_{22}} V_{i_{23} i_{24}} V_{i_{25} i_{26}} V_{i_{27} i_{28}]} -$$

$$- 435{,}891{,}456\, R_{[i_1 i_2|e|}{}^a R_{i_3 i_4|a|}{}^b R_{i_5 i_6|b|}{}^c R_{i_7 i_8|c|}{}^d R_{i_9 i_{10}|d|}{}^e R_{i_{11} i_{12}|j|}{}^f R_{i_{13} i_{14}|f|}{}^g R_{i_{15} i_{16}|g|}{}^h R_{i_{17} i_{18}|h|}{}^i R_{i_{19} i_{20}|i|}{}^j R_{i_{21} i_{22}|n|}{}^k R_{i_{23} i_{24}|k|}{}^l R_{i_{25} i_{26}|l|}{}^m R_{i_{27} i_{28}]}{}^n +$$

$$+ 581{,}188{,}608\, R_{[i_1 i_2|e|}{}^a R_{i_3 i_4|a|}{}^b R_{i_5 i_6|b|}{}^c R_{i_7 i_8|c|}{}^d R_{i_9 i_{10}|d|}{}^e R_{i_{11} i_{12}|j|}{}^f R_{i_{13} i_{14}|f|}{}^g R_{i_{15} i_{16}|g|}{}^h R_{i_{17} i_{18}|h|}{}^i R_{i_{19} i_{20}|i|}{}^j R_{i_{21} i_{22}|m|}{}^k R_{i_{23} i_{24}|k|}{}^l R_{i_{25} i_{26}|l|}{}^m V_{i_{27} i_{28}]} +$$

$$+ 217{,}945{,}728\, R_{[i_1 i_2|e|}{}^a R_{i_3 i_4|a|}{}^b R_{i_5 i_6|b|}{}^c R_{i_7 i_8|c|}{}^d R_{i_9 i_{10}|d|}{}^e R_{i_{11} i_{12}|j|}{}^f R_{i_{13} i_{14}|f|}{}^g R_{i_{15} i_{16}|g|}{}^h R_{i_{17} i_{18}|h|}{}^i R_{i_{19} i_{20}|i|}{}^j R_{i_{21} i_{22}|l|}{}^k R_{i_{23} i_{24}|k|}{}^l R_{i_{25} i_{26}|n|}{}^m R_{i_{27} i_{28}]}{}^m -$$

$$- 435{,}891{,}456\, R_{[i_1 i_2|e|}{}^a R_{i_3 i_4|a|}{}^b R_{i_5 i_6|b|}{}^c R_{i_7 i_8|c|}{}^d R_{i_9 i_{10}|d|}{}^e R_{i_{11} i_{12}|j|}{}^f R_{i_{13} i_{14}|f|}{}^g R_{i_{15} i_{16}|g|}{}^h R_{i_{17} i_{18}|h|}{}^i R_{i_{19} i_{20}|i|}{}^j R_{i_{21} i_{22}|l|}{}^k R_{i_{23} i_{24}|k|}{}^l V_{i_{25} i_{26}} V_{i_{27} i_{28}]} +$$

$$+ 72{,}648{,}576\, R_{[i_1 i_2|e|}{}^a R_{i_3 i_4|a|}{}^b R_{i_5 i_6|b|}{}^c R_{i_7 i_8|c|}{}^d R_{i_9 i_{10}|d|}{}^e R_{i_{11} i_{12}|j|}{}^f R_{i_{13} i_{14}|f|}{}^g R_{i_{15} i_{16}|g|}{}^h R_{i_{17} i_{18}|h|}{}^i R_{i_{19} i_{20}|i|}{}^j V_{i_{21} i_{22}} V_{i_{23} i_{24}} V_{i_{25} i_{26}} V_{i_{27} i_{28}]} +$$

$$+ 544{,}864{,}320\, R_{[i_1 i_2|e|}{}^a R_{i_3 i_4|a|}{}^b R_{i_5 i_6|b|}{}^c R_{i_7 i_8|c|}{}^d R_{i_9 i_{10}|d|}{}^e R_{i_{11} i_{12}|i|}{}^f R_{i_{13} i_{14}|f|}{}^g R_{i_{15} i_{16}|g|}{}^h R_{i_{17} i_{18}|h|}{}^i R_{i_{19} i_{20}|m|}{}^j R_{i_{21} i_{22}|j|}{}^k R_{i_{23} i_{24}|k|}{}^l R_{i_{25} i_{26}|l|}{}^m V_{i_{27} i_{28}]} +$$

$$+ 726{,}485{,}760\, R_{[i_1 i_2|e|}{}^a R_{i_3 i_4|a|}{}^b R_{i_5 i_6|b|}{}^c R_{i_7 i_8|c|}{}^d R_{i_9 i_{10}|d|}{}^e R_{i_{11} i_{12}|i|}{}^f R_{i_{13} i_{14}|f|}{}^g R_{i_{15} i_{16}|g|}{}^h R_{i_{17} i_{18}|h|}{}^i R_{i_{19} i_{20}|l|}{}^j R_{i_{21} i_{22}|j|}{}^k R_{i_{23} i_{24}|k|}{}^l R_{i_{25} i_{26}|n|}{}^m R_{i_{27} i_{28}]}{}^m -$$

$$- 726{,}485{,}760\, R_{[i_1 i_2|e|}{}^a R_{i_3 i_4|a|}{}^b R_{i_5 i_6|b|}{}^c R_{i_7 i_8|c|}{}^d R_{i_9 i_{10}|d|}{}^e R_{i_{11} i_{12}|i|}{}^f R_{i_{13} i_{14}|f|}{}^g R_{i_{15} i_{16}|g|}{}^h R_{i_{17} i_{18}|h|}{}^i R_{i_{19} i_{20}|l|}{}^j R_{i_{21} i_{22}|j|}{}^k R_{i_{23} i_{24}|k|}{}^l V_{i_{25} i_{26}} V_{i_{27} i_{28}]} -$$

$$- 544{,}864{,}320\, R_{[i_1 i_2|e|}{}^a R_{i_3 i_4|a|}{}^b R_{i_5 i_6|b|}{}^c R_{i_7 i_8|c|}{}^d R_{i_9 i_{10}|d|}{}^e R_{i_{11} i_{12}|i|}{}^f R_{i_{13} i_{14}|f|}{}^g R_{i_{15} i_{16}|g|}{}^h R_{i_{17} i_{18}|h|}{}^i R_{i_{19} i_{20}|k|}{}^j R_{i_{21} i_{22}|j|}{}^k R_{i_{23} i_{24}|m|}{}^l R_{i_{25} i_{26}|l|}{}^m V_{i_{27} i_{28}]} +$$

$$+ 363{,}242{,}880\, R_{[i_1 i_2|e|}{}^a R_{i_3 i_4|a|}{}^b R_{i_5 i_6|b|}{}^c R_{i_7 i_8|c|}{}^d R_{i_9 i_{10}|d|}{}^e R_{i_{11} i_{12}|i|}{}^f R_{i_{13} i_{14}|f|}{}^g R_{i_{15} i_{16}|g|}{}^h R_{i_{17} i_{18}|h|}{}^i R_{i_{19} i_{20}|k|}{}^j R_{i_{21} i_{22}|j|}{}^k V_{i_{23} i_{24}} V_{i_{25} i_{26}} V_{i_{27} i_{28}]} -$$

$$- 36{,}324{,}288\, R_{[i_1 i_2|e|}{}^a R_{i_3 i_4|a|}{}^b R_{i_5 i_6|b|}{}^c R_{i_7 i_8|c|}{}^d R_{i_9 i_{10}|d|}{}^e R_{i_{11} i_{12}|i|}{}^f R_{i_{13} i_{14}|f|}{}^g R_{i_{15} i_{16}|g|}{}^h R_{i_{17} i_{18}|h|}{}^i V_{i_{19} i_{20}} V_{i_{21} i_{22}} V_{i_{23} i_{24}} V_{i_{25} i_{26}} V_{i_{27} i_{28}]} +$$

$$+ 107{,}627{,}520\, R_{[i_1i_2|e|}{}^a R_{i_3i_4|a|}{}^b R_{i_5i_6|b|}{}^c R_{i_7i_8|c|}{}^d R_{i_9i_{10}|d|}{}^e R_{i_{11}i_{12}|h|}{}^f R_{i_{13}i_{14}|f|}{}^g R_{i_{15}i_{16}|g|}{}^h R_{i_{17}i_{18}|k|}{}^i R_{i_{19}i_{20}|i|}{}^j R_{i_{21}i_{22}|j|}{}^k R_{i_{23}i_{24}|n|}{}^l R_{i_{25}i_{26}|l|}{}^m R_{i_{27}i_{28}]m}{}^n -$$

$$- 484{,}323{,}840\, R_{[i_1i_2|e|}{}^a R_{i_3i_4|a|}{}^b R_{i_5i_6|b|}{}^c R_{i_7i_8|c|}{}^d R_{i_9i_{10}|d|}{}^e R_{i_{11}i_{12}|h|}{}^f R_{i_{13}i_{14}|f|}{}^g R_{i_{15}i_{16}|g|}{}^h R_{i_{17}i_{18}|k|}{}^i R_{i_{19}i_{20}|i|}{}^j R_{i_{21}i_{22}|j|}{}^k R_{i_{23}i_{24}|m|}{}^l R_{i_{25}i_{26}|l|}{}^m V_{i_{27}i_{28}]} +$$

$$+ 161{,}441{,}280\, R_{[i_1i_2|e|}{}^a R_{i_3i_4|a|}{}^b R_{i_5i_6|b|}{}^c R_{i_7i_8|c|}{}^d R_{i_9i_{10}|d|}{}^e R_{i_{11}i_{12}|h|}{}^f R_{i_{13}i_{14}|f|}{}^g R_{i_{15}i_{16}|g|}{}^h R_{i_{17}i_{18}|k|}{}^i R_{i_{19}i_{20}|i|}{}^j R_{i_{21}i_{22}|j|}{}^k V_{i_{23}i_{24}} V_{i_{25}i_{26}} V_{i_{27}i_{28}]} -$$

$$- 121{,}080{,}960\, R_{[i_1i_2|e|}{}^a R_{i_3i_4|a|}{}^b R_{i_5i_6|b|}{}^c R_{i_7i_8|c|}{}^d R_{i_9i_{10}|d|}{}^e R_{i_{11}i_{12}|h|}{}^f R_{i_{13}i_{14}|f|}{}^g R_{i_{15}i_{16}|g|}{}^h R_{i_{17}i_{18}|j|}{}^i R_{i_{19}i_{20}|i|}{}^j R_{i_{21}i_{22}|l|}{}^k R_{i_{23}i_{24}|k|}{}^l R_{i_{25}i_{26}|n|}{}^m R_{i_{27}i_{28}]m}{}^n +$$

$$+ 363{,}242{,}880\, R_{[i_1i_2|e|}{}^a R_{i_3i_4|a|}{}^b R_{i_5i_6|b|}{}^c R_{i_7i_8|c|}{}^d R_{i_9i_{10}|d|}{}^e R_{i_{11}i_{12}|h|}{}^f R_{i_{13}i_{14}|f|}{}^g R_{i_{15}i_{16}|g|}{}^h R_{i_{17}i_{18}|j|}{}^i R_{i_{19}i_{20}|i|}{}^j R_{i_{21}i_{22}|l|}{}^k R_{i_{23}i_{24}|k|}{}^l V_{i_{25}i_{26}} V_{i_{27}i_{28}]} -$$

$$- 121{,}080{,}960\, R_{[i_1i_2|e|}{}^a R_{i_3i_4|a|}{}^b R_{i_5i_6|b|}{}^c R_{i_7i_8|c|}{}^d R_{i_9i_{10}|d|}{}^e R_{i_{11}i_{12}|h|}{}^f R_{i_{13}i_{14}|f|}{}^g R_{i_{15}i_{16}|g|}{}^h R_{i_{17}i_{18}|j|}{}^i R_{i_{19}i_{20}|i|}{}^j V_{i_{21}i_{22}} V_{i_{23}i_{24}} V_{i_{25}i_{26}} V_{i_{27}i_{28}]} +$$

$$+ 8{,}072{,}064\, R_{[i_1i_2|e|}{}^a R_{i_3i_4|a|}{}^b R_{i_5i_6|b|}{}^c R_{i_7i_8|c|}{}^d R_{i_9i_{10}|d|}{}^e R_{i_{11}i_{12}|h|}{}^f R_{i_{13}i_{14}|f|}{}^g R_{i_{15}i_{16}|g|}{}^h V_{i_{17}i_{18}} V_{i_{19}i_{20}} V_{i_{21}i_{22}} V_{i_{23}i_{24}} V_{i_{25}i_{26}} V_{i_{27}i_{28}]} +$$

$$+ 45{,}405{,}360\, R_{[i_1i_2|e|}{}^a R_{i_3i_4|a|}{}^b R_{i_5i_6|b|}{}^c R_{i_7i_8|c|}{}^d R_{i_9i_{10}|d|}{}^e R_{i_{11}i_{12}|g|}{}^f R_{i_{13}i_{14}|f|}{}^g R_{i_{15}i_{16}|i|}{}^h R_{i_{17}i_{18}|h|}{}^i R_{i_{19}i_{20}|k|}{}^j R_{i_{21}i_{22}|j|}{}^k R_{i_{23}i_{24}|m|}{}^l R_{i_{25}i_{26}|l|}{}^m V_{i_{27}i_{28}]} -$$

$$- 60{,}540{,}480\, R_{[i_1i_2|e|}{}^a R_{i_3i_4|a|}{}^b R_{i_5i_6|b|}{}^c R_{i_7i_8|c|}{}^d R_{i_9i_{10}|d|}{}^e R_{i_{11}i_{12}|g|}{}^f R_{i_{13}i_{14}|f|}{}^g R_{i_{15}i_{16}|i|}{}^h R_{i_{17}i_{18}|h|}{}^i R_{i_{19}i_{20}|k|}{}^j R_{i_{21}i_{22}|j|}{}^k V_{i_{23}i_{24}} V_{i_{25}i_{26}} V_{i_{27}i_{28}]} +$$

$$+ 18{,}162{,}144\, R_{[i_1i_2|e|}{}^a R_{i_3i_4|a|}{}^b R_{i_5i_6|b|}{}^c R_{i_7i_8|c|}{}^d R_{i_9i_{10}|d|}{}^e R_{i_{11}i_{12}|g|}{}^f R_{i_{13}i_{14}|f|}{}^g R_{i_{15}i_{16}|i|}{}^h R_{i_{17}i_{18}|h|}{}^i V_{i_{19}i_{20}} V_{i_{21}i_{22}} V_{i_{23}i_{24}} V_{i_{25}i_{26}} V_{i_{27}i_{28}]} -$$

$$- 1{,}729{,}728\, R_{[i_1i_2|e|}{}^a R_{i_3i_4|a|}{}^b R_{i_5i_6|b|}{}^c R_{i_7i_8|c|}{}^d R_{i_9i_{10}|d|}{}^e R_{i_{11}i_{12}|g|}{}^f R_{i_{13}i_{14}|f|}{}^g V_{i_{15}i_{16}} V_{i_{17}i_{18}} V_{i_{19}i_{20}} V_{i_{21}i_{22}} V_{i_{23}i_{24}} V_{i_{25}i_{26}} V_{i_{27}i_{28}]} +$$

$$+ 48{,}048\, R_{[i_1i_2|e|}{}^a R_{i_3i_4|a|}{}^b R_{i_5i_6|b|}{}^c R_{i_7i_8|c|}{}^d R_{i_9i_{10}|d|}{}^e V_{i_{11}i_{12}} V_{i_{13}i_{14}} V_{i_{15}i_{16}} V_{i_{17}i_{18}} V_{i_{19}i_{20}} V_{i_{21}i_{22}} V_{i_{23}i_{24}} V_{i_{25}i_{26}} V_{i_{27}i_{28}]} +$$

$$+ 113{,}513{,}400\, R_{[i_1i_2|d|}{}^a R_{i_3i_4|a|}{}^b R_{i_5i_6|b|}{}^c R_{i_7i_8|c|}{}^d R_{i_9i_{10}|h|}{}^e R_{i_{11}i_{12}|e|}{}^f R_{i_{13}i_{14}|f|}{}^g R_{i_{15}i_{16}|g|}{}^h R_{i_{17}i_{18}|l|}{}^i R_{i_{19}i_{20}|i|}{}^j R_{i_{21}i_{22}|j|}{}^k R_{i_{23}i_{24}|k|}{}^l R_{i_{25}i_{26}|n|}{}^m R_{i_{27}i_{28}]m}{}^n -$$

$$- 113{,}513{,}400\, R_{[i_1i_2|d|}{}^a R_{i_3i_4|a|}{}^b R_{i_5i_6|b|}{}^c R_{i_7i_8|c|}{}^d R_{i_9i_{10}|h|}{}^e R_{i_{11}i_{12}|e|}{}^f R_{i_{13}i_{14}|f|}{}^g R_{i_{15}i_{16}|g|}{}^h R_{i_{17}i_{18}|l|}{}^i R_{i_{19}i_{20}|i|}{}^j R_{i_{21}i_{22}|j|}{}^k R_{i_{23}i_{24}|k|}{}^l V_{i_{25}i_{26}} V_{i_{27}i_{28}]} +$$

$$+ 151{,}351{,}200\, R_{[i_1i_2|d|}{}^a R_{i_3i_4|a|}{}^b R_{i_5i_6|b|}{}^c R_{i_7i_8|c|}{}^d R_{i_9i_{10}|h|}{}^e R_{i_{11}i_{12}|e|}{}^f R_{i_{13}i_{14}|f|}{}^g R_{i_{15}i_{16}|g|}{}^h R_{i_{17}i_{18}|k|}{}^i R_{i_{19}i_{20}|i|}{}^j R_{i_{21}i_{22}|j|}{}^k R_{i_{23}i_{24}|n|}{}^l R_{i_{25}i_{26}|l|}{}^m R_{i_{27}i_{28}]m}{}^n -$$

$$- 454{,}053{,}600\, R_{[i_1i_2|d|}{}^a R_{i_3i_4|a|}{}^b R_{i_5i_6|b|}{}^c R_{i_7i_8|c|}{}^d R_{i_9i_{10}|h|}{}^e R_{i_{11}i_{12}|e|}{}^f R_{i_{13}i_{14}|f|}{}^g R_{i_{15}i_{16}|g|}{}^h R_{i_{17}i_{18}|k|}{}^i R_{i_{19}i_{20}|i|}{}^j R_{i_{21}i_{22}|j|}{}^k R_{i_{23}i_{24}|m|}{}^l R_{i_{25}i_{26}|l|}{}^m V_{i_{27}i_{28}]} +$$

$$+ 151{,}351{,}200\, R_{[i_1i_2|d|}{}^a R_{i_3i_4|a|}{}^b R_{i_5i_6|b|}{}^c R_{i_7i_8|c|}{}^d R_{i_9i_{10}|h|}{}^e R_{i_{11}i_{12}|e|}{}^f R_{i_{13}i_{14}|f|}{}^g R_{i_{15}i_{16}|g|}{}^h R_{i_{17}i_{18}|k|}{}^i R_{i_{19}i_{20}|i|}{}^j R_{i_{21}i_{22}|j|}{}^k V_{i_{23}i_{24}} V_{i_{25}i_{26}} V_{i_{27}i_{28}]} -$$

$$- 56{,}756{,}700\, R_{[i_1i_2|d|}{}^a R_{i_3i_4|a|}{}^b R_{i_5i_6|b|}{}^c R_{i_7i_8|c|}{}^d R_{i_9i_{10}|h|}{}^e R_{i_{11}i_{12}|e|}{}^f R_{i_{13}i_{14}|f|}{}^g R_{i_{15}i_{16}|g|}{}^h R_{i_{17}i_{18}|j|}{}^i R_{i_{19}i_{20}|i|}{}^j R_{i_{21}i_{22}|l|}{}^k R_{i_{23}i_{24}|k|}{}^l R_{i_{25}i_{26}|n|}{}^m R_{i_{27}i_{28}]m}{}^n +$$

$$+ 170{,}270{,}100\, R_{[i_1i_2|d|}{}^a R_{i_3i_4|a|}{}^b R_{i_5i_6|b|}{}^c R_{i_7i_8|c|}{}^d R_{i_9i_{10}|h|}{}^e R_{i_{11}i_{12}|e|}{}^f R_{i_{13}i_{14}|f|}{}^g R_{i_{15}i_{16}|g|}{}^h R_{i_{17}i_{18}|j|}{}^i R_{i_{19}i_{20}|i|}{}^j R_{i_{21}i_{22}|l|}{}^k R_{i_{23}i_{24}|k|}{}^l V_{i_{25}i_{26}} V_{i_{27}i_{28}]} -$$

$$- 56{,}756{,}700\, R_{[i_1i_2|d|}{}^a R_{i_3i_4|a|}{}^b R_{i_5i_6|b|}{}^c R_{i_7i_8|c|}{}^d R_{i_9i_{10}|h|}{}^e R_{i_{11}i_{12}|e|}{}^f R_{i_{13}i_{14}|f|}{}^g R_{i_{15}i_{16}|g|}{}^h R_{i_{17}i_{18}|j|}{}^i R_{i_{19}i_{20}|i|}{}^j V_{i_{21}i_{22}} V_{i_{23}i_{24}} V_{i_{25}i_{26}} V_{i_{27}i_{28}]} +$$

$$+ 3{,}783{,}780\, R_{[i_1i_2|d|}{}^a R_{i_3i_4|a|}{}^b R_{i_5i_6|b|}{}^c R_{i_7i_8|c|}{}^d R_{i_9i_{10}|h|}{}^e R_{i_{11}i_{12}|e|}{}^f R_{i_{13}i_{14}|f|}{}^g R_{i_{15}i_{16}|g|}{}^h V_{i_{17}i_{18}} V_{i_{19}i_{20}} V_{i_{21}i_{22}} V_{i_{23}i_{24}} V_{i_{25}i_{26}} V_{i_{27}i_{28}]} -$$

$$- 134{,}534{,}400\, R_{[i_1i_2|d|}{}^a R_{i_3i_4|a|}{}^b R_{i_5i_6|b|}{}^c R_{i_7i_8|c|}{}^d R_{i_9i_{10}|g|}{}^e R_{i_{11}i_{12}|e|}{}^f R_{i_{13}i_{14}|f|}{}^g R_{i_{15}i_{16}|j|}{}^h R_{i_{17}i_{18}|h|}{}^i R_{i_{19}i_{20}|i|}{}^j R_{i_{21}i_{22}|m|}{}^k R_{i_{23}i_{24}|k|}{}^l R_{i_{25}i_{26}|l|}{}^m V_{i_{27}i_{28}]} -$$

$$- 151{,}351{,}200\, R_{[i_1i_2|d|}{}^a R_{i_3i_4|a|}{}^b R_{i_5i_6|b|}{}^c R_{i_7i_8|c|}{}^d R_{i_9i_{10}|g|}{}^e R_{i_{11}i_{12}|e|}{}^f R_{i_{13}i_{14}|f|}{}^g R_{i_{15}i_{16}|j|}{}^h R_{i_{17}i_{18}|h|}{}^i R_{i_{19}i_{20}|i|}{}^j R_{i_{21}i_{22}|j|}{}^k R_{i_{23}i_{24}|k|}{}^l R_{i_{25}i_{26}|n|}{}^m R_{i_{27}i_{28}]m}{}^n +$$

$$+ 302{,}702{,}400\, R_{[i_1i_2|d|}{}^a R_{i_3i_4|a|}{}^b R_{i_5i_6|b|}{}^c R_{i_7i_8|c|}{}^d R_{i_9i_{10}|g|}{}^e R_{i_{11}i_{12}|e|}{}^f R_{i_{13}i_{14}|f|}{}^g R_{i_{15}i_{16}|j|}{}^h R_{i_{17}i_{18}|h|}{}^i R_{i_{19}i_{20}|i|}{}^j R_{i_{21}i_{22}|l|}{}^k R_{i_{23}i_{24}|k|}{}^l V_{i_{25}i_{26}} V_{i_{27}i_{28}]} -$$

$$- 50{,}450{,}400\, R_{[i_1i_2|d|}{}^a R_{i_3i_4|a|}{}^b R_{i_5i_6|b|}{}^c R_{i_7i_8|c|}{}^d R_{i_9i_{10}|g|}{}^e R_{i_{11}i_{12}|e|}{}^f R_{i_{13}i_{14}|f|}{}^g R_{i_{15}i_{16}|j|}{}^h R_{i_{17}i_{18}|h|}{}^i R_{i_{19}i_{20}|i|}{}^j V_{i_{21}i_{22}} V_{i_{23}i_{24}} V_{i_{25}i_{26}} V_{i_{27}i_{28}]} +$$

$$+ 151{,}351{,}200\, R_{[i_1i_2|d|}{}^a R_{i_3i_4|a|}{}^b R_{i_5i_6|b|}{}^c R_{i_7i_8|c|}{}^d R_{i_9i_{10}|g|}{}^e R_{i_{11}i_{12}|e|}{}^f R_{i_{13}i_{14}|f|}{}^g R_{i_{15}i_{16}|i|}{}^h R_{i_{17}i_{18}|h|}{}^i R_{i_{19}i_{20}|k|}{}^j R_{i_{21}i_{22}|j|}{}^k R_{i_{23}i_{24}|m|}{}^l R_{i_{25}i_{26}|l|}{}^m V_{i_{27}i_{28}]} -$$

$$- 151{,}351{,}200\, R_{[i_1i_2|d|}{}^a R_{i_3i_4|a|}{}^b R_{i_5i_6|b|}{}^c R_{i_7i_8|c|}{}^d R_{i_9i_{10}|g|}{}^e R_{i_{11}i_{12}|e|}{}^f R_{i_{13}i_{14}|f|}{}^g R_{i_{15}i_{16}|i|}{}^h R_{i_{17}i_{18}|h|}{}^i R_{i_{19}i_{20}|k|}{}^j R_{i_{21}i_{22}|j|}{}^k V_{i_{23}i_{24}} V_{i_{25}i_{26}} V_{i_{27}i_{28}]} +$$

$$+ 30{,}270{,}240\, R_{[i_1i_2|d|}{}^a R_{i_3i_4|a|}{}^b R_{i_5i_6|b|}{}^c R_{i_7i_8|c|}{}^d R_{i_9i_{10}|g|}{}^e R_{i_{11}i_{12}|e|}{}^f R_{i_{13}i_{14}|f|}{}^g R_{i_{15}i_{16}|i|}{}^h R_{i_{17}i_{18}|h|}{}^i V_{i_{19}i_{20}} V_{i_{21}i_{22}} V_{i_{23}i_{24}} V_{i_{25}i_{26}} V_{i_{27}i_{28}]} -$$

$$- 1{,}441{,}440\, R_{[i_1i_2|d|}{}^a R_{i_3i_4|a|}{}^b R_{i_5i_6|b|}{}^c R_{i_7i_8|c|}{}^d R_{i_9i_{10}|g|}{}^e R_{i_{11}i_{12}|e|}{}^f R_{i_{13}i_{14}|f|}{}^g V_{i_{15}i_{16}} V_{i_{17}i_{18}} V_{i_{19}i_{20}} V_{i_{21}i_{22}} V_{i_{23}i_{24}} V_{i_{25}i_{26}} V_{i_{27}i_{28}]} +$$

$$+ 5{,}675{,}670\, R_{[i_1i_2|d|}{}^a R_{i_3i_4|a|}{}^b R_{i_5i_6|b|}{}^c R_{i_7i_8|c|}{}^d R_{i_9i_{10}|f|}{}^e R_{i_{11}i_{12}|e|}{}^f R_{i_{13}i_{14}|h|}{}^g R_{i_{15}i_{16}|g|}{}^h R_{i_{17}i_{18}|j|}{}^i R_{i_{19}i_{20}|i|}{}^j R_{i_{21}i_{22}|l|}{}^k R_{i_{23}i_{24}|k|}{}^l R_{i_{25}i_{26}|n|}{}^m R_{i_{27}i_{28}]m}{}^n -$$

$$- 28{,}378{,}350\, R_{[i_1i_2|d|}{}^a R_{i_3i_4|a|}{}^b R_{i_5i_6|b|}{}^c R_{i_7i_8|c|}{}^d R_{i_9i_{10}|f|}{}^e R_{i_{11}i_{12}|e|}{}^f R_{i_{13}i_{14}|h|}{}^g R_{i_{15}i_{16}|g|}{}^h R_{i_{17}i_{18}|j|}{}^i R_{i_{19}i_{20}|i|}{}^j R_{i_{21}i_{22}|l|}{}^k R_{i_{23}i_{24}|k|}{}^l V_{i_{25}i_{26}} V_{i_{27}i_{28}]} +$$

$$+ 18{,}918{,}900\, R_{[i_1i_2|d|}{}^a R_{i_3i_4|a|}{}^b R_{i_5i_6|b|}{}^c R_{i_7i_8|c|}{}^d R_{i_9i_{10}|f|}{}^e R_{i_{11}i_{12}|e|}{}^f R_{i_{13}i_{14}|h|}{}^g R_{i_{15}i_{16}|g|}{}^h R_{i_{17}i_{18}|j|}{}^i R_{i_{19}i_{20}|i|}{}^j V_{i_{21}i_{22}} V_{i_{23}i_{24}} V_{i_{25}i_{26}} V_{i_{27}i_{28}]} -$$

$$- 3{,}783{,}780\, R_{[i_1i_2|d|}{}^a R_{i_3i_4|a|}{}^b R_{i_5i_6|b|}{}^c R_{i_7i_8|c|}{}^d R_{i_9i_{10}|f|}{}^e R_{i_{11}i_{12}|e|}{}^f R_{i_{13}i_{14}|h|}{}^g R_{i_{15}i_{16}|g|}{}^h V_{i_{17}i_{18}} V_{i_{19}i_{20}} V_{i_{21}i_{22}} V_{i_{23}i_{24}} V_{i_{25}i_{26}} V_{i_{27}i_{28}]} +$$

$$+ 270{,}270\, R_{[i_1i_2|d|}{}^a R_{i_3i_4|a|}{}^b R_{i_5i_6|b|}{}^c R_{i_7i_8|c|}{}^d R_{i_9i_{10}|f|}{}^e R_{i_{11}i_{12}|e|}{}^f V_{i_{13}i_{14}} V_{i_{15}i_{16}} V_{i_{17}i_{18}} V_{i_{19}i_{20}} V_{i_{21}i_{22}} V_{i_{23}i_{24}} V_{i_{25}i_{26}} V_{i_{27}i_{28}]} -$$

$$- 6{,}006\, R_{[i_1i_2|d|}{}^a R_{i_3i_4|a|}{}^b R_{i_5i_6|b|}{}^c R_{i_7i_8|c|}{}^d V_{i_9i_{10}} V_{i_{11}i_{12}} V_{i_{13}i_{14}} V_{i_{15}i_{16}} V_{i_{17}i_{18}} V_{i_{19}i_{20}} V_{i_{21}i_{22}} V_{i_{23}i_{24}} V_{i_{25}i_{26}} V_{i_{27}i_{28}]} -$$

$$- 22{,}422{,}400\, R_{[i_1i_2|c|}{}^a R_{i_3i_4|a|}{}^b R_{i_5i_6|b|}{}^c R_{i_7i_8|f|}{}^d R_{i_9i_{10}|d|}{}^e R_{i_{11}i_{12}|e|}{}^f R_{i_{13}i_{14}|i|}{}^g R_{i_{15}i_{16}|g|}{}^h R_{i_{17}i_{18}|h|}{}^i R_{i_{19}i_{20}|l|}{}^j R_{i_{21}i_{22}|j|}{}^k R_{i_{23}i_{24}|k|}{}^l R_{i_{25}i_{26}|n|}{}^m R_{i_{27}i_{28}]m}{}^n +$$

$$+ 22{,}422{,}400\, R_{[i_1i_2|c|}{}^a R_{i_3i_4|a|}{}^b R_{i_5i_6|b|}{}^c R_{i_7i_8|f|}{}^d R_{i_9i_{10}|d|}{}^e R_{i_{11}i_{12}|e|}{}^f R_{i_{13}i_{14}|i|}{}^g R_{i_{15}i_{16}|g|}{}^h R_{i_{17}i_{18}|h|}{}^i R_{i_{19}i_{20}|l|}{}^j R_{i_{21}i_{22}|j|}{}^k R_{i_{23}i_{24}|k|}{}^l V_{i_{25}i_{26}} V_{i_{27}i_{28}]} +$$

$$+ 67{,}267{,}200\, R_{[i_1i_2|c|}{}^a R_{i_3i_4|a|}{}^b R_{i_5i_6|b|}{}^c R_{i_7i_8|f|}{}^d R_{i_9i_{10}|d|}{}^e R_{i_{11}i_{12}|e|}{}^f R_{i_{13}i_{14}|i|}{}^g R_{i_{15}i_{16}|g|}{}^h R_{i_{17}i_{18}|h|}{}^i R_{i_{19}i_{20}|k|}{}^j R_{i_{21}i_{22}|j|}{}^k R_{i_{23}i_{24}|m|}{}^l R_{i_{25}i_{26}|l|}{}^m V_{i_{27}i_{28}]} -$$

$$- 44{,}844{,}800\, R_{[i_1i_2|c|}{}^a R_{i_3i_4|a|}{}^b R_{i_5i_6|b|}{}^c R_{i_7i_8|f|}{}^d R_{i_9i_{10}|d|}{}^e R_{i_{11}i_{12}|e|}{}^f R_{i_{13}i_{14}|i|}{}^g R_{i_{15}i_{16}|g|}{}^h R_{i_{17}i_{18}|h|}{}^i R_{i_{19}i_{20}|k|}{}^j R_{i_{21}i_{22}|j|}{}^k V_{i_{23}i_{24}} V_{i_{25}i_{26}} V_{i_{27}i_{28}]} +$$

$$+ 4{,}484{,}480\, R_{[i_1i_2|c|}{}^a R_{i_3i_4|a|}{}^b R_{i_5i_6|b|}{}^c R_{i_7i_8|f|}{}^d R_{i_9i_{10}|d|}{}^e R_{i_{11}i_{12}|e|}{}^f R_{i_{13}i_{14}|i|}{}^g R_{i_{15}i_{16}|g|}{}^h R_{i_{17}i_{18}|h|}{}^i V_{i_{19}i_{20}} V_{i_{21}i_{22}} V_{i_{23}i_{24}} V_{i_{25}i_{26}} V_{i_{27}i_{28}]} +$$

$$+ 12{,}612{,}600\, R_{[i_1i_2|c|}{}^a R_{i_3i_4|a|}{}^b R_{i_5i_6|b|}{}^c R_{i_7i_8|f|}{}^d R_{i_9i_{10}|d|}{}^e R_{i_{11}i_{12}|e|}{}^f R_{i_{13}i_{14}|h|}{}^g R_{i_{15}i_{16}|g|}{}^h R_{i_{17}i_{18}|j|}{}^i R_{i_{19}i_{20}|i|}{}^j R_{i_{21}i_{22}|l|}{}^k R_{i_{23}i_{24}|k|}{}^l R_{i_{25}i_{26}|n|}{}^m R_{i_{27}i_{28}]m}{}^n -$$

$$- 50{,}450{,}400\, R_{[i_1i_2|c|}{}^a R_{i_3i_4|a|}{}^b R_{i_5i_6|b|}{}^c R_{i_7i_8|f|}{}^d R_{i_9i_{10}|d|}{}^e R_{i_{11}i_{12}|e|}{}^f R_{i_{13}i_{14}|h|}{}^g R_{i_{15}i_{16}|g|}{}^h R_{i_{17}i_{18}|j|}{}^i R_{i_{19}i_{20}|i|}{}^j R_{i_{21}i_{22}|l|}{}^k R_{i_{23}i_{24}|k|}{}^l V_{i_{25}i_{26}} V_{i_{27}i_{28}]} +$$

$$+ 25{,}225{,}200\, R_{[i_1i_2|c|}{}^a R_{i_3i_4|a|}{}^b R_{i_5i_6|b|}{}^c R_{i_7i_8|f|}{}^d R_{i_9i_{10}|d|}{}^e R_{i_{11}i_{12}|e|}{}^f R_{i_{13}i_{14}|h|}{}^g R_{i_{15}i_{16}|g|}{}^h R_{i_{17}i_{18}|j|}{}^i R_{i_{19}i_{20}|i|}{}^j V_{i_{21}i_{22}} V_{i_{23}i_{24}} V_{i_{25}i_{26}} V_{i_{27}i_{28}]} -$$

$$- 3{,}363{,}360\, R_{[i_1i_2|c|}{}^a R_{i_3i_4|a|}{}^b R_{i_5i_6|b|}{}^c R_{i_7i_8|f|}{}^d R_{i_9i_{10}|d|}{}^e R_{i_{11}i_{12}|e|}{}^f R_{i_{13}i_{14}|h|}{}^g R_{i_{15}i_{16}|g|}{}^h V_{i_{17}i_{18}} V_{i_{19}i_{20}} V_{i_{21}i_{22}} V_{i_{23}i_{24}} V_{i_{25}i_{26}} V_{i_{27}i_{28}]} +$$

$$+ 120{,}120\, R_{[i_1i_2|c|}{}^a R_{i_3i_4|a|}{}^b R_{i_5i_6|b|}{}^c R_{i_7i_8|f|}{}^d R_{i_9i_{10}|d|}{}^e R_{i_{11}i_{12}|e|}{}^f V_{i_{13}i_{14}} V_{i_{15}i_{16}} V_{i_{17}i_{18}} V_{i_{19}i_{20}} V_{i_{21}i_{22}} V_{i_{23}i_{24}} V_{i_{25}i_{26}} V_{i_{27}i_{28}]} -$$

$$- 7{,}567{,}560\, R_{[i_1i_2|c|}{}^a R_{i_3i_4|a|}{}^b R_{i_5i_6|b|}{}^c R_{i_7i_8|e|}{}^d R_{i_9i_{10}|d|}{}^e R_{i_{11}i_{12}|g|}{}^f R_{i_{13}i_{14}|f|}{}^g R_{i_{15}i_{16}|i|}{}^h R_{i_{17}i_{18}|h|}{}^i R_{i_{19}i_{20}|k|}{}^j R_{i_{21}i_{22}|j|}{}^k R_{i_{23}i_{24}|m|}{}^l R_{i_{25}i_{26}|l|}{}^m V_{i_{27}i_{28}]} +$$

$$+ 12{,}612{,}600\, R_{[i_1i_2|c|}{}^a R_{i_3i_4|a|}{}^b R_{i_5i_6|b|}{}^c R_{i_7i_8|e|}{}^d R_{i_9i_{10}|d|}{}^e R_{i_{11}i_{12}|g|}{}^f R_{i_{13}i_{14}|f|}{}^g R_{i_{15}i_{16}|i|}{}^h R_{i_{17}i_{18}|h|}{}^i R_{i_{19}i_{20}|k|}{}^j R_{i_{21}i_{22}|j|}{}^k V_{i_{23}i_{24}} V_{i_{25}i_{26}} V_{i_{27}i_{28}]} -$$

$$- 5{,}045{,}040\, R_{[i_1i_2|c|}{}^a R_{i_3i_4|a|}{}^b R_{i_5i_6|b|}{}^c R_{i_7i_8|e|}{}^d R_{i_9i_{10}|d|}{}^e R_{i_{11}i_{12}|g|}{}^f R_{i_{13}i_{14}|f|}{}^g R_{i_{15}i_{16}|i|}{}^h R_{i_{17}i_{18}|h|}{}^i V_{i_{19}i_{20}} V_{i_{21}i_{22}} V_{i_{23}i_{24}} V_{i_{25}i_{26}} V_{i_{27}i_{28}]} +$$

$$+ 720{,}720\, R_{[i_1i_2|c|}{}^a R_{i_3i_4|a|}{}^b R_{i_5i_6|b|}{}^c R_{i_7i_8|e|}{}^d R_{i_9i_{10}|d|}{}^e R_{i_{11}i_{12}|g|}{}^f R_{i_{13}i_{14}|f|}{}^g V_{i_{15}i_{16}} V_{i_{17}i_{18}} V_{i_{19}i_{20}} V_{i_{21}i_{22}} V_{i_{23}i_{24}} V_{i_{25}i_{26}} V_{i_{27}i_{28}]} -$$

$$- 40{,}040\, R_{[i_1i_2|c|}{}^a R_{i_3i_4|a|}{}^b R_{i_5i_6|b|}{}^c R_{i_7i_8|e|}{}^d R_{i_9i_{10}|d|}{}^e V_{i_{11}i_{12}} V_{i_{13}i_{14}} V_{i_{15}i_{16}} V_{i_{17}i_{18}} V_{i_{19}i_{20}} V_{i_{21}i_{22}} V_{i_{23}i_{24}} V_{i_{25}i_{26}} V_{i_{27}i_{28}]} +$$

$$+ 728\, R_{[i_1i_2|c|}{}^a R_{i_3i_4|a|}{}^b R_{i_5i_6|b|}{}^c V_{i_7i_8} V_{i_9i_{10}} V_{i_{11}i_{12}} V_{i_{13}i_{14}} V_{i_{15}i_{16}} V_{i_{17}i_{18}} V_{i_{19}i_{20}} V_{i_{21}i_{22}} V_{i_{23}i_{24}} V_{i_{25}i_{26}} V_{i_{27}i_{28}]} -$$

$$- 135{,}135\, R_{[i_1i_2|b|}{}^a R_{i_3i_4|a|}{}^b R_{i_5i_6|d|}{}^c R_{i_7i_8|c|}{}^d R_{i_9i_{10}|f|}{}^e R_{i_{11}i_{12}|e|}{}^f R_{i_{13}i_{14}|h|}{}^g R_{i_{15}i_{16}|g|}{}^h R_{i_{17}i_{18}|j|}{}^i R_{i_{19}i_{20}|i|}{}^j R_{i_{21}i_{22}|l|}{}^k R_{i_{23}i_{24}|k|}{}^l R_{i_{25}i_{26}|n|}{}^m R_{i_{27}i_{28}]m}{}^n +$$

$$+ 945{,}945\, R_{[i_1i_2|b|}{}^a R_{i_3i_4|a|}{}^b R_{i_5i_6|d|}{}^c R_{i_7i_8|c|}{}^d R_{i_9i_{10}|f|}{}^e R_{i_{11}i_{12}|e|}{}^f R_{i_{13}i_{14}|h|}{}^g R_{i_{15}i_{16}|g|}{}^h R_{i_{17}i_{18}|j|}{}^i R_{i_{19}i_{20}|i|}{}^j R_{i_{21}i_{22}|l|}{}^k R_{i_{23}i_{24}|k|}{}^l V_{i_{25}i_{26}} V_{i_{27}i_{28}]} -$$

$$- 945{,}945\, R_{[i_1i_2|b|}{}^a R_{i_3i_4|a|}{}^b R_{i_5i_6|d|}{}^c R_{i_7i_8|c|}{}^d R_{i_9i_{10}|f|}{}^e R_{i_{11}i_{12}|e|}{}^f R_{i_{13}i_{14}|h|}{}^g R_{i_{15}i_{16}|g|}{}^h R_{i_{17}i_{18}|j|}{}^i R_{i_{19}i_{20}|i|}{}^j V_{i_{21}i_{22}} V_{i_{23}i_{24}} V_{i_{25}i_{26}} V_{i_{27}i_{28}]} +$$

$$+ 315{,}315\, R_{[i_1i_2|b|}{}^a R_{i_3i_4|a|}{}^b R_{i_5i_6|d|}{}^c R_{i_7i_8|c|}{}^d R_{i_9i_{10}|f|}{}^e R_{i_{11}i_{12}|e|}{}^f R_{i_{13}i_{14}|h|}{}^g R_{i_{15}i_{16}|g|}{}^h V_{i_{17}i_{18}} V_{i_{19}i_{20}} V_{i_{21}i_{22}} V_{i_{23}i_{24}} V_{i_{25}i_{26}} V_{i_{27}i_{28}]} -$$

$$- 45{,}045\, R_{[i_1i_2|b|}{}^a R_{i_3i_4|a|}{}^b R_{i_5i_6|d|}{}^c R_{i_7i_8|c|}{}^d R_{i_9i_{10}|f|}{}^e R_{i_{11}i_{12}|e|}{}^f V_{i_{13}i_{14}} V_{i_{15}i_{16}} V_{i_{17}i_{18}} V_{i_{19}i_{20}} V_{i_{21}i_{22}} V_{i_{23}i_{24}} V_{i_{25}i_{26}} V_{i_{27}i_{28}]} +$$

$$+ 3{,}003\, R_{[i_1i_2|b|}{}^a R_{i_3i_4|a|}{}^b R_{i_5i_6|d|}{}^c R_{i_7i_8|c|}{}^d V_{i_9i_{10}} V_{i_{11}i_{12}} V_{i_{13}i_{14}} V_{i_{15}i_{16}} V_{i_{17}i_{18}} V_{i_{19}i_{20}} V_{i_{21}i_{22}} V_{i_{23}i_{24}} V_{i_{25}i_{26}} V_{i_{27}i_{28}]} -$$

$$- 91\, R_{[i_1i_2|b|}{}^a R_{i_3i_4|a|}{}^b V_{i_5i_6} V_{i_7i_8} V_{i_9i_{10}} V_{i_{11}i_{12}} V_{i_{13}i_{14}} V_{i_{15}i_{16}} V_{i_{17}i_{18}} V_{i_{19}i_{20}} V_{i_{21}i_{22}} V_{i_{23}i_{24}} V_{i_{25}i_{26}} V_{i_{27}i_{28}]} +$$

$$+ V_{[i_1i_2} V_{i_3i_4} V_{i_5i_6} V_{i_7i_8} V_{i_9i_{10}} V_{i_{11}i_{12}} V_{i_{13}i_{14}} V_{i_{15}i_{16}} V_{i_{17}i_{18}} V_{i_{19}i_{20}} V_{i_{21}i_{22}} V_{i_{23}i_{24}} V_{i_{25}i_{26}} V_{i_{27}i_{28}]})$$

$$= \frac{i^{14}}{2^{28}\pi^{14} 14!}\Big(-6{,}227{,}020{,}800\, P^{(28)}{}_{i_1i_2i_3i_4i_5i_6i_7i_8i_9i_{10}i_{11}i_{12}i_{13}i_{14}i_{15}i_{16}i_{17}i_{18}i_{19}i_{20}i_{21}i_{22}i_{23}i_{24}i_{25}i_{26}i_{27}i_{28}} +$$

$$+ 6{,}706{,}022{,}400\, P^{(26)}{}_{[i_1i_2i_3i_4i_5i_6i_7i_8i_9i_{10}i_{11}i_{12}i_{13}i_{14}i_{15}i_{16}i_{17}i_{18}i_{19}i_{20}i_{21}i_{22}i_{23}i_{24}i_{25}i_{26}}\, P^{(2)}{}_{i_{27}i_{28}]} +$$

$$+ 3{,}632{,}428{,}800\, P^{(24)}{}_{[i_1i_2i_3i_4i_5i_6i_7i_8i_9i_{10}i_{11}i_{12}i_{13}i_{14}i_{15}i_{16}i_{17}i_{18}i_{19}i_{20}i_{21}i_{22}i_{23}i_{24}}\, P^{(4)}{}_{i_{25}i_{26}i_{27}i_{28}]} -$$

$$- 3{,}632{,}428{,}800\, P^{(24)}{}_{[i_1i_2i_3i_4i_5i_6i_7i_8i_9i_{10}i_{11}i_{12}i_{13}i_{14}i_{15}i_{16}i_{17}i_{18}i_{19}i_{20}i_{21}i_{22}i_{23}i_{24}}\, P^{(2)}{}_{i_{25}i_{26}}\, P^{(2)}{}_{i_{27}i_{28}]} +$$

$$+ 2{,}641{,}766{,}400\, P^{(22)}{}_{[i_1i_2i_3i_4i_5i_6i_7i_8i_9i_{10}i_{11}i_{12}i_{13}i_{14}i_{15}i_{16}i_{17}i_{18}i_{19}i_{20}i_{21}i_{22}}\, P^{(6)}{}_{i_{23}i_{24}i_{25}i_{26}i_{27}i_{28}]} -$$

$$- 3{,}962{,}649{,}600\, P^{(22)}{}_{[i_1i_2i_3i_4i_5i_6i_7i_8i_9i_{10}i_{11}i_{12}i_{13}i_{14}i_{15}i_{16}i_{17}i_{18}i_{19}i_{20}i_{21}i_{22}}\, P^{(4)}{}_{i_{23}i_{24}i_{25}i_{26}}\, P^{(2)}{}_{i_{27}i_{28}]} +$$

$$+ 1{,}320{,}883{,}200\, P^{(22)}{}_{[i_1i_2i_3i_4i_5i_6i_7i_8i_9i_{10}i_{11}i_{12}i_{13}i_{14}i_{15}i_{16}i_{17}i_{18}i_{19}i_{20}i_{21}i_{22}}\, P^{(2)}{}_{i_{23}i_{24}}\, P^{(2)}{}_{i_{25}i_{26}}\, P^{(2)}{}_{i_{27}i_{28}]} +$$

$$+ 2{,}179{,}457{,}280\, P^{(20)}{}_{[i_1i_2i_3i_4i_5i_6i_7i_8i_9i_{10}i_{11}i_{12}i_{13}i_{14}i_{15}i_{16}i_{17}i_{18}i_{19}i_{20}}\, P^{(8)}{}_{i_{21}i_{22}i_{23}i_{24}i_{25}i_{26}i_{27}i_{28}]} -$$

$$- 2{,}905{,}943{,}040\, P^{(20)}{}_{[i_1i_2i_3i_4i_5i_6i_7i_8i_9i_{10}i_{11}i_{12}i_{13}i_{14}i_{15}i_{16}i_{17}i_{18}i_{19}i_{20}}\, P^{(6)}{}_{i_{21}i_{22}i_{23}i_{24}i_{25}i_{26}}\, P^{(2)}{}_{i_{27}i_{28}]} -$$

$$- 1{,}089{,}728{,}640\, P^{(20)}{}_{[i_1i_2i_3i_4i_5i_6i_7i_8i_9i_{10}i_{11}i_{12}i_{13}i_{14}i_{15}i_{16}i_{17}i_{18}i_{19}i_{20}}\, P^{(4)}{}_{i_{21}i_{22}i_{23}i_{24}}\, P^{(4)}{}_{i_{25}i_{26}i_{27}i_{28}]} +$$

$$+ 2{,}179{,}457{,}280\, P^{(20)}{}_{[i_1i_2i_3i_4i_5i_6i_7i_8i_9i_{10}i_{11}i_{12}i_{13}i_{14}i_{15}i_{16}i_{17}i_{18}i_{19}i_{20}}\, P^{(4)}{}_{i_{21}i_{22}i_{23}i_{24}}\, P^{(2)}{}_{i_{25}i_{26}}\, P^{(2)}{}_{i_{27}i_{28}]} -$$

$$- 363{,}242{,}880\, P^{(20)}{}_{[i_1i_2i_3i_4i_5i_6i_7i_8i_9i_{10}i_{11}i_{12}i_{13}i_{14}i_{15}i_{16}i_{17}i_{18}i_{19}i_{20}}\, P^{(2)}{}_{i_{21}i_{22}}\, P^{(2)}{}_{i_{23}i_{24}}\, P^{(2)}{}_{i_{25}i_{26}}\, P^{(2)}{}_{i_{27}i_{28}]} +$$

$$+ 1{,}937{,}295{,}360\, P^{(18)}{}_{[i_1i_2i_3i_4i_5i_6i_7i_8i_9i_{10}i_{11}i_{12}i_{13}i_{14}i_{15}i_{16}i_{17}i_{18}}\, P^{(10)}{}_{i_{19}i_{20}i_{21}i_{22}i_{23}i_{24}i_{25}i_{26}i_{27}i_{28}]} -$$

$$- 2{,}421{,}619{,}200\, P^{(18)}{}_{[i_1i_2i_3i_4i_5i_6i_7i_8i_9i_{10}i_{11}i_{12}i_{13}i_{14}i_{15}i_{16}i_{17}i_{18}}\, P^{(8)}{}_{i_{19}i_{20}i_{21}i_{22}i_{23}i_{24}i_{25}i_{26}}\, P^{(2)}{}_{i_{27}i_{28}]} -$$

$$- 1{,}614{,}412{,}800\, P^{(18)}{}_{[i_1i_2i_3i_4i_5i_6i_7i_8i_9i_{10}i_{11}i_{12}i_{13}i_{14}i_{15}i_{16}i_{17}i_{18}}\, P^{(6)}{}_{i_{19}i_{20}i_{21}i_{22}i_{23}i_{24}}\, P^{(4)}{}_{i_{25}i_{26}i_{27}i_{28}]} +$$

$$+ 1{,}614{,}412{,}800\, P^{(18)}{}_{[i_1i_2i_3i_4i_5i_6i_7i_8i_9i_{10}i_{11}i_{12}i_{13}i_{14}i_{15}i_{16}i_{17}i_{18}}\, P^{(6)}{}_{i_{19}i_{20}i_{21}i_{22}i_{23}i_{24}}\, P^{(2)}{}_{i_{25}i_{26}}\, P^{(2)}{}_{i_{27}i_{28}]} +$$

$$+ 1{,}210{,}809{,}600\, P^{(18)}{}_{[i_1i_2i_3i_4i_5i_6i_7i_8i_9i_{10}i_{11}i_{12}i_{13}i_{14}i_{15}i_{16}i_{17}i_{18}}\, P^{(4)}{}_{i_{19}i_{20}i_{21}i_{22}}\, P^{(4)}{}_{i_{23}i_{24}i_{25}i_{26}}\, P^{(2)}{}_{i_{27}i_{28}]} -$$

$$- 807{,}206{,}400\, P^{(18)}{}_{[i_1i_2i_3i_4i_5i_6i_7i_8i_9i_{10}i_{11}i_{12}i_{13}i_{14}i_{15}i_{16}i_{17}i_{18}}\, P^{(4)}{}_{i_{19}i_{20}i_{21}i_{22}}\, P^{(2)}{}_{i_{23}i_{24}}\, P^{(2)}{}_{i_{25}i_{26}}\, P^{(2)}{}_{i_{27}i_{28}]} +$$

$$+ 80{,}720{,}640\, P^{(18)}{}_{[i_1i_2i_3i_4i_5i_6i_7i_8i_9i_{10}i_{11}i_{12}i_{13}i_{14}i_{15}i_{16}i_{17}i_{18}}\, P^{(2)}{}_{i_{19}i_{20}}\, P^{(2)}{}_{i_{21}i_{22}}\, P^{(2)}{}_{i_{23}i_{24}}\, P^{(2)}{}_{i_{25}i_{26}}\, P^{(2)}{}_{i_{27}i_{28}]} +$$

$$+ 1{,}816{,}214{,}400\, P^{(16)}{}_{[i_1i_2i_3i_4i_5i_6i_7i_8i_9i_{10}i_{11}i_{12}i_{13}i_{14}i_{15}i_{16}}\, P^{(12)}{}_{i_{17}i_{18}i_{19}i_{20}i_{21}i_{22}i_{23}i_{24}i_{25}i_{26}i_{27}i_{28}]} -$$

$$- 2{,}179{,}457{,}280\, P^{(16)}{}_{[i_1i_2i_3i_4i_5i_6i_7i_8i_9i_{10}i_{11}i_{12}i_{13}i_{14}i_{15}i_{16}}\, P^{(10)}{}_{i_{17}i_{18}i_{19}i_{20}i_{21}i_{22}i_{23}i_{24}i_{25}i_{26}}\, P^{(2)}{}_{i_{27}i_{28}]} -$$

$$- 1{,}362{,}160{,}800\, P^{(16)}{}_{[i_1i_2i_3i_4i_5i_6i_7i_8i_9i_{10}i_{11}i_{12}i_{13}i_{14}i_{15}i_{16}}\, P^{(8)}{}_{i_{17}i_{18}i_{19}i_{20}i_{21}i_{22}i_{23}i_{24}}\, P^{(4)}{}_{i_{25}i_{26}i_{27}i_{28}]} +$$

$$+ 1{,}362{,}160{,}800\, P^{(16)}{}_{[i_1i_2i_3i_4i_5i_6i_7i_8i_9i_{10}i_{11}i_{12}i_{13}i_{14}i_{15}i_{16}}\, P^{(8)}{}_{i_{17}i_{18}i_{19}i_{20}i_{21}i_{22}i_{23}i_{24}}\, P^{(2)}{}_{i_{25}i_{26}}\, P^{(2)}{}_{i_{27}i_{28}]} -$$

$- 605{,}404{,}800 \; P^{(16)}_{[i_1i_2i_3i_4i_5i_6i_7i_8i_9i_{10}i_{11}i_{12}i_{13}i_{14}i_{15}i_{16}} \, P^{(6)}_{i_{17}i_{18}i_{19}i_{20}i_{21}i_{22}} \, P^{(6)}_{i_{23}i_{24}i_{25}i_{26}i_{27}i_{28}]} +$

$+ 1{,}816{,}214{,}400 \; P^{(16)}_{[i_1i_2i_3i_4i_5i_6i_7i_8i_9i_{10}i_{11}i_{12}i_{13}i_{14}i_{15}i_{16}} \, P^{(6)}_{i_{17}i_{18}i_{19}i_{20}i_{21}i_{22}} \, P^{(4)}_{i_{23}i_{24}i_{25}i_{26}} \, P^{(2)}_{i_{27}i_{28}]} -$

$- 605{,}404{,}800 \; P^{(16)}_{[i_1i_2i_3i_4i_5i_6i_7i_8i_9i_{10}i_{11}i_{12}i_{13}i_{14}i_{15}i_{16}} \, P^{(6)}_{i_{17}i_{18}i_{19}i_{20}i_{21}i_{22}} \, P^{(2)}_{i_{23}i_{24}} \, P^{(2)}_{i_{25}i_{26}} \, P^{(2)}_{i_{27}i_{28}]} +$

$+ 227{,}026{,}800 \; P^{(16)}_{[i_1i_2i_3i_4i_5i_6i_7i_8i_9i_{10}i_{11}i_{12}i_{13}i_{14}i_{15}i_{16}} \, P^{(4)}_{i_{17}i_{18}i_{19}i_{20}} \, P^{(4)}_{i_{21}i_{22}i_{23}i_{24}} \, P^{(4)}_{i_{25}i_{26}i_{27}i_{28}]} -$

$- 681{,}080{,}400 \; P^{(16)}_{[i_1i_2i_3i_4i_5i_6i_7i_8i_9i_{10}i_{11}i_{12}i_{13}i_{14}i_{15}i_{16}} \, P^{(4)}_{i_{17}i_{18}i_{19}i_{20}} \, P^{(4)}_{i_{21}i_{22}i_{23}i_{24}} \, P^{(2)}_{i_{25}i_{26}} \, P^{(2)}_{i_{27}i_{28}]} +$

$+ 227{,}026{,}800 \; P^{(16)}_{[i_1i_2i_3i_4i_5i_6i_7i_8i_9i_{10}i_{11}i_{12}i_{13}i_{14}i_{15}i_{16}} \, P^{(4)}_{i_{17}i_{18}i_{19}i_{20}} \, P^{(2)}_{i_{21}i_{22}} \, P^{(2)}_{i_{23}i_{24}} \, P^{(2)}_{i_{25}i_{26}} \, P^{(2)}_{i_{27}i_{28}]} -$

$- 15{,}135{,}120 \; P^{(16)}_{[i_1i_2i_3i_4i_5i_6i_7i_8i_9i_{10}i_{11}i_{12}i_{13}i_{14}i_{15}i_{16}} \, P^{(2)}_{i_{17}i_{18}} \, P^{(2)}_{i_{19}i_{20}} \, P^{(2)}_{i_{21}i_{22}} \, P^{(2)}_{i_{23}i_{24}} \, P^{(2)}_{i_{25}i_{26}} \, P^{(2)}_{i_{27}i_{28}]} +$

$+ 889{,}574{,}400 \; P^{(14)}_{[i_1i_2i_3i_4i_5i_6i_7i_8i_9i_{10}i_{11}i_{12}i_{13}i_{14}} \, P^{(14)}_{i_{15}i_{16}i_{17}i_{18}i_{19}i_{20}i_{21}i_{22}i_{23}i_{24}i_{25}i_{26}i_{27}i_{28}]} -$

$- 2{,}075{,}673{,}600 \; P^{(14)}_{[i_1i_2i_3i_4i_5i_6i_7i_8i_9i_{10}i_{11}i_{12}i_{13}i_{14}} \, P^{(12)}_{i_{15}i_{16}i_{17}i_{18}i_{19}i_{20}i_{21}i_{22}i_{23}i_{24}i_{25}i_{26}} \, P^{(2)}_{i_{27}i_{28}]} -$

$- 1{,}245{,}404{,}160 \; P^{(14)}_{[i_1i_2i_3i_4i_5i_6i_7i_8i_9i_{10}i_{11}i_{12}i_{13}i_{14}} \, P^{(10)}_{i_{15}i_{16}i_{17}i_{18}i_{19}i_{20}i_{21}i_{22}i_{23}i_{24}} \, P^{(4)}_{i_{25}i_{26}i_{27}i_{28}]} +$

$+ 1{,}245{,}404{,}160 \; P^{(14)}_{[i_1i_2i_3i_4i_5i_6i_7i_8i_9i_{10}i_{11}i_{12}i_{13}i_{14}} \, P^{(10)}_{i_{15}i_{16}i_{17}i_{18}i_{19}i_{20}i_{21}i_{22}i_{23}i_{24}} \, P^{(2)}_{i_{25}i_{26}} \, P^{(2)}_{i_{27}i_{28}]} -$

$- 1{,}037{,}836{,}800 \; P^{(14)}_{[i_1i_2i_3i_4i_5i_6i_7i_8i_9i_{10}i_{11}i_{12}i_{13}i_{14}} \, P^{(8)}_{i_{15}i_{16}i_{17}i_{18}i_{19}i_{20}i_{21}i_{22}} \, P^{(6)}_{i_{23}i_{24}i_{25}i_{26}i_{27}i_{28}]} +$

$+ 1{,}556{,}755{,}200 \; P^{(14)}_{[i_1i_2i_3i_4i_5i_6i_7i_8i_9i_{10}i_{11}i_{12}i_{13}i_{14}} \, P^{(8)}_{i_{15}i_{16}i_{17}i_{18}i_{19}i_{20}i_{21}i_{22}} \, P^{(4)}_{i_{23}i_{24}i_{25}i_{26}} \, P^{(2)}_{i_{27}i_{28}]} -$

$- 518{,}918{,}400 \; P^{(14)}_{[i_1i_2i_3i_4i_5i_6i_7i_8i_9i_{10}i_{11}i_{12}i_{13}i_{14}} \, P^{(8)}_{i_{15}i_{16}i_{17}i_{18}i_{19}i_{20}i_{21}i_{22}} \, P^{(2)}_{i_{23}i_{24}} \, P^{(2)}_{i_{25}i_{26}} \, P^{(2)}_{i_{27}i_{28}]} +$

$+ 691{,}891{,}200 \; P^{(14)}_{[i_1i_2i_3i_4i_5i_6i_7i_8i_9i_{10}i_{11}i_{12}i_{13}i_{14}} \, P^{(6)}_{i_{15}i_{16}i_{17}i_{18}i_{19}i_{20}} \, P^{(6)}_{i_{21}i_{22}i_{23}i_{24}i_{25}i_{26}} \, P^{(2)}_{i_{27}i_{28}]} +$

$+ 518{,}918{,}400 \; P^{(14)}_{[i_1i_2i_3i_4i_5i_6i_7i_8i_9i_{10}i_{11}i_{12}i_{13}i_{14}} \, P^{(6)}_{i_{15}i_{16}i_{17}i_{18}i_{19}i_{20}} \, P^{(4)}_{i_{21}i_{22}i_{23}i_{24}} \, P^{(4)}_{i_{25}i_{26}i_{27}i_{28}]} -$

$- 1{,}037{,}836{,}800 \; P^{(14)}_{[i_1i_2i_3i_4i_5i_6i_7i_8i_9i_{10}i_{11}i_{12}i_{13}i_{14}} \, P^{(6)}_{i_{15}i_{16}i_{17}i_{18}i_{19}i_{20}} \, P^{(4)}_{i_{21}i_{22}i_{23}i_{24}} \, P^{(2)}_{i_{25}i_{26}} \, P^{(2)}_{i_{27}i_{28}]} +$

$+ 172{,}972{,}800 \; P^{(14)}_{[i_1i_2i_3i_4i_5i_6i_7i_8i_9i_{10}i_{11}i_{12}i_{13}i_{14}} \, P^{(6)}_{i_{15}i_{16}i_{17}i_{18}i_{19}i_{20}} \, P^{(2)}_{i_{21}i_{22}} \, P^{(2)}_{i_{23}i_{24}} \, P^{(2)}_{i_{25}i_{26}} \, P^{(2)}_{i_{27}i_{28}]} -$

$- 259{,}459{,}200 \; P^{(14)}_{[i_1i_2i_3i_4i_5i_6i_7i_8i_9i_{10}i_{11}i_{12}i_{13}i_{14}} \, P^{(4)}_{i_{15}i_{16}i_{17}i_{18}} \, P^{(4)}_{i_{19}i_{20}i_{21}i_{22}} \, P^{(4)}_{i_{23}i_{24}i_{25}i_{26}} \, P^{(2)}_{i_{27}i_{28}]} +$

$+ 259{,}459{,}200 \; P^{(14)}_{[i_1i_2i_3i_4i_5i_6i_7i_8i_9i_{10}i_{11}i_{12}i_{13}i_{14}} \, P^{(4)}_{i_{15}i_{16}i_{17}i_{18}} \, P^{(4)}_{i_{19}i_{20}i_{21}i_{22}} \, P^{(2)}_{i_{23}i_{24}} \, P^{(2)}_{i_{25}i_{26}} \, P^{(2)}_{i_{27}i_{28}]} -$

$- 51{,}891{,}840 \; P^{(14)}_{[i_1i_2i_3i_4i_5i_6i_7i_8i_9i_{10}i_{11}i_{12}i_{13}i_{14}} \, P^{(4)}_{i_{15}i_{16}i_{17}i_{18}} \, P^{(2)}_{i_{19}i_{20}} \, P^{(2)}_{i_{21}i_{22}} \, P^{(2)}_{i_{23}i_{24}} \, P^{(2)}_{i_{25}i_{26}} \, P^{(2)}_{i_{27}i_{28}]} +$

$+ 2{,}471{,}040 \; P^{(14)}_{[i_1i_2i_3i_4i_5i_6i_7i_8i_9i_{10}i_{11}i_{12}i_{13}i_{14}} \, P^{(2)}_{i_{15}i_{16}} \, P^{(2)}_{i_{17}i_{18}} \, P^{(2)}_{i_{19}i_{20}} \, P^{(2)}_{i_{21}i_{22}} \, P^{(2)}_{i_{23}i_{24}} \, P^{(2)}_{i_{25}i_{26}} \, P^{(2)}_{i_{27}i_{28}]} -$

$- 605{,}404{,}800 \; P^{(12)}_{[i_1i_2i_3i_4i_5i_6i_7i_8i_9i_{10}i_{11}i_{12}} \, P^{(12)}_{i_{13}i_{14}i_{15}i_{16}i_{17}i_{18}i_{19}i_{20}i_{21}i_{22}i_{23}i_{24}} \, P^{(4)}_{i_{25}i_{26}i_{27}i_{28}]} +$

$+ 605{,}404{,}800 \; P^{(12)}_{[i_1i_2i_3i_4i_5i_6i_7i_8i_9i_{10}i_{11}i_{12}} \, P^{(12)}_{i_{13}i_{14}i_{15}i_{16}i_{17}i_{18}i_{19}i_{20}i_{21}i_{22}i_{23}i_{24}} \, P^{(2)}_{i_{25}i_{26}} \, P^{(2)}_{i_{27}i_{28}]} -$

$- 968{,}647{,}680 \; P^{(12)}_{[i_1i_2i_3i_4i_5i_6i_7i_8i_9i_{10}i_{11}i_{12}} \, P^{(10)}_{i_{13}i_{14}i_{15}i_{16}i_{17}i_{18}i_{19}i_{20}i_{21}i_{22}} \, P^{(6)}_{i_{23}i_{24}i_{25}i_{26}i_{27}i_{28}]} +$

$+ 1{,}452{,}971{,}520 \; P^{(12)}_{[i_1i_2i_3i_4i_5i_6i_7i_8i_9i_{10}i_{11}i_{12}} \, P^{(10)}_{i_{13}i_{14}i_{15}i_{16}i_{17}i_{18}i_{19}i_{20}i_{21}i_{22}} \, P^{(4)}_{i_{23}i_{24}i_{25}i_{26}} \, P^{(2)}_{i_{27}i_{28}]} -$

$- 484{,}323{,}840 \; P^{(12)}_{[i_1i_2i_3i_4i_5i_6i_7i_8i_9i_{10}i_{11}i_{12}} \, P^{(10)}_{i_{13}i_{14}i_{15}i_{16}i_{17}i_{18}i_{19}i_{20}i_{21}i_{22}} \, P^{(2)}_{i_{23}i_{24}} \, P^{(2)}_{i_{25}i_{26}} \, P^{(2)}_{i_{27}i_{28}]} -$

$- 454{,}053{,}600 \; P^{(12)}_{[i_1i_2i_3i_4i_5i_6i_7i_8i_9i_{10}i_{11}i_{12}} \, P^{(8)}_{i_{13}i_{14}i_{15}i_{16}i_{17}i_{18}i_{19}i_{20}} \, P^{(8)}_{i_{21}i_{22}i_{23}i_{24}i_{25}i_{26}i_{27}i_{28}]} +$

$+ 1{,}210{,}809{,}600 \; P^{(12)}_{[i_1i_2i_3i_4i_5i_6i_7i_8i_9i_{10}i_{11}i_{12}} \, P^{(8)}_{i_{13}i_{14}i_{15}i_{16}i_{17}i_{18}i_{19}i_{20}} \, P^{(6)}_{i_{21}i_{22}i_{23}i_{24}i_{25}i_{26}} \, P^{(2)}_{i_{27}i_{28}]} +$

$+ 454{,}053{,}600 \; P^{(12)}_{[i_1i_2i_3i_4i_5i_6i_7i_8i_9i_{10}i_{11}i_{12}} \, P^{(8)}_{i_{13}i_{14}i_{15}i_{16}i_{17}i_{18}i_{19}i_{20}} \, P^{(4)}_{i_{21}i_{22}i_{23}i_{24}} \, P^{(4)}_{i_{25}i_{26}i_{27}i_{28}]} -$

$- 908{,}107{,}200 \; P^{(12)}_{[i_1i_2i_3i_4i_5i_6i_7i_8i_9i_{10}i_{11}i_{12}} \, P^{(8)}_{i_{13}i_{14}i_{15}i_{16}i_{17}i_{18}i_{19}i_{20}} \, P^{(4)}_{i_{21}i_{22}i_{23}i_{24}} \, P^{(2)}_{i_{25}i_{26}} \, P^{(2)}_{i_{27}i_{28}]} +$

$+ 151{,}351{,}200 \; P^{(12)}_{[i_1i_2i_3i_4i_5i_6i_7i_8i_9i_{10}i_{11}i_{12}} \, P^{(8)}_{i_{13}i_{14}i_{15}i_{16}i_{17}i_{18}i_{19}i_{20}} \, P^{(2)}_{i_{21}i_{22}} \, P^{(2)}_{i_{23}i_{24}} \, P^{(2)}_{i_{25}i_{26}} \, P^{(2)}_{i_{27}i_{28}]} +$

$+ 403{,}603{,}200 \; P^{(12)}_{[i_1i_2i_3i_4i_5i_6i_7i_8i_9i_{10}i_{11}i_{12}} \, P^{(6)}_{i_{13}i_{14}i_{15}i_{16}i_{17}i_{18}} \, P^{(6)}_{i_{19}i_{20}i_{21}i_{22}i_{23}i_{24}} \, P^{(4)}_{i_{25}i_{26}i_{27}i_{28}]} -$

$- 403{,}603{,}200 \; P^{(12)}_{[i_1i_2i_3i_4i_5i_6i_7i_8i_9i_{10}i_{11}i_{12}} \, P^{(6)}_{i_{13}i_{14}i_{15}i_{16}i_{17}i_{18}} \, P^{(6)}_{i_{19}i_{20}i_{21}i_{22}i_{23}i_{24}} \, P^{(2)}_{i_{25}i_{26}} \, P^{(2)}_{i_{27}i_{28}]} -$

$- 605{,}404{,}800 \; P^{(12)}_{[i_1i_2i_3i_4i_5i_6i_7i_8i_9i_{10}i_{11}i_{12}} \, P^{(6)}_{i_{13}i_{14}i_{15}i_{16}i_{17}i_{18}} \, P^{(4)}_{i_{19}i_{20}i_{21}i_{22}} \, P^{(4)}_{i_{23}i_{24}i_{25}i_{26}} \, P^{(2)}_{i_{27}i_{28}]} +$

$+ 403{,}603{,}200 \; P^{(12)}_{[i_1i_2i_3i_4i_5i_6i_7i_8i_9i_{10}i_{11}i_{12}} \, P^{(6)}_{i_{13}i_{14}i_{15}i_{16}i_{17}i_{18}} \, P^{(4)}_{i_{19}i_{20}i_{21}i_{22}} \, P^{(2)}_{i_{23}i_{24}} \, P^{(2)}_{i_{25}i_{26}} \, P^{(2)}_{i_{27}i_{28}]} -$

$- 40{,}360{,}320 \; P^{(12)}_{[i_1i_2i_3i_4i_5i_6i_7i_8i_9i_{10}i_{11}i_{12}} \, P^{(6)}_{i_{13}i_{14}i_{15}i_{16}i_{17}i_{18}} \, P^{(2)}_{i_{19}i_{20}} \, P^{(2)}_{i_{21}i_{22}} \, P^{(2)}_{i_{23}i_{24}} \, P^{(2)}_{i_{25}i_{26}} \, P^{(2)}_{i_{27}i_{28}]} -$

$- 37{,}837{,}800 \; P^{(12)}_{[i_1i_2i_3i_4i_5i_6i_7i_8i_9i_{10}i_{11}i_{12}} \, P^{(4)}_{i_{13}i_{14}i_{15}i_{16}} \, P^{(4)}_{i_{17}i_{18}i_{19}i_{20}} \, P^{(4)}_{i_{21}i_{22}i_{23}i_{24}} \, P^{(4)}_{i_{25}i_{26}i_{27}i_{28}]} +$

$+ 151{,}351{,}200 \; P^{(12)}_{[i_1i_2i_3i_4i_5i_6i_7i_8i_9i_{10}i_{11}i_{12}} \, P^{(4)}_{i_{13}i_{14}i_{15}i_{16}} \, P^{(4)}_{i_{17}i_{18}i_{19}i_{20}} \, P^{(4)}_{i_{21}i_{22}i_{23}i_{24}} \, P^{(2)}_{i_{25}i_{26}} \, P^{(2)}_{i_{27}i_{28}]} -$

$- 75{,}675{,}600 \; P^{(12)}_{[i_1i_2i_3i_4i_5i_6i_7i_8i_9i_{10}i_{11}i_{12}} \, P^{(4)}_{i_{13}i_{14}i_{15}i_{16}} \, P^{(4)}_{i_{17}i_{18}i_{19}i_{20}} \, P^{(2)}_{i_{21}i_{22}} \, P^{(2)}_{i_{23}i_{24}} \, P^{(2)}_{i_{25}i_{26}} \, P^{(2)}_{i_{27}i_{28}]} +$

$+ 10{,}090{,}080 \; P^{(12)}_{[i_1i_2i_3i_4i_5i_6i_7i_8i_9i_{10}i_{11}i_{12}} \, P^{(4)}_{i_{13}i_{14}i_{15}i_{16}} \, P^{(2)}_{i_{17}i_{18}} \, P^{(2)}_{i_{19}i_{20}} \, P^{(2)}_{i_{21}i_{22}} \, P^{(2)}_{i_{23}i_{24}} \, P^{(2)}_{i_{25}i_{26}} \, P^{(2)}_{i_{27}i_{28}]} -$

$$- 360{,}360\, P^{(12)}{}_{[i_1i_2i_3i_4i_5i_6i_7i_8i_9i_{10}i_{11}i_{12}} P^{(2)}{}_{i_{13}i_{14}} P^{(2)}{}_{i_{15}i_{16}} P^{(2)}{}_{i_{17}i_{18}} P^{(2)}{}_{i_{19}i_{20}} P^{(2)}{}_{i_{21}i_{22}} P^{(2)}{}_{i_{23}i_{24}} P^{(2)}{}_{i_{25}i_{26}} P^{(2)}{}_{i_{27}i_{28}]} -$$

$$- 435{,}891{,}456\, P^{(10)}{}_{[i_1i_2i_3i_4i_5i_6i_7i_8i_9i_{10}} P^{(10)}{}_{i_{11}i_{12}i_{13}i_{14}i_{15}i_{16}i_{17}i_{18}i_{19}i_{20}} P^{(8)}{}_{i_{21}i_{22}i_{23}i_{24}i_{25}i_{26}i_{27}i_{28}]} +$$

$$+ 581{,}188{,}608\, P^{(10)}{}_{[i_1i_2i_3i_4i_5i_6i_7i_8i_9i_{10}} P^{(10)}{}_{i_{11}i_{12}i_{13}i_{14}i_{15}i_{16}i_{17}i_{18}i_{19}i_{20}} P^{(6)}{}_{i_{21}i_{22}i_{23}i_{24}i_{25}i_{26}} P^{(2)}{}_{i_{27}i_{28}]} +$$

$$+ 217{,}945{,}728\, P^{(10)}{}_{[i_1i_2i_3i_4i_5i_6i_7i_8i_9i_{10}} P^{(10)}{}_{i_{11}i_{12}i_{13}i_{14}i_{15}i_{16}i_{17}i_{18}i_{19}i_{20}} P^{(4)}{}_{i_{21}i_{22}i_{23}i_{24}} P^{(4)}{}_{i_{25}i_{26}i_{27}i_{28}]} -$$

$$- 435{,}891{,}456\, P^{(10)}{}_{[i_1i_2i_3i_4i_5i_6i_7i_8i_9i_{10}} P^{(10)}{}_{i_{11}i_{12}i_{13}i_{14}i_{15}i_{16}i_{17}i_{18}i_{19}i_{20}} P^{(4)}{}_{i_{21}i_{22}i_{23}i_{24}} P^{(2)}{}_{i_{25}i_{26}} P^{(2)}{}_{i_{27}i_{28}]} +$$

$$+ 72{,}648{,}576\, P^{(10)}{}_{[i_1i_2i_3i_4i_5i_6i_7i_8i_9i_{10}} P^{(10)}{}_{i_{11}i_{12}i_{13}i_{14}i_{15}i_{16}i_{17}i_{18}i_{19}i_{20}} P^{(2)}{}_{i_{21}i_{22}} P^{(2)}{}_{i_{23}i_{24}} P^{(2)}{}_{i_{25}i_{26}} P^{(2)}{}_{i_{27}i_{28}]} +$$

$$+ 544{,}864{,}320\, P^{(10)}{}_{[i_1i_2i_3i_4i_5i_6i_7i_8i_9i_{10}} P^{(8)}{}_{i_{11}i_{12}i_{13}i_{14}i_{15}i_{16}i_{17}i_{18}} P^{(8)}{}_{i_{19}i_{20}i_{21}i_{22}i_{23}i_{24}i_{25}i_{26}} P^{(2)}{}_{i_{27}i_{28}]} +$$

$$+ 726{,}485{,}760\, P^{(10)}{}_{[i_1i_2i_3i_4i_5i_6i_7i_8i_9i_{10}} P^{(8)}{}_{i_{11}i_{12}i_{13}i_{14}i_{15}i_{16}i_{17}i_{18}} P^{(6)}{}_{i_{19}i_{20}i_{21}i_{22}i_{23}i_{24}} P^{(4)}{}_{i_{25}i_{26}i_{27}i_{28}]} -$$

$$- 726{,}485{,}760\, P^{(10)}{}_{[i_1i_2i_3i_4i_5i_6i_7i_8i_9i_{10}} P^{(8)}{}_{i_{11}i_{12}i_{13}i_{14}i_{15}i_{16}i_{17}i_{18}} P^{(6)}{}_{i_{19}i_{20}i_{21}i_{22}i_{23}i_{24}} P^{(2)}{}_{i_{25}i_{26}} P^{(2)}{}_{i_{27}i_{28}]} -$$

$$- 544{,}864{,}320\, P^{(10)}{}_{[i_1i_2i_3i_4i_5i_6i_7i_8i_9i_{10}} P^{(8)}{}_{i_{11}i_{12}i_{13}i_{14}i_{15}i_{16}i_{17}i_{18}} P^{(4)}{}_{i_{19}i_{20}i_{21}i_{22}} P^{(4)}{}_{i_{23}i_{24}i_{25}i_{26}} P^{(2)}{}_{i_{27}i_{28}]} +$$

$$+ 363{,}242{,}880\, P^{(10)}{}_{[i_1i_2i_3i_4i_5i_6i_7i_8i_9i_{10}} P^{(8)}{}_{i_{11}i_{12}i_{13}i_{14}i_{15}i_{16}i_{17}i_{18}} P^{(4)}{}_{i_{19}i_{20}i_{21}i_{22}} P^{(2)}{}_{i_{23}i_{24}} P^{(2)}{}_{i_{25}i_{26}} P^{(2)}{}_{i_{27}i_{28}]} -$$

$$- 36{,}324{,}288\, P^{(10)}{}_{[i_1i_2i_3i_4i_5i_6i_7i_8i_9i_{10}} P^{(8)}{}_{i_{11}i_{12}i_{13}i_{14}i_{15}i_{16}i_{17}i_{18}} P^{(2)}{}_{i_{19}i_{20}} P^{(2)}{}_{i_{21}i_{22}} P^{(2)}{}_{i_{23}i_{24}} P^{(2)}{}_{i_{25}i_{26}} P^{(2)}{}_{i_{27}i_{28}]} +$$

$$+ 107{,}627{,}520\, P^{(10)}{}_{[i_1i_2i_3i_4i_5i_6i_7i_8i_9i_{10}} P^{(6)}{}_{i_{11}i_{12}i_{13}i_{14}i_{15}i_{16}} P^{(6)}{}_{i_{17}i_{18}i_{19}i_{20}i_{21}i_{22}} P^{(6)}{}_{i_{23}i_{24}i_{25}i_{26}i_{27}i_{28}]} -$$

$$- 484{,}323{,}840\, P^{(10)}{}_{[i_1i_2i_3i_4i_5i_6i_7i_8i_9i_{10}} P^{(6)}{}_{i_{11}i_{12}i_{13}i_{14}i_{15}i_{16}} P^{(6)}{}_{i_{17}i_{18}i_{19}i_{20}i_{21}i_{22}} P^{(4)}{}_{i_{23}i_{24}i_{25}i_{26}} P^{(2)}{}_{i_{27}i_{28}]} +$$

$$+ 161{,}441{,}280\, P^{(10)}{}_{[i_1i_2i_3i_4i_5i_6i_7i_8i_9i_{10}} P^{(6)}{}_{i_{11}i_{12}i_{13}i_{14}i_{15}i_{16}} P^{(6)}{}_{i_{17}i_{18}i_{19}i_{20}i_{21}i_{22}} P^{(2)}{}_{i_{23}i_{24}} P^{(2)}{}_{i_{25}i_{26}} P^{(2)}{}_{i_{27}i_{28}]} -$$

$$- 121{,}080{,}960\, P^{(10)}{}_{[i_1i_2i_3i_4i_5i_6i_7i_8i_9i_{10}} P^{(6)}{}_{i_{11}i_{12}i_{13}i_{14}i_{15}i_{16}} P^{(4)}{}_{i_{17}i_{18}i_{19}i_{20}} P^{(4)}{}_{i_{21}i_{22}i_{23}i_{24}} P^{(4)}{}_{i_{25}i_{26}i_{27}i_{28}]} +$$

$$+ 363{,}242{,}880\, P^{(10)}{}_{[i_1i_2i_3i_4i_5i_6i_7i_8i_9i_{10}} P^{(6)}{}_{i_{11}i_{12}i_{13}i_{14}i_{15}i_{16}} P^{(4)}{}_{i_{17}i_{18}i_{19}i_{20}} P^{(4)}{}_{i_{21}i_{22}i_{23}i_{24}} P^{(2)}{}_{i_{25}i_{26}} P^{(2)}{}_{i_{27}i_{28}]} -$$

$$- 121{,}080{,}960\, P^{(10)}{}_{[i_1i_2i_3i_4i_5i_6i_7i_8i_9i_{10}} P^{(6)}{}_{i_{11}i_{12}i_{13}i_{14}i_{15}i_{16}} P^{(4)}{}_{i_{17}i_{18}i_{19}i_{20}} P^{(2)}{}_{i_{21}i_{22}} P^{(2)}{}_{i_{23}i_{24}} P^{(2)}{}_{i_{25}i_{26}} P^{(2)}{}_{i_{27}i_{28}]} +$$

$$+ 8{,}072{,}064\, P^{(10)}{}_{[i_1i_2i_3i_4i_5i_6i_7i_8i_9i_{10}} P^{(6)}{}_{i_{11}i_{12}i_{13}i_{14}i_{15}i_{16}} P^{(2)}{}_{i_{17}i_{18}} P^{(2)}{}_{i_{19}i_{20}} P^{(2)}{}_{i_{21}i_{22}} P^{(2)}{}_{i_{23}i_{24}} P^{(2)}{}_{i_{25}i_{26}} P^{(2)}{}_{i_{27}i_{28}]} +$$

$$+ 45{,}405{,}360\, P^{(10)}{}_{[i_1i_2i_3i_4i_5i_6i_7i_8i_9i_{10}} P^{(4)}{}_{i_{11}i_{12}i_{13}i_{14}} P^{(4)}{}_{i_{15}i_{16}i_{17}i_{18}} P^{(4)}{}_{i_{19}i_{20}i_{21}i_{22}} P^{(4)}{}_{i_{23}i_{24}i_{25}i_{26}} P^{(2)}{}_{i_{27}i_{28}]} -$$

$$- 60{,}540{,}480\, P^{(10)}{}_{[i_1i_2i_3i_4i_5i_6i_7i_8i_9i_{10}} P^{(4)}{}_{i_{11}i_{12}i_{13}i_{14}} P^{(4)}{}_{i_{15}i_{16}i_{17}i_{18}} P^{(4)}{}_{i_{19}i_{20}i_{21}i_{22}} P^{(2)}{}_{i_{23}i_{24}} P^{(2)}{}_{i_{25}i_{26}} P^{(2)}{}_{i_{27}i_{28}]} +$$

$$+ 18{,}162{,}144\, P^{(10)}{}_{[i_1i_2i_3i_4i_5i_6i_7i_8i_9i_{10}} P^{(4)}{}_{i_{11}i_{12}i_{13}i_{14}} P^{(4)}{}_{i_{15}i_{16}i_{17}i_{18}} P^{(2)}{}_{i_{19}i_{20}} P^{(2)}{}_{i_{21}i_{22}} P^{(2)}{}_{i_{23}i_{24}} P^{(2)}{}_{i_{25}i_{26}} P^{(2)}{}_{i_{27}i_{28}]} -$$

$$- 1{,}729{,}728\, P^{(10)}{}_{[i_1i_2i_3i_4i_5i_6i_7i_8i_9i_{10}} P^{(4)}{}_{i_{11}i_{12}i_{13}i_{14}} P^{(2)}{}_{i_{15}i_{16}} P^{(2)}{}_{i_{17}i_{18}} P^{(2)}{}_{i_{19}i_{20}} P^{(2)}{}_{i_{21}i_{22}} P^{(2)}{}_{i_{23}i_{24}} P^{(2)}{}_{i_{25}i_{26}} P^{(2)}{}_{i_{27}i_{28}]} +$$

$$+ 48{,}048\, P^{(10)}{}_{[i_1i_2i_3i_4i_5i_6i_7i_8i_9i_{10}} P^{(2)}{}_{i_{11}i_{12}} P^{(2)}{}_{i_{13}i_{14}} P^{(2)}{}_{i_{15}i_{16}} P^{(2)}{}_{i_{17}i_{18}} P^{(2)}{}_{i_{19}i_{20}} P^{(2)}{}_{i_{21}i_{22}} P^{(2)}{}_{i_{23}i_{24}} P^{(2)}{}_{i_{25}i_{26}} P^{(2)}{}_{i_{27}i_{28}]} +$$

$$+ 113{,}513{,}400\, P^{(8)}{}_{[i_1i_2i_3i_4i_5i_6i_7i_8} P^{(8)}{}_{i_9i_{10}i_{11}i_{12}i_{13}i_{14}i_{15}i_{16}} P^{(8)}{}_{i_{17}i_{18}i_{19}i_{20}i_{21}i_{22}i_{23}i_{24}} P^{(4)}{}_{i_{25}i_{26}i_{27}i_{28}]} -$$

$$- 113{,}513{,}400\, P^{(8)}{}_{[i_1i_2i_3i_4i_5i_6i_7i_8} P^{(8)}{}_{i_9i_{10}i_{11}i_{12}i_{13}i_{14}i_{15}i_{16}} P^{(8)}{}_{i_{17}i_{18}i_{19}i_{20}i_{21}i_{22}i_{23}i_{24}} P^{(2)}{}_{i_{25}i_{26}} P^{(2)}{}_{i_{27}i_{28}]} +$$

$$+ 151{,}351{,}200\, P^{(8)}{}_{[i_1i_2i_3i_4i_5i_6i_7i_8} P^{(8)}{}_{i_9i_{10}i_{11}i_{12}i_{13}i_{14}i_{15}i_{16}} P^{(6)}{}_{i_{17}i_{18}i_{19}i_{20}i_{21}i_{22}} P^{(6)}{}_{i_{23}i_{24}i_{25}i_{26}i_{27}i_{28}]} -$$

$$- 454{,}053{,}600\, P^{(8)}{}_{[i_1i_2i_3i_4i_5i_6i_7i_8} P^{(8)}{}_{i_9i_{10}i_{11}i_{12}i_{13}i_{14}i_{15}i_{16}} P^{(6)}{}_{i_{17}i_{18}i_{19}i_{20}i_{21}i_{22}} P^{(4)}{}_{i_{23}i_{24}i_{25}i_{26}} P^{(2)}{}_{i_{27}i_{28}]} +$$

$$+ 151{,}351{,}200\, P^{(8)}{}_{[i_1i_2i_3i_4i_5i_6i_7i_8} P^{(8)}{}_{i_9i_{10}i_{11}i_{12}i_{13}i_{14}i_{15}i_{16}} P^{(6)}{}_{i_{17}i_{18}i_{19}i_{20}i_{21}i_{22}} P^{(2)}{}_{i_{23}i_{24}} P^{(2)}{}_{i_{25}i_{26}} P^{(2)}{}_{i_{27}i_{28}]} -$$

$$- 56{,}756{,}700\, P^{(8)}{}_{[i_1i_2i_3i_4i_5i_6i_7i_8} P^{(8)}{}_{i_9i_{10}i_{11}i_{12}i_{13}i_{14}i_{15}i_{16}} P^{(4)}{}_{i_{17}i_{18}i_{19}i_{20}} P^{(4)}{}_{i_{21}i_{22}i_{23}i_{24}} P^{(4)}{}_{i_{25}i_{26}i_{27}i_{28}]} +$$

$$+ 170{,}270{,}100\, P^{(8)}{}_{[i_1i_2i_3i_4i_5i_6i_7i_8} P^{(8)}{}_{i_9i_{10}i_{11}i_{12}i_{13}i_{14}i_{15}i_{16}} P^{(4)}{}_{i_{17}i_{18}i_{19}i_{20}} P^{(4)}{}_{i_{21}i_{22}i_{23}i_{24}} P^{(2)}{}_{i_{25}i_{26}} P^{(2)}{}_{i_{27}i_{28}]} -$$

$$- 56{,}756{,}700\, P^{(8)}{}_{[i_1i_2i_3i_4i_5i_6i_7i_8} P^{(8)}{}_{i_9i_{10}i_{11}i_{12}i_{13}i_{14}i_{15}i_{16}} P^{(4)}{}_{i_{17}i_{18}i_{19}i_{20}} P^{(2)}{}_{i_{21}i_{22}} P^{(2)}{}_{i_{23}i_{24}} P^{(2)}{}_{i_{25}i_{26}} P^{(2)}{}_{i_{27}i_{28}]} +$$

$$+ 3{,}783{,}780\, P^{(8)}{}_{[i_1i_2i_3i_4i_5i_6i_7i_8} P^{(8)}{}_{i_9i_{10}i_{11}i_{12}i_{13}i_{14}i_{15}i_{16}} P^{(2)}{}_{i_{17}i_{18}} P^{(2)}{}_{i_{19}i_{20}} P^{(2)}{}_{i_{21}i_{22}} P^{(2)}{}_{i_{23}i_{24}} P^{(2)}{}_{i_{25}i_{26}} P^{(2)}{}_{i_{27}i_{28}]} -$$

$$- 134{,}534{,}400\, P^{(8)}{}_{[i_1i_2i_3i_4i_5i_6i_7i_8} P^{(6)}{}_{i_9i_{10}i_{11}i_{12}i_{13}i_{14}} P^{(6)}{}_{i_{15}i_{16}i_{17}i_{18}i_{19}i_{20}} P^{(6)}{}_{i_{21}i_{22}i_{23}i_{24}i_{25}i_{26}} P^{(2)}{}_{i_{27}i_{28}]} -$$

$$- 151{,}351{,}200\, P^{(8)}{}_{[i_1i_2i_3i_4i_5i_6i_7i_8} P^{(6)}{}_{i_9i_{10}i_{11}i_{12}i_{13}i_{14}} P^{(6)}{}_{i_{15}i_{16}i_{17}i_{18}i_{19}i_{20}} P^{(4)}{}_{i_{21}i_{22}i_{23}i_{24}} P^{(4)}{}_{i_{25}i_{26}i_{27}i_{28}]} +$$

$$+ 302{,}702{,}400\, P^{(8)}{}_{[i_1i_2i_3i_4i_5i_6i_7i_8} P^{(6)}{}_{i_9i_{10}i_{11}i_{12}i_{13}i_{14}} P^{(6)}{}_{i_{15}i_{16}i_{17}i_{18}i_{19}i_{20}} P^{(4)}{}_{i_{21}i_{22}i_{23}i_{24}} P^{(2)}{}_{i_{25}i_{26}} P^{(2)}{}_{i_{27}i_{28}]} -$$

$$- 50{,}450{,}400\, P^{(8)}{}_{[i_1i_2i_3i_4i_5i_6i_7i_8} P^{(6)}{}_{i_9i_{10}i_{11}i_{12}i_{13}i_{14}} P^{(6)}{}_{i_{15}i_{16}i_{17}i_{18}i_{19}i_{20}} P^{(2)}{}_{i_{21}i_{22}} P^{(2)}{}_{i_{23}i_{24}} P^{(2)}{}_{i_{25}i_{26}} P^{(2)}{}_{i_{27}i_{28}]} +$$

$$+ 151{,}351{,}200\, P^{(8)}{}_{[i_1i_2i_3i_4i_5i_6i_7i_8} P^{(6)}{}_{i_9i_{10}i_{11}i_{12}i_{13}i_{14}} P^{(4)}{}_{i_{15}i_{16}i_{17}i_{18}} P^{(4)}{}_{i_{19}i_{20}i_{21}i_{22}} P^{(4)}{}_{i_{23}i_{24}i_{25}i_{26}} P^{(2)}{}_{i_{27}i_{28}]} -$$

$$- 151{,}351{,}200\, P^{(8)}{}_{[i_1i_2i_3i_4i_5i_6i_7i_8} P^{(6)}{}_{i_9i_{10}i_{11}i_{12}i_{13}i_{14}} P^{(4)}{}_{i_{15}i_{16}i_{17}i_{18}} P^{(4)}{}_{i_{19}i_{20}i_{21}i_{22}} P^{(2)}{}_{i_{23}i_{24}} P^{(2)}{}_{i_{25}i_{26}} P^{(2)}{}_{i_{27}i_{28}]} +$$

$$+ 30{,}270{,}240\, P^{(8)}{}_{[i_1i_2i_3i_4i_5i_6i_7i_8} P^{(6)}{}_{i_9i_{10}i_{11}i_{12}i_{13}i_{14}} P^{(4)}{}_{i_{15}i_{16}i_{17}i_{18}} P^{(2)}{}_{i_{19}i_{20}} P^{(2)}{}_{i_{21}i_{22}} P^{(2)}{}_{i_{23}i_{24}} P^{(2)}{}_{i_{25}i_{26}} P^{(2)}{}_{i_{27}i_{28}]} -$$

$$- 1{,}441{,}440\, P^{(8)}{}_{[i_1i_2i_3i_4i_5i_6i_7i_8} P^{(6)}{}_{i_9i_{10}i_{11}i_{12}i_{13}i_{14}} P^{(2)}{}_{i_{15}i_{16}} P^{(2)}{}_{i_{17}i_{18}} P^{(2)}{}_{i_{19}i_{20}} P^{(2)}{}_{i_{21}i_{22}} P^{(2)}{}_{i_{23}i_{24}} P^{(2)}{}_{i_{25}i_{26}} P^{(2)}{}_{i_{27}i_{28}]} +$$

$$
\begin{aligned}
&+ 5{,}675{,}670 \, P^{(8)}_{[i_1i_2i_3i_4i_5i_6i_7i_8} P^{(4)}_{i_9i_{10}i_{11}i_{12}} P^{(4)}_{i_{13}i_{14}i_{15}i_{16}} P^{(4)}_{i_{17}i_{18}i_{19}i_{20}} P^{(4)}_{i_{21}i_{22}i_{23}i_{24}} P^{(4)}_{i_{25}i_{26}i_{27}i_{28}]} - \\
&- 28{,}378{,}350 \, P^{(8)}_{[i_1i_2i_3i_4i_5i_6i_7i_8} P^{(4)}_{i_9i_{10}i_{11}i_{12}} P^{(4)}_{i_{13}i_{14}i_{15}i_{16}} P^{(4)}_{i_{17}i_{18}i_{19}i_{20}} P^{(4)}_{i_{21}i_{22}i_{23}i_{24}} P^{(2)}_{i_{25}i_{26}} P^{(2)}_{i_{27}i_{28}]} + \\
&+ 18{,}918{,}900 \, P^{(8)}_{[i_1i_2i_3i_4i_5i_6i_7i_8} P^{(4)}_{i_9i_{10}i_{11}i_{12}} P^{(4)}_{i_{13}i_{14}i_{15}i_{16}} P^{(4)}_{i_{17}i_{18}i_{19}i_{20}} P^{(2)}_{i_{21}i_{22}} P^{(2)}_{i_{23}i_{24}} P^{(2)}_{i_{25}i_{26}} P^{(2)}_{i_{27}i_{28}]} - \\
&- 3{,}783{,}780 \, P^{(8)}_{[i_1i_2i_3i_4i_5i_6i_7i_8} P^{(4)}_{i_9i_{10}i_{11}i_{12}} P^{(4)}_{i_{13}i_{14}i_{15}i_{16}} P^{(2)}_{i_{17}i_{18}} P^{(2)}_{i_{19}i_{20}} P^{(2)}_{i_{21}i_{22}} P^{(2)}_{i_{23}i_{24}} P^{(2)}_{i_{25}i_{26}} P^{(2)}_{i_{27}i_{28}]} + \\
&+ 270{,}270 \, P^{(8)}_{[i_1i_2i_3i_4i_5i_6i_7i_8} P^{(4)}_{i_9i_{10}i_{11}i_{12}} P^{(2)}_{i_{13}i_{14}} P^{(2)}_{i_{15}i_{16}} P^{(2)}_{i_{17}i_{18}} P^{(2)}_{i_{19}i_{20}} P^{(2)}_{i_{21}i_{22}} P^{(2)}_{i_{23}i_{24}} P^{(2)}_{i_{25}i_{26}} P^{(2)}_{i_{27}i_{28}]} - \\
&- 6{,}006 \, P^{(8)}_{[i_1i_2i_3i_4i_5i_6i_7i_8} P^{(2)}_{i_9i_{10}} P^{(2)}_{i_{11}i_{12}} P^{(2)}_{i_{13}i_{14}} P^{(2)}_{i_{15}i_{16}} P^{(2)}_{i_{17}i_{18}} P^{(2)}_{i_{19}i_{20}} P^{(2)}_{i_{21}i_{22}} P^{(2)}_{i_{23}i_{24}} P^{(2)}_{i_{25}i_{26}} P^{(2)}_{i_{27}i_{28}]} - \\
&- 22{,}422{,}400 \, P^{(6)}_{[i_1i_2i_3i_4i_5i_6} P^{(6)}_{i_7i_8i_9i_{10}i_{11}i_{12}} P^{(6)}_{i_{13}i_{14}i_{15}i_{16}i_{17}i_{18}} P^{(6)}_{i_{19}i_{20}i_{21}i_{22}i_{23}i_{24}} P^{(4)}_{i_{25}i_{26}i_{27}i_{28}]} + \\
&+ 22{,}422{,}400 \, P^{(6)}_{[i_1i_2i_3i_4i_5i_6} P^{(6)}_{i_7i_8i_9i_{10}i_{11}i_{12}} P^{(6)}_{i_{13}i_{14}i_{15}i_{16}i_{17}i_{18}} P^{(6)}_{i_{19}i_{20}i_{21}i_{22}i_{23}i_{24}} P^{(2)}_{i_{25}i_{26}} P^{(2)}_{i_{27}i_{28}]} + \\
&+ 67{,}267{,}200 \, P^{(6)}_{[i_1i_2i_3i_4i_5i_6} P^{(6)}_{i_7i_8i_9i_{10}i_{11}i_{12}} P^{(6)}_{i_{13}i_{14}i_{15}i_{16}i_{17}i_{18}} P^{(4)}_{i_{19}i_{20}i_{21}i_{22}} P^{(4)}_{i_{23}i_{24}i_{25}i_{26}} P^{(2)}_{i_{27}i_{28}]} - \\
&- 44{,}844{,}800 \, P^{(6)}_{[i_1i_2i_3i_4i_5i_6} P^{(6)}_{i_7i_8i_9i_{10}i_{11}i_{12}} P^{(6)}_{i_{13}i_{14}i_{15}i_{16}i_{17}i_{18}} P^{(4)}_{i_{19}i_{20}i_{21}i_{22}} P^{(2)}_{i_{23}i_{24}} P^{(2)}_{i_{25}i_{26}} P^{(2)}_{i_{27}i_{28}]} + \\
&+ 4{,}484{,}480 \, P^{(6)}_{[i_1i_2i_3i_4i_5i_6} P^{(6)}_{i_7i_8i_9i_{10}i_{11}i_{12}} P^{(6)}_{i_{13}i_{14}i_{15}i_{16}i_{17}i_{18}} P^{(2)}_{i_{19}i_{20}} P^{(2)}_{i_{21}i_{22}} P^{(2)}_{i_{23}i_{24}} P^{(2)}_{i_{25}i_{26}} P^{(2)}_{i_{27}i_{28}]} + \\
&+ 12{,}612{,}600 \, P^{(6)}_{[i_1i_2i_3i_4i_5i_6} P^{(6)}_{i_7i_8i_9i_{10}i_{11}i_{12}} P^{(4)}_{i_{13}i_{14}i_{15}i_{16}} P^{(4)}_{i_{17}i_{18}i_{19}i_{20}} P^{(4)}_{i_{21}i_{22}i_{23}i_{24}} P^{(4)}_{i_{25}i_{26}i_{27}i_{28}]} - \\
&- 50{,}450{,}400 \, P^{(6)}_{[i_1i_2i_3i_4i_5i_6} P^{(6)}_{i_7i_8i_9i_{10}i_{11}i_{12}} P^{(4)}_{i_{13}i_{14}i_{15}i_{16}} P^{(4)}_{i_{17}i_{18}i_{19}i_{20}} P^{(4)}_{i_{21}i_{22}i_{23}i_{24}} P^{(2)}_{i_{25}i_{26}} P^{(2)}_{i_{27}i_{28}]} + \\
&+ 25{,}225{,}200 \, P^{(6)}_{[i_1i_2i_3i_4i_5i_6} P^{(6)}_{i_7i_8i_9i_{10}i_{11}i_{12}} P^{(4)}_{i_{13}i_{14}i_{15}i_{16}} P^{(4)}_{i_{17}i_{18}i_{19}i_{20}} P^{(2)}_{i_{21}i_{22}} P^{(2)}_{i_{23}i_{24}} P^{(2)}_{i_{25}i_{26}} P^{(2)}_{i_{27}i_{28}]} - \\
&- 3{,}363{,}360 \, P^{(6)}_{[i_1i_2i_3i_4i_5i_6} P^{(6)}_{i_7i_8i_9i_{10}i_{11}i_{12}} P^{(4)}_{i_{13}i_{14}i_{15}i_{16}} P^{(2)}_{i_{17}i_{18}} P^{(2)}_{i_{19}i_{20}} P^{(2)}_{i_{21}i_{22}} P^{(2)}_{i_{23}i_{24}} P^{(2)}_{i_{25}i_{26}} P^{(2)}_{i_{27}i_{28}]} + \\
&+ 120{,}120 \, P^{(6)}_{[i_1i_2i_3i_4i_5i_6} P^{(6)}_{i_7i_8i_9i_{10}i_{11}i_{12}} P^{(2)}_{i_{13}i_{14}} P^{(2)}_{i_{15}i_{16}} P^{(2)}_{i_{17}i_{18}} P^{(2)}_{i_{19}i_{20}} P^{(2)}_{i_{21}i_{22}} P^{(2)}_{i_{23}i_{24}} P^{(2)}_{i_{25}i_{26}} P^{(2)}_{i_{27}i_{28}]} - \\
&- 7{,}567{,}560 \, P^{(6)}_{[i_1i_2i_3i_4i_5i_6} P^{(4)}_{i_7i_8i_9i_{10}} P^{(4)}_{i_{11}i_{12}i_{13}i_{14}} P^{(4)}_{i_{15}i_{16}i_{17}i_{18}} P^{(4)}_{i_{19}i_{20}i_{21}i_{22}} P^{(4)}_{i_{23}i_{24}i_{25}i_{26}} P^{(2)}_{i_{27}i_{28}]} + \\
&+ 12{,}612{,}600 \, P^{(6)}_{[i_1i_2i_3i_4i_5i_6} P^{(4)}_{i_7i_8i_9i_{10}} P^{(4)}_{i_{11}i_{12}i_{13}i_{14}} P^{(4)}_{i_{15}i_{16}i_{17}i_{18}} P^{(4)}_{i_{19}i_{20}i_{21}i_{22}} P^{(2)}_{i_{23}i_{24}} P^{(2)}_{i_{25}i_{26}} P^{(2)}_{i_{27}i_{28}]} - \\
&- 5{,}045{,}040 \, P^{(6)}_{[i_1i_2i_3i_4i_5i_6} P^{(4)}_{i_7i_8i_9i_{10}} P^{(4)}_{i_{11}i_{12}i_{13}i_{14}} P^{(4)}_{i_{15}i_{16}i_{17}i_{18}} P^{(2)}_{i_{19}i_{20}} P^{(2)}_{i_{21}i_{22}} P^{(2)}_{i_{23}i_{24}} P^{(2)}_{i_{25}i_{26}} P^{(2)}_{i_{27}i_{28}]} + \\
&+ 720{,}720 \, P^{(6)}_{[i_1i_2i_3i_4i_5i_6} P^{(4)}_{i_7i_8i_9i_{10}} P^{(4)}_{i_{11}i_{12}i_{13}i_{14}} P^{(2)}_{i_{15}i_{16}} P^{(2)}_{i_{17}i_{18}} P^{(2)}_{i_{19}i_{20}} P^{(2)}_{i_{21}i_{22}} P^{(2)}_{i_{23}i_{24}} P^{(2)}_{i_{25}i_{26}} P^{(2)}_{i_{27}i_{28}]} - \\
&- 40{,}040 \, P^{(6)}_{[i_1i_2i_3i_4i_5i_6} P^{(4)}_{i_7i_8i_9i_{10}} P^{(2)}_{i_{11}i_{12}} P^{(2)}_{i_{13}i_{14}} P^{(2)}_{i_{15}i_{16}} P^{(2)}_{i_{17}i_{18}} P^{(2)}_{i_{19}i_{20}} P^{(2)}_{i_{21}i_{22}} P^{(2)}_{i_{23}i_{24}} P^{(2)}_{i_{25}i_{26}} P^{(2)}_{i_{27}i_{28}]} + \\
&+ 728 \, P^{(6)}_{[i_1i_2i_3i_4i_5i_6} P^{(2)}_{i_7i_8} P^{(2)}_{i_9i_{10}} P^{(2)}_{i_{11}i_{12}} P^{(2)}_{i_{13}i_{14}} P^{(2)}_{i_{15}i_{16}} P^{(2)}_{i_{17}i_{18}} P^{(2)}_{i_{19}i_{20}} P^{(2)}_{i_{21}i_{22}} P^{(2)}_{i_{23}i_{24}} P^{(2)}_{i_{25}i_{26}} P^{(2)}_{i_{27}i_{28}]} - \\
&- 135{,}135 \, P^{(4)}_{[i_1i_2i_3i_4} P^{(4)}_{i_5i_6i_7i_8} P^{(4)}_{i_9i_{10}i_{11}i_{12}} P^{(4)}_{i_{13}i_{14}i_{15}i_{16}} P^{(4)}_{i_{17}i_{18}i_{19}i_{20}} P^{(4)}_{i_{21}i_{22}i_{23}i_{24}} P^{(4)}_{i_{25}i_{26}i_{27}i_{28}]} + \\
&+ 945{,}945 \, P^{(4)}_{[i_1i_2i_3i_4} P^{(4)}_{i_5i_6i_7i_8} P^{(4)}_{i_9i_{10}i_{11}i_{12}} P^{(4)}_{i_{13}i_{14}i_{15}i_{16}} P^{(4)}_{i_{17}i_{18}i_{19}i_{20}} P^{(4)}_{i_{21}i_{22}i_{23}i_{24}} P^{(2)}_{i_{25}i_{26}} P^{(2)}_{i_{27}i_{28}]} - \\
&- 945{,}945 \, P^{(4)}_{[i_1i_2i_3i_4} P^{(4)}_{i_5i_6i_7i_8} P^{(4)}_{i_9i_{10}i_{11}i_{12}} P^{(4)}_{i_{13}i_{14}i_{15}i_{16}} P^{(4)}_{i_{17}i_{18}i_{19}i_{20}} P^{(2)}_{i_{21}i_{22}} P^{(2)}_{i_{23}i_{24}} P^{(2)}_{i_{25}i_{26}} P^{(2)}_{i_{27}i_{28}]} + \\
&+ 315{,}315 \, P^{(4)}_{[i_1i_2i_3i_4} P^{(4)}_{i_5i_6i_7i_8} P^{(4)}_{i_9i_{10}i_{11}i_{12}} P^{(4)}_{i_{13}i_{14}i_{15}i_{16}} P^{(2)}_{i_{17}i_{18}} P^{(2)}_{i_{19}i_{20}} P^{(2)}_{i_{21}i_{22}} P^{(2)}_{i_{23}i_{24}} P^{(2)}_{i_{25}i_{26}} P^{(2)}_{i_{27}i_{28}]} - \\
&- 45{,}045 \, P^{(4)}_{[i_1i_2i_3i_4} P^{(4)}_{i_5i_6i_7i_8} P^{(4)}_{i_9i_{10}i_{11}i_{12}} P^{(2)}_{i_{13}i_{14}} P^{(2)}_{i_{15}i_{16}} P^{(2)}_{i_{17}i_{18}} P^{(2)}_{i_{19}i_{20}} P^{(2)}_{i_{21}i_{22}} P^{(2)}_{i_{23}i_{24}} P^{(2)}_{i_{25}i_{26}} P^{(2)}_{i_{27}i_{28}]} + \\
&+ 3{,}003 \, P^{(4)}_{[i_1i_2i_3i_4} P^{(4)}_{i_5i_6i_7i_8} P^{(2)}_{i_9i_{10}} P^{(2)}_{i_{11}i_{12}} P^{(2)}_{i_{13}i_{14}} P^{(2)}_{i_{15}i_{16}} P^{(2)}_{i_{17}i_{18}} P^{(2)}_{i_{19}i_{20}} P^{(2)}_{i_{21}i_{22}} P^{(2)}_{i_{23}i_{24}} P^{(2)}_{i_{25}i_{26}} P^{(2)}_{i_{27}i_{28}]} - \\
&- 91 \, P^{(4)}_{[i_1i_2i_3i_4} P^{(2)}_{i_5i_6} P^{(2)}_{i_7i_8} P^{(2)}_{i_9i_{10}} P^{(2)}_{i_{11}i_{12}} P^{(2)}_{i_{13}i_{14}} P^{(2)}_{i_{15}i_{16}} P^{(2)}_{i_{17}i_{18}} P^{(2)}_{i_{19}i_{20}} P^{(2)}_{i_{21}i_{22}} P^{(2)}_{i_{23}i_{24}} P^{(2)}_{i_{25}i_{26}} P^{(2)}_{i_{27}i_{28}]} + \\
&+ P^{(2)}_{[i_1i_2} P^{(2)}_{i_3i_4} P^{(2)}_{i_5i_6} P^{(2)}_{i_7i_8} P^{(2)}_{i_9i_{10}} P^{(2)}_{i_{11}i_{12}} P^{(2)}_{i_{13}i_{14}} P^{(2)}_{i_{15}i_{16}} P^{(2)}_{i_{17}i_{18}} P^{(2)}_{i_{19}i_{20}} P^{(2)}_{i_{21}i_{22}} P^{(2)}_{i_{23}i_{24}} P^{(2)}_{i_{25}i_{26}} P^{(2)}_{i_{27}i_{28}]})
\end{aligned}
$$

## Concluding Remark

For a check, note that the magnitudes of the numerical factors in the preceding expressions for $c_{(14)i_1i_2i_3i_4i_5i_6i_7i_8i_9i_{10}i_{11}i_{12}i_{13}i_{14}i_{15}i_{16}i_{17}i_{18}i_{19}i_{20}i_{21}i_{22}i_{23}i_{24}i_{25}i_{26}i_{27}i_{28}}$ in Eq. (9) add up—aside from the respective overall numerical factors—to $14! = 87{,}178{,}291{,}200$.